\documentclass[aps,prx,reprint,superscriptaddress,longbibliography,floatfix,amsmath,amssymb,nofootinbib]{revtex4-2}
\usepackage[utf8]{inputenc}
\usepackage{graphicx}
\usepackage{xcolor}
\usepackage{bm}
\usepackage{booktabs}
\usepackage{enumitem}
\usepackage{placeins}
\usepackage{tikz}
\usetikzlibrary{positioning,fit,arrows.meta}
\usepackage[colorlinks=true,allcolors=blue]{hyperref}
\graphicspath{{figs/}}
\usepackage{titlesec}
\titleformat{\subsubsection}[runin]{\normalfont\itshape}{\thesection.\thesubsection.\thesubsubsection}{0.5em}{}[:]
\titlespacing*{\subsubsection}{0pt}{1.5ex plus .5ex minus .2ex}{0.5em}
\titleformat{\subsection}{\small\bfseries\filcenter}{\thesection.\thesubsection}{0.5em}{}

\newcommand{\bn}{\langle|\mathbf{B}\cdot\hat{\mathbf n}|\rangle/\langle|\mathbf B|\rangle}

\newcommand{\best}[1]{{\boldmath\bfseries #1}}
\begin{document}

\title{Single-Shot Single-Stage Stellarators}

\author{Pedro F. Gil}
\affiliation{William E. Boeing Department of Aeronautics and Astronautics, University of Washington, Seattle, Washington 98195, USA}
\author{Lanke Fu}
\affiliation{Courant Institute of Mathematical Sciences, New York University, New York, New York 10012, USA}
\author{Rogério Jorge}
\affiliation{Department of Physics, University of Wisconsin-Madison, WI 53706, USA}
\author{Alan A. Kaptanoglu}
\affiliation{William E. Boeing Department of Aeronautics and Astronautics, University of Washington, Seattle, Washington 98195, USA}
\affiliation{Courant Institute of Mathematical Sciences, New York University, New York, New York 10012, USA}

\begin{abstract}
In stellarator design, the plasma and the coils are so geometrically complex that they are usually designed separately, so that a high-performance plasma may be impossible to reproduce with coils that can actually be manufactured and assembled. We present a new method to close this gap. In this framework, the plasma shape is optimized while, for every candidate shape, near-optimal coils that satisfy hard engineering limits are found. Without any manually-tuned weights, the method steers the plasma towards shapes that real coils can produce. Moreover, we demonstrate the first single-shot optimization runs starting from tokamak ``cold-starts''.
From a single-shot cold-start, we design a finite-$\beta$ stellarator comparable to reactor-scaled NCSX, as well as a compact aspect-ratio 2.5 and $\langle\beta\rangle=2\%$ quasi-axisymmetric stellarator with one field period and self-consistent bootstrap current. Starting from a boundary derived from W7-X, we design a quasi-isodynamic stellarator with reactor-relevant stability properties. We then use this to warm-start an optimization for a single coil-set solution that produces a high-performance QI stellarator at every pressure value from vacuum to $\langle\beta\rangle=4\%$, addressing directly the ``start-up'' phase of a reactor. The self-consistent free-boundary solutions for these stellarators typically have better coils than a traditional separate-design approach, at comparable or better plasma performance,
illustrating that robust single-stage methods are possible and providing a direct route to the design of reactor-scale stellarators.
\end{abstract}

\maketitle

\section{Introduction}
The stellarator, a toroidal magnetic confinement device, is a promising candidate for achieving controlled nuclear fusion~\cite{spitzer, Helander_2014, Najmabadi01102008, Goodman2025reactor}. Unlike the tokamak, the stellarator relies on a three-dimensional magnetic field produced mostly by external coils, and therefore does not require large plasma currents that can lead to disruptions~\cite{Boozer1998, Helander_2012, Schuller_1995, Hender2007}. This allows for steady-state operation, which is a clear advantage for a future power plant~\cite{Helander_2012, Beidler2021}. However, this advantage does not come for free. A generic three-dimensional field does not confine the plasma well. Without the continuous symmetry of the tokamak, the orbits of trapped particles drift out of the plasma, which leads to large transport losses~\cite{Helander_2014, Nemov1999} and to the loss of the energetic alpha particles born in deuterium--tritium (D--T) fusion reactions before they can heat the plasma~\cite{Bader2019, SIMPLE_albert}. 
The confining field must be carefully shaped, and the coils that produce such a field, which are complicated three-dimensional curves, must also be shaped so that they can be built~\cite{Strickler01032002, Williamson2005, zhu2017new, Wechsung_2022}.

Recent advances in stellarator optimization have shown that the magnetic field can be shaped with remarkable precision. Quasisymmetry (QS), a property of the field strength that confines collisionless trapped particles, can now be achieved to a very high degree across the plasma volume~\cite{LandremanPaul, Landreman_2022}. Similarly, quasi-isodynamic (QI) configurations, in which the contours of the field strength close poloidally, are being pursued both in research and in industry because of their low bootstrap current and good fast-particle confinement~\cite{goodman_prx, Dudt_Goodman_Conlin_Panici_Kolemen_2024, LION2025114868}. These results are, however, obtained for the plasma alone. Whether a feasible set of coils can reproduce such a field is a separate question.

\label{sec:stage1_stage2_background}
Traditionally, stellarator design is subdivided into two stages. In stage I, the plasma boundary is optimized for physics properties by solving the magnetohydrodynamic (MHD) equations of force balance in fixed boundary~\cite{hirshman_steepestdescent_1983-1, LandremanPaul}. In stage II, coils are found that reproduce this boundary~\cite{Merkel_1987, Landreman_2017, landreman_simsopt_2021}. This two-stage procedure has been used for essentially all stellarators built to date, including Wendelstein 7-X~\cite{w7x_scientific_basis, KLINGER2013461}, Wendelstein 7-AS~\cite{Sapper1990}, HSX~\cite{anderson_almagri_anderson_matthews_talmadge_shohet_1995,Almagri1999}, LHD~\cite{Iiyoshi1999}. The two stages are coupled only through the target plasma boundary. A boundary that is excellent for the plasma properties may be hard to produce with coils of limited length, curvature and clearance, while small, physics-neutral changes of the boundary might make the coil problem much easier. Moreover, the coil problem is an ill-posed magnetostatic inverse problem, and most of its solutions are impractical from an engineering standpoint~\cite{stellarator_intro}. The consequences can be severe. Coil complexity and tight tolerances were responsible for most of the cost growth that led to the cancellation of NCSX, and for long delays in the construction of Wendelstein 7-X~\cite{ncsx, KLINGER2013461, Bosch_2017}.

Robust joint optimization (single-stage) methods therefore have the capacity to dramatically improve stellarator design. 
Single-stage methods optimize the plasma boundary and the coils simultaneously. Historically, they have often required ``warm-starts'' (initializing from a plasma and associated coils that are already highly similar to the desired design point), heavy computation, and the assumption of vacuum stellarators~\cite{henneberg_single_stage, giuliani_single_stage, Jorge_2023, Jorge_2024, smiet_single_stage, Gil_stochastic_2026}. Single-stage optimization is increasingly used to design real experiments. The Columbia
Stellarator eXperiment (CSX) and Electrons and Positrons in an Optimized Stellarator (EPOS) devices were designed with both the fixed-boundary and the Boozer-surface
single-stage methods~\cite{Giuliani2022boozer}, with constraints on the strain of its
superconducting
tape~\cite{Baillod2025csx,Baillod2026csx,Huslage2024strain, epos_2026}. Moreover, for some of these methods, the coil degrees of freedom and the plasma boundary degrees of freedom are \emph{both} optimized over a single cost function, in which the squared flux, the coil engineering penalties and the plasma targets are balanced by weights that are usually tuned by hand. As many as 15--20 such weights are typically used, and there is no a priori knowledge of their optimal values since the class of ``good'' weights often depend strongly on the particular plasma design. The constrained formulation of the coil problem with an augmented Lagrangian method automatically tunes these weights~\cite{auglag_prl, auglag_pre}, avoiding manual weight selection and tuning, and coil sets obtained in this way typically outperform published coils in both field accuracy and engineering metrics.

A second line of work keeps the two stages separate and steers the boundary optimization towards shapes that are easier to produce with coils, by adding a proxy of coil complexity to stage I. Such proxies are computed either from a current potential on a winding surface, as in REGCOIL, whose current density or coil--plasma distance can be targeted in fixed-boundary optimization~\cite{Landreman_2017, Drevlak2019rose} and differentiated with respect to the boundary~\cite{Hudson2018, CarltonJones2021}, or from the plasma field alone. In the latter case, the shortest scale length of the gradient of $\mathbf B$ on the boundary correlates with the achievable coil--plasma distance~\cite{Kappel2024, Kappel2026}, and the geodesic torsion of the field lines with the non-planarity of the coils~\cite{Pavone2026}. These proxies are cheap and have been adopted in design studies, but they are correlations with the coils and give no direct link to them. A boundary with a favourable proxy is not guaranteed to be produced with a good fit by filamentary coils that satisfy given limits on length, curvature and clearance, and the coils must still be found, and the fit checked, in a second stage.

The quasi-single-stage (QSS) method directly optimizes only the plasma degrees of freedom, by treating the coil optimization as a subproblem that can be deterministically solved and differentiated with respect to the plasma boundary. Yu \emph{et al.} introduced the QSS method originally for designing permanent-magnet stellarators~\cite{Yu2024pm}, and recently extended it for designing modular coils~\cite{Yu2026qss}. Coil feasibility enters the boundary optimization through a surrogate; the maximum normalized normal field on a winding surface~\cite{Landreman_2017} at a uniform offset from the plasma. Fu \emph{et al.} built a differentiable proxy on the QUADCOIL winding-surface method~\cite{Fu_2025, Fu2026proxy}, which estimates the coil complexity and its gradient with respect to the boundary under constraints on curvature, force and coil spacing. 

In both cases, the quantity that enters the plasma optimization is evaluated on a current sheet or an offset surface rather than with a more realistic set of filamentary coils. This means that the coils still have to be designed afterwards, and the self-consistent match with the plasma design, as well as the coil engineering limits, are known only then.
However, they have a major advantage which we adopt here; the coil degrees of freedom, i.e. the coil shapes and coil currents, never become variables of the optimization. For every candidate boundary, they are the solution of a coil problem in which the engineering requirements are imposed as hard constraints. 

\section{Contributions of this work}
\label{sec:contributions}

In this manuscript, we introduce a novel QSS formulation where the plasma is optimized and a full-scale filamentary coil optimization is solved as a deterministic subproblem. 
The main contributions of this work are the following advances:
\begin{itemize}[leftmargin=*]
\item \textbf{Single-stage optimization of the plasma and filamentary coils at finite plasma pressure.} The plasma boundary is optimized while the coils remain (locally) optimal under hard engineering limits for every candidate boundary. 
\item \textbf{Single-shot performance from simple starting points.} Whether the optimization begins from an axisymmetric circular torus with circular coils (hereafter referred to as the ``cold-start''), from a published optimized configuration (a warm-start), or from a completely different stellarator variety, our QSS method seems to robustly find precise quasi-axisymmetric, quasi-helical and quasi-isodynamic designs, despite no prior knowledge or biasing of the final shape. As far as we are aware, finding high-performance stellarators together with their coils from such simple axisymmetric starts, or from a design of a different symmetry class, has not been demonstrated before in single-stage optimization. 
\item \textbf{Automatic weight tuning.} Traditionally all of the 15--20 objective weights in the optimization problem are manually tuned to achieve engineering tolerances on the coils. It requires substantial effort and many runs to obtain optimal solutions this way. In contrast, in our work all of the weights for the plasma and coil objectives are handled automatically by our QSS algorithm to achieve the prescribed hard engineering limits. 
\item \textbf{Speed-up.} A complete QSS design of a vacuum configuration from a cold-start takes about one GPU-hour. Finite-pressure and reactor-relevant designs run in continuation stages for about 5 to 10 GPU-hours. Appendix~\ref{sec:speedup} compares them with the cost of other methods.
\item \textbf{Extensive validation.} Every design is validated after the optimization. We report all the validation metrics from high-resolution converged free-boundary equilibria computed with the DESC code~\cite{conlin2024high}, and from guiding-center alpha-particle orbits from a reactor D--T birth profile in equilibria that resolve the ripple of the discrete coils.
\item \textbf{Finding high-performance finite-$\beta$ reactor-scale stellarators in single-shot runs.} We present two reactor-relevant designs. The first is a quasi-axisymmetric design at $\langle\beta\rangle=4\%$ that is comparable to NCSX at reactor scale. It is obtained in a single run from a cold-start and uses 18 modular coils, whereas the NCSX design needs 78 coils for a converged free boundary solution. The second is a quasi-isodynamic design obtained in a single run that is warm-started from a W7-X-derived boundary. Its plasma is then designed together with a single set of coils that can produce a reactor-relevant plasma at every pressure from vacuum to $\langle\beta\rangle=4\%$, with only the coil currents changing. Both designs have plasma and coil metrics comparable to existing published designs and real-world stellarator experiments. 
\end{itemize}

\begin{table*}[!t]
\caption{Capabilities of single-stage optimization methods. AD denotes automatic differentiation and FD finite differences.}
\label{tab:capabilities}
\begin{ruledtabular}
\begin{tabular}{lccccc}
 & Jorge \emph{et al.}~\cite{Jorge_2023} & Giuliani \emph{et al.}~\cite{Giuliani2023direct} & Yu \emph{et al.}~\cite{Yu2026qss} & Fu \emph{et al.}~\cite{Fu2026proxy} & This work \\
\hline
Coils in the optimization loop & filaments & filaments & current sheet & current sheet & filaments \\
\hline
Hard engineering constraints & $\times$ & $\times$ & $\times$ & limited& $\checkmark$ \\
\hline
Free of hand-tuned weights & $\times$ & $\times$ & $\times$ & $\times$ & $\checkmark$ \\
\hline
Gradient of the coil term & FD & adjoint & not reported & adjoint, AD & envelope \\
\hline
Finite plasma pressure & $\times$ & $\times$ & $\times$ & $\checkmark$& $\checkmark$ \\
\hline
Fast-particle confinement & $\times$ & $\times$ & $\times$ & $\times$ & alpha losses \\
\hline
Starting points & \begin{tabular}[t]{@{}c@{}}optimized\\equilibrium \&  coils\end{tabular} & \begin{tabular}[t]{@{}c@{}}optimized equilibrium\\or near-axis design\end{tabular} & rotating ellipse & \begin{tabular}[t]{@{}c@{}}optimized\\equilibrium\end{tabular} & \begin{tabular}[t]{@{}c@{}}axisymmetric torus\\or published design\end{tabular} \\
\end{tabular}
\end{ruledtabular}
\end{table*}

We compare the method in Table~\ref{tab:capabilities} with the four single-stage approaches closest to it. Our work greatly benefited from adapting ideas from this literature, and their work has allowed us to identify and address the remaining gaps. For instance, all four previous methods: are restricted to vacuum fields, do not perform post-optimization free-boundary equilibria or fast-particle orbits validation, and balance physics and coils with weights that must be tuned by hand. In the single-stage method of Jorge \emph{et al.}, the engineering limits of the filamentary coils are quadratic penalties and the equilibrium gradients are computed by finite differences. Giuliani \emph{et al.} represent the plasma by Boozer surfaces of the vacuum field of the coils, with adjoint gradients, but also impose the coil limits as quadratic penalties. In the methods of Yu \emph{et al.} and Fu \emph{et al.}, the plasma optimization sees only a current sheet on a winding surface, and the filamentary coils are found afterwards, in a separate stage at fixed plasma shape, which can introduce errors that the plasma optimization never sees.

Our method, in contrast, keeps the filamentary coils in the optimization loop under hard engineering constraints, needs no hand-tuned weights, obtains the gradient of the coil term from the envelope theorem, works at finite plasma pressure, and works from simple axisymmetric starting configurations. This work then validates the coil sets with high-resolution free-boundary DESC equilibria and alpha-particle orbits.

The article is organized as follows. Section~\ref{sec:methods} describes the nested optimization problem, the coil solver, the gradient that links plasma and coils, the equilibrium and the outer optimizer. Section~\ref{sec:cold} presents the cold-started designs, Sec.~\ref{sec:ncsx78} the NCSX-like design, and Sec.~\ref{sec:reactor} the reactor-like quasi-isodynamic design, its trade-offs and the single coil set for the pressure ramp. The appendices give the constraints, the warm-started designs from the quasi-axisymmetric configuration of Landreman and Paul~\cite{LandremanPaul}, called LP QA below, the gradient derivations and their Taylor tests, the free-boundary and alpha-particle calculations, and the benchmarks against single-stage optimization~\cite{Jorge_2023} and augmented Lagrangian coils~\cite{auglag_prl, auglag_pre}.

\section{Optimization Method}
\label{sec:methods}
We now describe the formulation and algorithmic solution of our optimization problem.

\subsection{Nested optimization formulation}
\label{sec:nested}
\subsubsection{The physics problem} Its variable is the plasma boundary, often referred to as the last closed flux surface (LCFS), which is the boundary between the region of nested magnetic surfaces and that of ergodic field lines. It is represented by a stellarator-symmetric Fourier series in cylindrical coordinates,
\begin{equation}
\begin{aligned}
R(\theta,\phi)&=\sum_{m,n}r_{mn}\cos(m\theta-nN_{\mathrm{fp}}\phi),\\
Z(\theta,\phi)&=\sum_{m,n}z_{mn}\sin(m\theta-nN_{\mathrm{fp}}\phi),
\end{aligned}
\end{equation}
where $\theta$ is the poloidal angle, $\phi$ the toroidal angle and $N_{\mathrm{fp}}$ the number of field periods. The free coefficients $r_{mn}$ and $z_{mn}$ form the boundary vector $\mathbf s\in\mathbb R^{n_s}$, with $r_{00}$ fixed to set the major radius.
From these degrees of freedom, we define the plasma optimization problem as
\begin{equation}
\begin{aligned}
\min_{\mathbf s}\;&f(\mathbf s)\\
\text{subject to}\;&\mathbf h(\mathbf s)\le 0,
\quad J_f^\star(\mathbf s)\le\epsilon .
\end{aligned}
\label{eq:outer}
\end{equation}
Here $f$ is the physics objective. It is the quasisymmetry (QS) error $f_{\mathrm{QS}}$ of Refs.~\cite{Rodríguez_Paul_Bhattacharjee_2022, LandremanPaul} for the quasi-axisymmetric and quasi-helically symmetric configurations, and the quasi-isodynamicity (QI) error $f_{\mathrm{QI}}$, which follows the construction of Goodman \emph{et al.}~\cite{Goodman_Camacho}, for the quasi-isodynamic ones. Both are defined in Appendix~\ref{app:errors}, together with the other physics constraints used in this work. The vector $\mathbf h$ collects the other physics constraints, each written as $h_i(\mathbf s)\le0$. Typical physics constraints, all defined in Appendix~\ref{app:errors}, are an upper bound on the aspect ratio $A$, a target value of the mean rotational transform $\bar\iota$, imposed as two inequalities, or bands on the rotational transform on axis and at the edge, an upper bound on the maximum elongation $E$ of the cross-sections, an upper bound on the mirror ratio $\Delta$ for the quasi-isodynamic configurations, and a lower bound on the magnetic well $W$. The last constraint bounds $J_f^\star(\mathbf s)$, the squared flux of the locally optimal coils for the boundary $\mathbf s$, which is obtained from the coil problem described next.

\subsubsection{The engineering problem} 
\label{sec:the_engineering_problem}
Its variables are the shapes and currents of the three-dimensional electromagnetic coils, which are filamentary curves, each described by a truncated Fourier series in Cartesian coordinates,
\begin{align}
x^i(t)=\sum_{l=0}^{M}x^i_{c,l}\cos(lt)+\sum_{l=1}^{M}x^i_{s,l}\sin(lt),
\end{align}
and similarly for $y^i$ and $z^i$, with the Fourier order $M=5$ for every design of this paper except the reactor-like design of Sec.~\ref{sec:reactor}, which uses $M=8$. Note that this is lower order than most published coil solutions; this can easily be extended to higher order but we find high-performance solutions in this work regardless. We also observe empirically that keeping the Fourier order $M$ relatively low reduces the amount of coil ripple in the final solution, with the downstream consequence that free boundary calculations tend to converge at lower resolutions. For instance, the published $M=16$ QUASR coils that we compare with in Sec.~\ref{sec:qa_nfp1}, the free boundary calculation and $\alpha$-losses do not converge until very high boundary resolution, around $n=m=28$. The resolution effects are severe; the first calculation at $n=m=16$ reported $20\%$ $\alpha$-losses, and at $n=m=28$ the $\alpha$-loss has converged down to $4.5\%$.

Each half field period carries $N_c$ base coils. The Fourier coefficients of the base coils and their currents form the coil vector $\mathbf c$, and the full set of $2N_{\mathrm{fp}}N_c$ coils is generated by the field-period symmetry and stellarator symmetry. The total current of the base coils is fixed before the optimization begins, so that only $N_c-1$ currents are degrees of freedom and no equality constraint is needed.

\begin{table}[!t]
\caption{Engineering constraints of the coil problem, imposed as hard constraints without margins. Here $\mathbf r_i(t)$ is the centre line of coil $i$ and $\kappa=|\mathbf r_i'\times\mathbf r_i''|/|\mathbf r_i'|^3$ its curvature. Throughout the paper, $\kappa_{\max}$, $\kappa_\mathrm{MSC}$, $\bar\tau$ and $\gamma_c$ are reported as their maximum over all coils, $L_i$ as its range over the coils, and $d_{cc}$ and $d_{cs}$ are minima by definition. The weighted torsion and the nonconvexity are not often used within optimization, but we report them for comparison. The last row is not a constraint but a gauge term added to the objective with a weight of $10^{-3}$.}
\label{tab:coil_limits}
\begin{ruledtabular}
\begin{tabular}{p{0.28\columnwidth}p{0.58\columnwidth}}
Constraint & Definition \\
\hline
Length $L_i$ [m] & $L_i=\oint|\mathbf r_i'(t)|\,dt$, the length of each coil, which sets the amount of conductor. The total per half field period is $L_\mathrm{total}=\sum_{i=1}^{N_c}L_i$. \\
Maximum curvature $\kappa_{\max}$  [m$^{-1}$] & $\kappa_{\max,i}=\max_t\kappa_i(t)$, the inverse of the tightest bending radius along a coil, limited by how far a superconducting cable can be bent \\
Mean-squared curvature $\kappa_\mathrm{MSC}$  [m$^{-2}$] & $\kappa_{\mathrm{MSC},i}=L_i^{-1}\oint\kappa_i^2\,d\ell$, which penalizes coils that wind back and forth even where no single bend is sharp. \\
Coil--coil distance $d_{cc}$  [m] & $d_{cc}=\min_{i\ne j,\,t,t'}|\mathbf r_i(t)-\mathbf r_j(t')|$, room for the winding packs and their support structure, and an indirect limit on the forces between neighbouring coils. \\
Coil--surface distance $d_{cs}$  [m] & $d_{cs}=\min_{i,\,t,\,\mathbf x\in\Gamma}|\mathbf r_i(t)-\mathbf x|$ over the plasma boundary $\Gamma$, room for the vacuum vessel, blanket and shielding. \\
Weighted torsion $\bar\tau$ [m$^{-1}$] & $\bar\tau_i=\big(\oint\kappa_i^2\tau_i^2\,d\ell/\oint\kappa_i^2\,d\ell\big)^{1/2}$, the curvature-weighted root-mean-square torsion $\tau_i$ of a coil, which measures how far the coil leaves a plane where it is strongly bent. \\
Nonconvexity $\gamma_c$  [m$^{-2}$] & $\gamma_{c,i}=L_i^{-1}\oint\max(0,-\sigma_i k_{s,i})^2\,d\ell$, with $k_{s,i}$ the curvature of the coil projected on its best-fit plane and $\sigma_i$ the sign of its mean, which measures the inward bends of a coil. \\
\hline
Arclength gauge & $R_{\mathrm{arc}}=\frac{1}{N_c N_q}\sum_{i,q}\big(|\mathbf r_i'(t_q)|/\bar\ell_i-1\big)^2$, with $\bar\ell_i$ the mean of $|\mathbf r_i'|$ over the $N_q$ quadrature points of each of the $N_c$ base coils, the normalized variance of the arclength along each coil. It fixes the parametrization of the coils, which does not change their shape. \\
\end{tabular}
\end{ruledtabular}
\end{table}

For a given boundary, $J_f^\star$ is the result of the coil problem
\begin{equation}
J_f^\star(\mathbf s)=\min_{\mathbf c}\; J_f(\mathbf c,\mathbf s)\quad\text{subject to}\quad \mathbf g(\mathbf c,\mathbf s)\le 0,
\label{eq:inner}
\end{equation}
where $J_f$ is the local squared flux. It is the surface integral of the squared normal field over the boundary, normalized by the field strength,
\begin{align}
\label{eq:squared_flux_continuous_def}
J_f=\frac{1}{2}\int_\Gamma\frac{(\mathbf B\cdot\hat{\mathbf n})^2}{|\mathbf B|^2}\,d\Gamma.
\end{align}
The normal component of the magnetic field must vanish on the plasma boundary $\Gamma$, and no finite set of coils can achieve this exactly. This single scalar can then encode the coil information into the plasma optimization, without the coil degrees of freedom entering it.
The integral is evaluated by quadrature on a uniform angular grid over the unique half-field-period of the boundary. Here $\mathbf B$ is the field of the coils computed with the Biot--Savart law and $\hat{\mathbf n}$ the unit normal of the plasma boundary. 
The vector $\mathbf g$ collects the engineering limits defined in Table~\ref{tab:coil_limits}, which ensure that the coils can be manufactured and installed. We call $J_f^\star(\mathbf s)$ the optimal squared flux of the boundary $\mathbf s$ because it is the squared flux of locally optimal coils that satisfy all the engineering limits for this plasma.

\subsubsection{The coupling} 
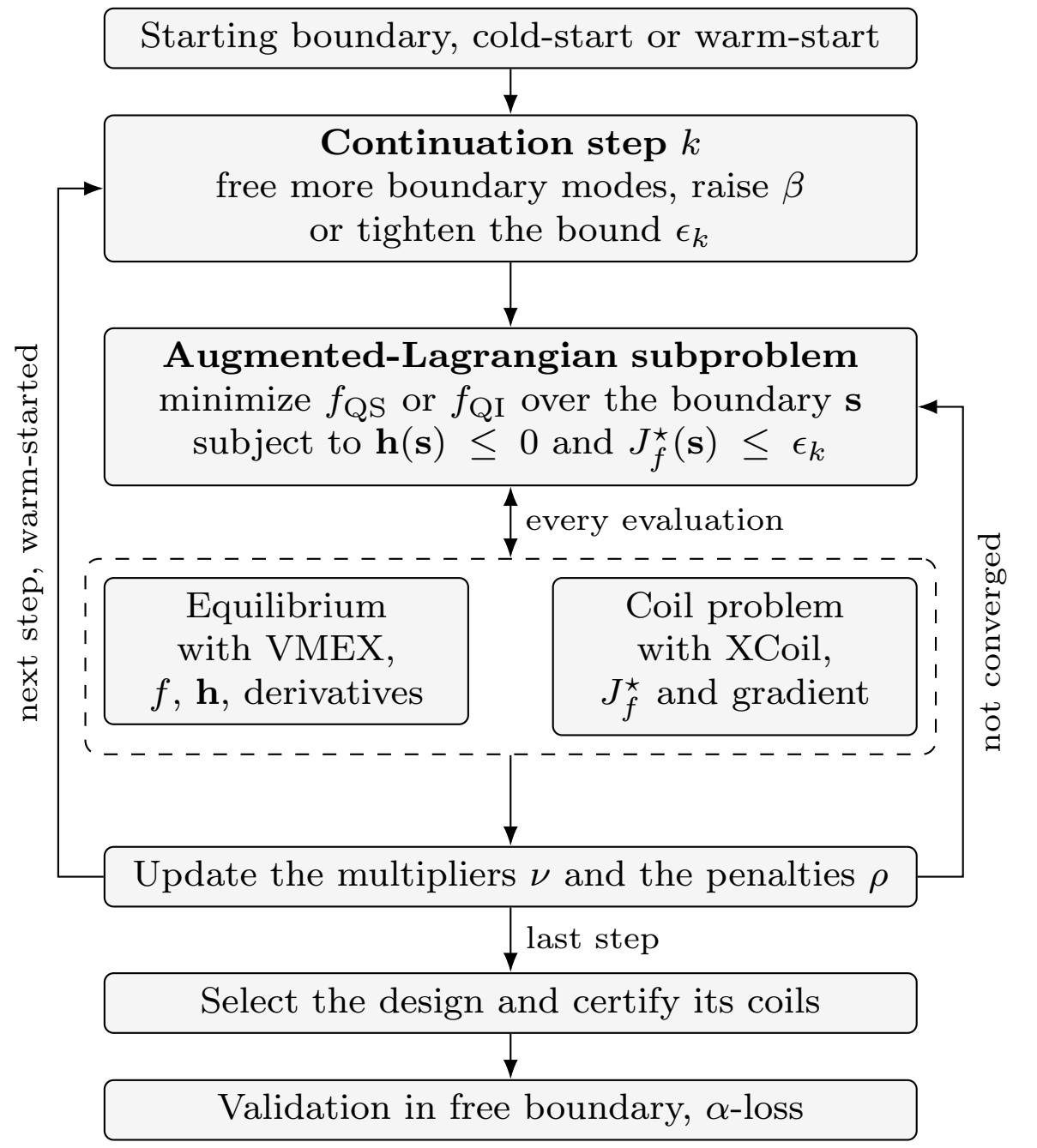
\begin{figure}[!t]
\centering
\begin{tikzpicture}[font=\footnotesize, >={Latex[length=1.5mm]},
  box/.style={draw, rounded corners=2pt, align=center, inner sep=3pt, fill=black!4, text width=#1},
  box/.default=0.70\columnwidth]
\node[box] (start) {Starting boundary, cold-start or warm-start};
\node[box, below=3.5mm of start] (cont) {\textbf{Continuation step $k$}\\free more boundary modes, raise $\beta$\\or tighten the bound $\epsilon_k$};
\node[box, below=5mm of cont] (opt) {\textbf{Augmented-Lagrangian subproblem}\\minimize $f_{\mathrm{QS}}$ or $f_{\mathrm{QI}}$ over the boundary $\mathbf s$\\subject to $\mathbf h(\mathbf s)\le0$ and $J_f^\star(\mathbf s)\le\epsilon_k$};
\node[box=0.30\columnwidth, below=7mm of opt.south west, anchor=north west] (eq) {Equilibrium\\with VMEX,\\$f$, $\mathbf h$, derivatives};
\node[box=0.30\columnwidth, below=7mm of opt.south east, anchor=north east] (coil) {Coil problem\\with XCoil,\\$J_f^\star$ and gradient};
\node[draw, dashed, rounded corners=3pt, fit=(eq)(coil), inner sep=4pt] (ev) {};
\node[box, below=7mm of ev.south, anchor=north] (upd) {Update the multipliers $\nu$ and the penalties $\rho$};
\node[box, below=5mm of upd] (sel) {Select the design and certify its coils};
\node[box, below=3.5mm of sel] (val) {Validation in free boundary, $\alpha$-loss};
\draw[->] (start) -- (cont);
\draw[->] (cont) -- (opt);
\draw[<->] (opt.south) -- node[right, font=\scriptsize] {every evaluation} (ev.north);
\draw[->] (ev.south) -- (upd.north);
\draw[->] (upd.east) -- ++(3.5mm,0) |- node[pos=0.25, below, sloped, font=\scriptsize] {not converged} (opt.east);
\draw[->] (upd.west) -- ++(-3.5mm,0) |- node[pos=0.25, above, sloped, font=\scriptsize] {next step, warm-started} (cont.west);
\draw[->] (upd) -- node[right, font=\scriptsize] {last step} (sel);
\draw[->] (sel) -- (val);
\end{tikzpicture}
\caption{Structure of the quasi-single-stage method. Every function evaluation of a subproblem, the dashed box, solves a VMEX equilibrium and the coil problem of Eq.~\eqref{eq:inner}, and each continuation step is warm-started from the previous one.}
\label{fig:flowchart}
\end{figure}

The bound $\epsilon$ in Eq.~\eqref{eq:outer} couples the two problems, and Fig.~\ref{fig:flowchart} summarizes how the resulting nested algorithm is organized. In the two-stage approach, Eq.~\eqref{eq:outer} is solved without the flux constraint and Eq.~\eqref{eq:inner} is solved once, on the resulting boundary. In the QSS approach, the optimal squared flux is evaluated by solving Eq.~\eqref{eq:inner} at every function evaluation of the plasma problem. The plasma optimizer therefore sees the squared flux of coils that are (locally) optimal for the current boundary, rather than that of a fixed coil set fitted to an earlier one. The bound is a constraint rather than a weighted objective so that it represents the normal field the designer is willing to accept.

\subsection{The inner coil problem}
\label{sec:methods_coils}
We begin with the innermost problem, the coil problem of Eq.~\eqref{eq:inner}, which is solved at every function evaluation of the plasma problem. In vacuum, $J_f$ is invariant to the scale of the currents, so the choice of the current magnitude does not affect the optimization. At finite pressure, the plasma currents produce their own field $\mathbf B_{\mathrm{pl}}$, which we compute on the boundary from the converged VMEX state by virtual casing~\cite{VMEX}. The plasma field enters the coil problem divided by $\lambda_G=I_G/I$, the ratio of the net poloidal current of the equilibrium, $I_G=2\pi|G|/\mu_0$ at the LCFS with $G$ the poloidal current function of Appendix~\ref{app:alpha}, to the summed current $I$ of all coils, which puts the fixed total coil current on the scale of the equilibrium. The coil residual then becomes $(\mathbf B_{\mathrm{coil}}+\sigma\mathbf B_{\mathrm{pl}})\cdot\hat{\mathbf n}/|\mathbf B_{\mathrm{coil}}+\sigma\mathbf B_{\mathrm{pl}}|$, where the sign $\sigma=\mathrm{sign}\langle\mathbf B_{\mathrm{coil}}\cdot\mathbf B_{\mathrm{pl}}\rangle$ orients the plasma field with the coils.

The coil problem is solved by XCoil, a new coil solver written in JAX and built for speed, that will be the subject of a separate forthcoming publication. On a single NVIDIA H200 GPU, a coil solve typically takes $\sim0.2$~s in vacuum and $\sim 1$~s at finite pressure, less than the equilibrium solve that runs in parallel with it. It uses an augmented Lagrangian method in the spirit of Ref.~\cite{auglag_prl} so that the objective weights are automatically tuned to achieve the hard constraint limits.

\subsubsection{Initialization} Since the coil problem must be solved thousands of times during a plasma optimization, the way in which it is started matters. A cold solve starts from surface-following coils, obtained by offsetting the boundary along its normal and scaling them to 85\% of the length limit, with up to two fallback starts (tilted, shorter or clearance-limited offsets). For the cold-starts from axisymmetric circular tokamaks, this produces circular axisymmetric coils. Along the plasma optimization, the coils, multipliers and penalty of the previous solve are reused as a warm-start whenever the boundary moves by less than $10^{-5}$ in every coefficient, and otherwise the solve starts cold. 

\subsubsection{Coil certification}
The coil metrics reported in this paper come from a final, more thorough validation stage of the coils. We test if the final coil solution obtained is path-dependent or not, by resolving a cold-start stage-two coil optimization with tighter tolerances, until the certified violation is below $10^{-5}$, and the coil set with the lowest Lagrangian value is kept. 
The normal field of these coils is evaluated on grids over the full torus that are refined until two successive grids agree to $10^{-3}$. It turns out that our solutions are not path-dependent; the certified coil sets also agree with the in-loop optimization coil sets to within 1\% of the constraints and targets, meaning that certified and in-loop coils consistently lie at the same local minimum.

Essentially every final coil solution presented in this work satisfies all of the prescribed constraints and targets to very high accuracy; each maximum normalized constraint violation is within $10^{-6}$, and often closer to $10^{-9}$. The one exception is the shared coil set of Sec.~\ref{sec:startup}, whose maximum curvature only exceeds its limit by 0.7\%.

The coil solve provides the value of $J_f^\star$ for the current boundary. To use a gradient-based optimizer on Eq.~\eqref{eq:outer}, we need the derivative of the optimal squared flux with respect to the plasma boundary.
The derivation of a very efficient gradient using the envelope theorem is given in Appendix~\ref{app:envelope}.
The vacuum and finite-$\beta$ gradients are extensively verified by Taylor tests in Appendix~\ref{app:taylor_tests}.

\subsection{Equilibrium and its derivatives}
\label{sec:equilibrium}
Both the physical quantities of the plasma problem and, at finite pressure, the plasma magnetic field contribution, come from the equilibrium of the given plasma boundary. An ideal magnetohydrodynamic (MHD) equilibrium is the state in which the plasma pressure gradient is balanced by the magnetic force, $\nabla p=\mathbf J\times\mathbf B$, inside the given boundary. It determines the nested magnetic surfaces and the field strength on them, and therefore every physics quantity of the plasma problem, such as the quasisymmetry error or the rotational transform. In a fixed-boundary equilibrium, the shape of the outermost surface is prescribed and the equilibrium is solved inside it. Throughout this paper, the magnetic surfaces are labelled by the normalized toroidal flux $s=\psi/\psi_{\mathrm{LCFS}}$, where $2\pi\psi$ is the toroidal flux enclosed by a surface, so that $s=0$ on the magnetic axis and $s=1$ on the LCFS. 

\subsubsection{Equilibrium settings}
The optimization uses fixed-boundary equilibria, and the final designs are validated with free-boundary ones (Sec.~\ref{sec:validation}). Fixed-boundary equilibria are computed with VMEX, a JAX reimplementation of VMEC~\cite{hirshman_steepestdescent_1983-1, VMEX}, with the settings of VMEC2000 given in Appendix~\ref{app:vmex_settings}.

Together with the gradient $\nabla_\mathbf{s}J_f^*$ from Appendix~\ref{app:envelope}, this provides every value and derivative that Eq.~\eqref{eq:outer} requires.

\subsection{Outer optimizer}
\label{sec:outer}
We now describe how the plasma problem is solved for a fixed bound $\epsilon$. It is solved with a Powell--Hestenes--Rockafellar (PHR) augmented Lagrangian method~\cite{NocedalWright2006}. The physics objective is a sum of squares, $f(\mathbf s)=\|\mathbf q(\mathbf s)\|^2$. The residual vector $\mathbf q$ collects the residuals of the QS error, or of the QI error of Eq.~\eqref{eq:qi}, on all evaluation surfaces. The physics constraints are $h_i(\mathbf s)\le0$ for $i=1,\dots,n_h$, as in Eq.~\eqref{eq:outer}. The flux constraint is written as $c_f(\mathbf s)=S\,[J_f^\star(\mathbf s)-\epsilon]\le0$, where the scale $S$ brings the flux row to the size of the other rows and does not change the feasible set. Since $J_f^\star$ can only be kept small by boundaries that coils within the engineering limits can produce, the flux constraint also acts as a very strong regularization of the outer problem, which keeps the boundary away from shapes that no admissible coils can reproduce. The outer multipliers are $\nu_i\ge0$ and $\nu_c\ge0$, and the penalties are $\rho_i>0$ and $\rho_c>0$. For fixed multipliers and penalties, each subproblem is the nonlinear least-squares problem
\begin{equation}
\begin{aligned}
\min_{\mathbf s}\;&\|\mathbf q(\mathbf s)\|^2+\sum_{i=1}^{n_h}\frac{1}{2\rho_i}\max\big(0,\,\nu_i+\rho_i h_i(\mathbf s)\big)^2\\
&+\frac{1}{2\rho_c}\max\big(0,\,\nu_c+\rho_c c_f(\mathbf s)\big)^2 ,
\end{aligned}
\label{eq:phr}
\end{equation}
which is the subproblem of Fig.~\ref{fig:flowchart}. It is solved by a trust-region reflective least-squares method~\cite{BranchColemanLi1999, Virtanen2020SciPy} with at most 30 function evaluations per outer iteration for all designs presented here. 
After each subproblem, the multipliers are updated as $\nu_i\leftarrow\min(\max(\nu_i+\rho_i h_i,0),10^3)$, and likewise for $\nu_c$. A penalty $\rho_i$ is doubled, up to $10^4$, only if its violation exceeds $10^{-4}$ and a quarter of the previous violation, which keeps the subproblems well conditioned. After every outer iteration, the feasibility, complementarity and stationarity of the current point are measured, with the multipliers of the active constraints fitted by bounded least squares. Each function evaluation of the subproblem therefore requires one VMEX equilibrium with its exact Jacobian and one coil solve, as shown by the dashed box in Fig.~\ref{fig:flowchart}. We call a point certified when its feasibility is below $10^{-4}$ and its relative stationarity below 0.1. 
This optimizer solves the plasma problem for one value of $\epsilon$ and one boundary resolution. The continuation that links successive problems is described next.

We will also make comparisons throughout this manuscript against a baseline that performs the traditional two-stage version of the optimization problem, using the same optimizer without the flux row. 

\subsection{Continuation step}
\label{sec:homotopy}
It is well known that the performance of stellarator optimization can be sensitive to the scaling of the optimization variables, the Fourier resolution of the plasma boundary or coil shapes, and the self-consistency requirements of the plasma and coils. 
Following this literature, we solve the plasma problem along a sequence of problems, each warm-started from the previous one with its multipliers (Fig.~\ref{fig:flowchart}). In a \emph{flux-bound continuation}, the boundary resolution is fixed and the bound of Eq.~\eqref{eq:outer} is tightened, $\epsilon_k=r^k J_{f,0}$, where $k$ is the index of the continuation point, $r<1$ the ratio between successive bounds, 0.85 by default, and $J_{f,0}$ the optimal squared flux of the starting boundary, as evaluated by the in-loop coil solve. Each point runs six outer iterations in vacuum and four at finite pressure, and the continuation stops when the flux target is missed twice in a row or when $\rho_c$ reaches its cap. Continuation with small steps appears to make a significant difference. We found that single runs that jump directly to a tight bound can end at stationary points with 3.5 times higher QS error. In a \emph{boundary-mode continuation}, the maximum poloidal and toroidal mode numbers of the free coefficients of the boundary grow instead, $m,|n|\le1,2,\dots$, with the flux bound set at the start of each step, typically to half of the optimal squared flux of the current boundary, so that the large-scale shape of the plasma is settled before finer details are free to change. This is a common technique~\cite{LandremanPaul, jang2025exponential,panici2023desc}. Similarly to the DESC code, we occasionally use \emph{pressure continuation}, where a sequence of problems are solved at increasing $\beta$. 

The flux-bound continuation suits a start that is already a good design, such as a published one. It traces the trade-off between physics and squared flux from that point and gives a curve of designs to choose from (Appendix~\ref{sec:protocol}). The boundary-mode and pressure continuations suit cold-starts that are far from a good design or large changes of shape, since each step stays close to the previous solution and the coils follow the boundary as it develops.

\subsection{Validation}
\label{sec:validation}
Every selected design is checked independently of the optimization. The mean normal field $\bn$ of its coils is evaluated on the converged grid of the final coil stage. Since a low normal field on the target boundary does not guarantee that the coils reproduce the target plasma, we also compute the free-boundary equilibrium of every coil set, at a resolution that resolves the ripple of the discrete coils. We use the free-boundary solver of DESC~\cite{desc_code, conlin2024high} with a radial resolution $l_\rho=16$ and maximum poloidal and toroidal mode numbers $m_\theta=n_\phi=16$, with the profiles of the design and, at finite pressure, the coil currents scaled by $\lambda_G$. At this resolution, these calculations resolve the ripple of the discrete coils up to $n=16$ per field period. We chose DESC free boundary calculations because they converge robustly in free boundary at these resolutions, which we have not observed with other solvers such as VMEC. To support that claim, we report very low free boundary residuals for all results in this work in Table~\ref{tab:desc}.

All the final plasma properties of this paper are evaluated directly from these free-boundary calculations. These are the quasisymmetry or QI error, $f_{\mathrm{QS}}$ or $f_{\mathrm{QI}}$, the rotational transform $\iota$ on axis and at the edge, the aspect ratio $A$, the maximum elongation $E$ of the cross-sections, the mirror ratio $\Delta$ on the innermost flux surface, the magnetic well $W$, the normalized Mercier criterion $\hat D_\mathrm{Merc}$, which is positive for stability, the $\alpha$-loss, which is the fraction of fusion-born alphas lost within 0.2~s, and the mean and maximum distance $\delta_\mathrm{LCFS}$ between the free-boundary LCFS and the target. Their definitions are given in Appendix~\ref{app:errors}. The existing stellarator designs that we compare with throughout this work (HSX, W7-X, NCSX, etc.) are evaluated in exactly the same way. We recompute free-boundary equilibria from their own coil sets, and we do not quote published free-boundary values. Every plasma quantity quoted in this paper, in the text and in the tables, is therefore the value of the converged free-boundary equilibrium of the certified coils, unless it is explicitly labelled fixed-boundary. The $\alpha$-loss is obtained from collisionless guiding-center orbits of fusion-born alphas with a reactor D--T birth profile, followed for 0.2~s in the free-boundary equilibrium, as described in Appendix~\ref{app:alpha}.

\subsubsection{Scaling}
All configurations in this work, including the published designs, are scaled to the size of the ARIES-CS reactor~\cite{Najmabadi01102008}, with a minor radius of 1.704~m and a volume-averaged field of 5.865~T. All dimensional quantities in this paper refer to this size.

\section{Cold Starts and Robustness}
\label{sec:cold}
\begin{figure*}[!t]
\includegraphics[width=0.96\textwidth]{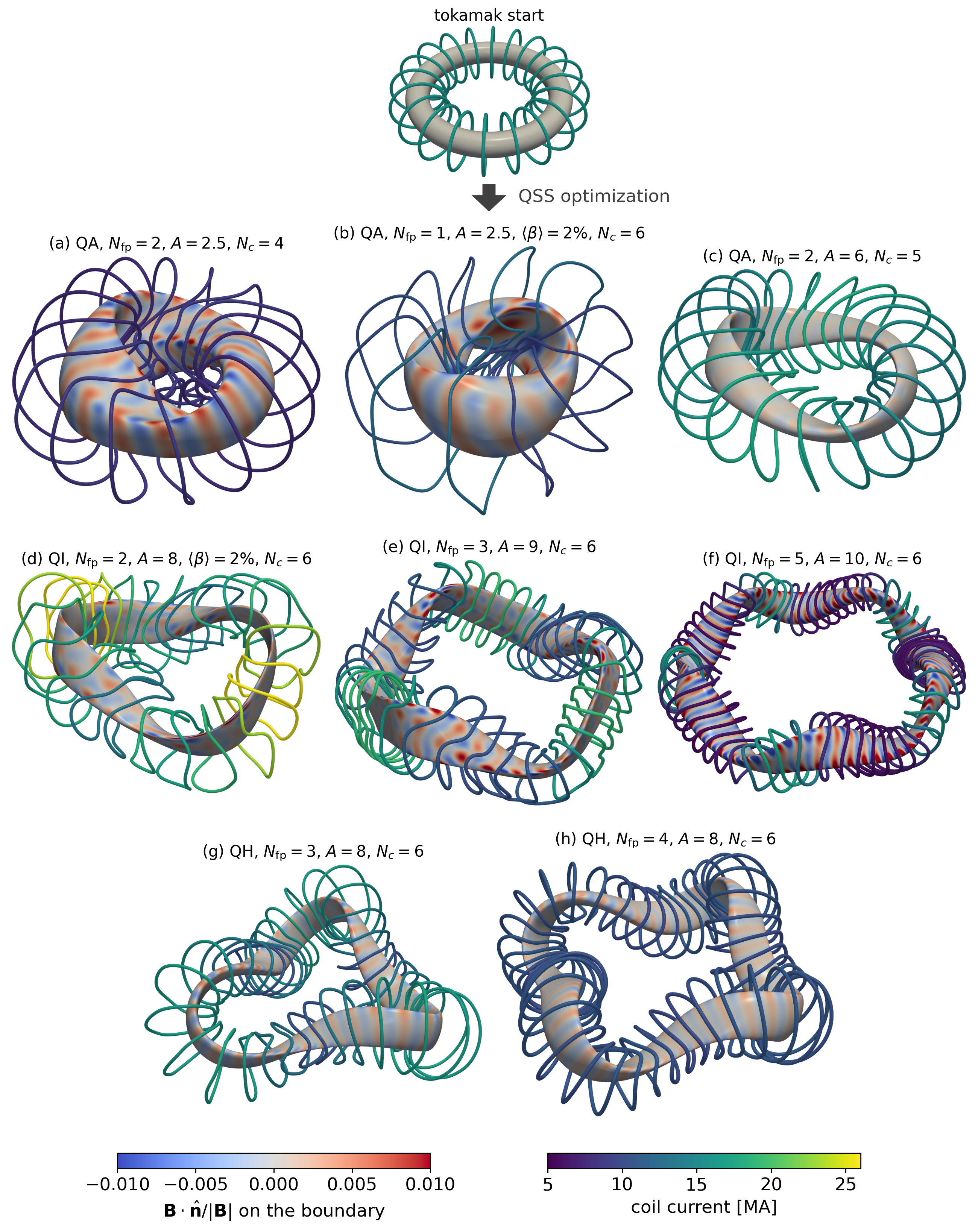}
\caption{Stellarators optimized from a cold-start, a circular torus with circular coils, from which every design below is obtained. Panels (a)--(h) show the designs at ARIES-CS scale with their certified coils. The normal field includes the plasma field in (b) and (d). Both color scales are common to all panels. (a) Quasi-axisymmetric, $N_{\mathrm{fp}}=2$, $A=2.5$, $N_c=4$. (b) Quasi-axisymmetric, $N_{\mathrm{fp}}=1$, $A=2.5$, $\langle\beta\rangle=2\%$ with self-consistent bootstrap current, $N_c=6$. (c) Quasi-axisymmetric, $N_{\mathrm{fp}}=2$, $N_c=5$. (d) Quasi-isodynamic, $N_{\mathrm{fp}}=2$, $\langle\beta\rangle=2\%$, $N_c=6$. (e) Quasi-isodynamic, $N_{\mathrm{fp}}=3$, $N_c=6$. (f) Quasi-isodynamic, $N_{\mathrm{fp}}=5$, $N_c=6$. (g) Quasi-helically symmetric, $N_{\mathrm{fp}}=3$, $N_c=6$. (h) Quasi-helically symmetric, $N_{\mathrm{fp}}=4$, $N_c=6$.}
\label{fig:cold}
\end{figure*}

This completes the description of the method. We now apply it to designs that initialize from a cold-start in order to show the flexibility and the robustness of the method. Designs warm-started from a published quasi-axisymmetric configuration are given in Appendix~\ref{app:warm}. The method is \emph{flexible} because it reaches very diverse targets, across symmetry classes, aspect ratios, coil counts and plasma pressures, and \emph{robust}, because it reaches them from very general starts. Fig.~\ref{fig:cold} shows the designs, and Table~\ref{tab:coils_all} gives their coils, free-boundary errors, normal fields and alpha losses. They cover the three classes of omnigenous fields, quasi-axisymmetry, quasi-helical symmetry and quasi-isodynamicity, with one to five field periods, aspect ratios from 2.5 to 10 and $N_c$ from four to six, some in vacuum, some at finite pressure. Every run starts from a circular torus with a small symmetry-breaking perturbation and circular coils. None of them needs a well-designed initial shape, tuned weights or a separate coil stage, and every one ends with feasible coils. We culminate the demonstration of cold-start performance by running a continuous optimization from a cold state through all three symmetry classes back to the start.

\subsection{Quasi-axisymmetric stellarators}
\label{sec:qa_cold}
\subsubsection{Low aspect ratio}
\label{sec:lowA_tokamak}
Design (a) of Fig.~\ref{fig:cold}, at $A=2.5$ with $N_c=4$ is one of very few $N_\text{fp}=2$ quasi-axisymmetric configuration at such a low aspect ratio with certified modular coils. Compact quasi-axisymmetric plasmas in this range have mostly been explored for the plasma alone, because it is very challenging to fit any coils through the center. The exception is that modular coils at this aspect ratio were first found for the
MHH2 family of the ARIES-CS study, which has two field periods, $A\approx2.6$ and 16
modular coils at $\langle\beta\rangle=5\%$ with a
bootstrap-like current~\cite{Garabedian2005,Ku2005mhh2}. Unfortunately, we could not find the original files and verify these calculations directly. More recently, compact
quasi-axisymmetric plasmas with two to five field periods were optimized at
$A\approx2.4$~\cite{schuett2024exploring}. In these, the plasma current provides most of the
rotational transform, the three-dimensional shaping adds an external transform of about
0.05 to 0.2, and the field is produced by tokamak coils complemented by banana
coils~\cite{henneberg2024compact,Zettl2026} or by modular coils~\cite{schuett2024exploring}.
Design (a) carries no current, and its rotational transform of about 0.4 comes entirely from the three-dimensional shaping.

Design (a) has two field periods and starts from a cold-start with an elongation of 1.3, since a circular seed at this aspect ratio has no converged VMEX equilibrium. Its physics constraints and the limits of its four base coils are given in Table~\ref{tab:coils_all}. Note that these coil limits would not work at reactor-scale (or only shielding, no neutron-absorbing blanket, would be used in the inboard side), possibly only for an experiment scale stellarator as HSX which has similar spacings at 1.2~m major radius~\cite{auglag_pre}. The design meets all constraints with certified coils. In free-boundary, the quasisymmetry error rises several-fold, while the LCFS stays close to its target. Table~\ref{tab:coils_all} gives its normal field, quasisymmetry error and alpha loss. At this aspect ratio, four coils per half field period leave a ripple that matters for fast particles. Nonetheless, this is surprisingly good solution for such a challenging low-aspect-ratio regime; the alpha losses are tolerable at $8\%$, the quasisymmetry is reasonably good, and despite being very short, the coils match the plasma boundary well.

\subsubsection{One field period with self-consistent bootstrap current}
Design (b)\label{sec:qa_nfp1} of Fig.~\ref{fig:cold} extends the compact quasi-axisymmetric designs to one field period and to a finite-pressure plasma that is optimized with a self-consistent bootstrap current. This is probably the most demanding geometry for modular coils among the designs of this paper. The design has six base coils per half field period and an aspect ratio of 2.5, with the constraints given in Table~\ref{tab:qa_nfp1}. The final boundary of design (b) exhibits an average vacuum rotational transform of 0.24, and the bootstrap current contributes rotational transform of up to 0.14 at the edge.

The design is reached in a single-shot optimization from a cold-start, but broken down into five continuation stages. It is first optimized in vacuum from a cold-start. Its pressure is then raised to $\langle\beta\rangle=1\%$ and 2\%, with $p\propto1-s$ and no net toroidal current. The third stage improves this solution by raising the limits on the coil--coil and coil--surface distances, while the bound on the magnetic well is raised in three steps and the curvature limit is halved, to the values given in Table~\ref{tab:qa_nfp1}. The fourth stage adds weak penalties on the torsion and the nonconvexity of the coils to the coil objective.

A plasma at this pressure carries a bootstrap current, which in a compact quasi-axisymmetric configuration changes the rotational transform substantially. The fifth stage adds it. The bootstrap current is computed with the drift-kinetic solver yancc~\cite{yancc}, which gives $\langle\mathbf J\cdot\mathbf B\rangle$ on eight flux surfaces. The density and temperature profiles are those of Ref.~\cite{LandremanBullerDrevlak2022}, scaled to keep the volume-averaged $\beta$ of the design near 2\%. The bootstrap current calculation is iterated with fixed-boundary solves until it changes by less than 2\% between rounds. 

Table~\ref{tab:qa_nfp1} lists the plasma and coil properties of the design after this pass. 
As far as we are aware, a one-field-period quasi-axisymmetric stellarator at finite pressure with a self-consistent bootstrap current and modular coils has not been reported. We compare in the table to the most similar solution we could find: an aspect ratio 2.86 vacuum QA stellarator from QUASR~\cite{Giuliani_2024}. The QUASR design has a lower quasisymmetry error, a lower normal field and lower alpha losses. Our design exhibits a lower aspect ratio, finite pressure with a bootstrap current, a magnetic well, and its coils are shorter and of a much lower Fourier order. In our experiments, we found that optimizing for Mercier stability tended to dramatically reduce the quasi-symmetry, so we did not directly target it in the end. The most similar finite-$\langle\beta\rangle$ device in the literature where we could find a plasma-coil solution is the $N_{\mathrm{fp}} = 2$ device at aspect ratio 3.4~\cite{henneberg2019properties} using 8 coils per half-field-period (32 modular coils) with an additional four poloidal field coils and still exhibiting substantially larger $\bn$ errors than we report here, with comparable alpha losses. Henneberg \emph{et al.} have a number of other promising finite-$\langle\beta\rangle$ $N_{\mathrm{fp}} > 1$ plasma-coil designs but we have omitted them here because we were unable to find all the final coil metrics to compare with~\cite{henneberg2024compact,schuett2024exploring}. 

Here, and elsewhere, we make no claims of optimality for our solutions; the emphasis here is that XCoil allows us to quickly find plausible single-shot solutions even in very challenging parameter spaces, e.g. at very low aspect ratio, finite-pressure, and self-consistent bootstrap current.

\begin{table*}[!t]
\caption{Reactor-scale quasi-axisymmetric design (b) of Fig.~\ref{fig:cold} compared with the vacuum configuration 134070 of QUASR~\cite{Giuliani_2024} and its coils. The constraint column gives the limits of design (b) used in its final stage. Throughout this paper, bold marks the clearly better value of a comparison. Values within a few percent of each other, or within twice the sampling error of the alpha losses, are not bolded. As discussed in Sec.~\ref{sec:validation}, the equilibrium of QUASR 134070 uses 28 poloidal and toroidal modes to resolve the ripple of its coils.}
\label{tab:qa_nfp1}
\begin{ruledtabular}
\begin{tabular}{lccc}
 & constraint & design (b) & QUASR 134070 \\
\hline
$A$ / $E$ & 2.5 / $\le6$ & 2.50 / 4.25 & 2.86 / 3.80 \\
$\langle\beta\rangle$ [\%] & & 2.06 & 0 \\
bootstrap current [MA] & & $-3.46$ & 0 \\
$\iota(0)$ / $\iota(1)$ & $0.27\le\bar\iota\le0.33$ & 0.24 / 0.39 & 0.31 / 0.29 \\
$W$ [\%] & $\ge1$ & \best{4.4} & $-14.5$ \\
$\hat D_\mathrm{Merc}$ [$10^{-3}$] & & $-100$ & \best{$-1.5$} \\
$f_{\mathrm{QS}}$ [$10^{-3}$] & & 1.3 & \best{0.18} \\
$\delta_\mathrm{LCFS}$ mean / max [mm] & & 6.1 / 24 & \best{1.0} / \best{3.9} \\
$\alpha$-loss [\%] & & $7.1\pm0.3$ & \best{$4.5\pm0.2$} \\
\hline
$N_c$ / total coils / Fourier order $M$ & & 6 / 12 / 5 & 6 / 12 / 16 \\
$L_i$ [m] & $\le35.7$ & \best{32.2--35.6} & 39.5--44.1 \\
$L_\mathrm{total}$ [m] & & \best{206.3} & 247.7 \\
$\kappa_{\max}$ / $\kappa_\mathrm{MSC}$ [m$^{-1}$ / m$^{-2}$] & $\le0.89$ / $\le0.51$ & 0.89 / 0.16 & \best{0.75} / \best{0.11} \\
$\bar\tau$ / $\gamma_c$ [m$^{-1}$ / $10^{-3}$~m$^{-2}$] & & \best{0.48} / 6.0 & 0.61 / \best{4.4} \\
$d_{cc}$ / $d_{cs}$ [m] & $\ge0.49$ / $\ge1.28$ & 0.49 / 1.29 & 0.48 / \best{1.42} \\
$\bn$ [\%] & & 0.17 & \best{0.053} \\
\end{tabular}
\end{ruledtabular}
\end{table*}

\subsubsection{Landreman and Paul QA from a cold-start}
Design (c)\label{sec:cold_start_lpqa} of Fig.~\ref{fig:cold} repeats the vacuum LP QA problem of Appendix~\ref{sec:lpqa} from a cold-start, to test whether the method works without the prior knowledge of a published design, and we compare it in detail with a two-stage design from the same start. It starts from a circular tokamak with circular coils with $A=6$ and the toroidal flux of LP QA, with the same physics constraints and $N_c=5$. 
In less than two GPU-hours, the optimization reaches a low normal field, with all physics constraints met and a low fixed-boundary quasisymmetry error. The QSS method thus leads from a cold-start to a LP-QA-like design with excellent coils. In the free-boundary equilibrium, its LCFS lies close to the target, and it loses only slightly more alphas than the QSS design warm-started from LP QA in Appendix~\ref{sec:lpqa}, which has a similar quasisymmetry error and a similar $\bn$. In fact, the two LCFS lie on average 1.5\% of the minor radius apart, and their Fourier coefficients differ by $4\%$. The coils agree as well. They lie on average 0.2\% of their length apart, and their currents differ by 0.5\% on average. QSS is essentially producing the same stellarator regardless of a warm-start or a cold-start. 

The same start using stage-1 then stage-2 ends in a clearly worse solution. With the same seed, mode schedule and physics constraints, but without the coil terms, the boundary reaches a much lower fixed-boundary quasisymmetry error, but its certified coils leave a larger normal field, which leads to a free-boundary quasisymmetry error an order of magnitude worse than that of the QSS solution. In fact, the LCFS deviates far from the target, and it loses more than twice as many alphas as the QSS design. Its plasma is essentially that of LP QA, and it loses as many alphas as the LP QA two-stage reference. Table~\ref{tab:qa_all} of Appendix~\ref{sec:lpqa} lists the plasma and coil properties of these two designs together with those of the warm-started designs of Appendix~\ref{sec:lpqa}. 

\begin{table*}[!t]
\caption{Quasi-helically symmetric designs (g) and (h) of Fig.~\ref{fig:cold} compared
with LP QH~\cite{LandremanPaul} with the coils of Wiedman \emph{et
al.}~\cite{Wiedman_Buller_Landreman_2024} and with the HSX design from Geiger et
al.~\cite{Geiger2026HSXdata} at reactor-scale. The LP QH coils are those published with
Ref.~\cite{Wiedman_Buller_Landreman_2024}. We compute all plasma and coil metrics below
from the published coil and plasma files, with the defaults of
Sec.~\ref{sec:validation}.
$^\dagger$Reference~\cite{Wiedman_Buller_Landreman_2024} reports 0.56 in these units,
but the value degrades when the coil ripple is resolved. $^\ddagger$The fixed-boundary
VMEC equilibrium of the HSX team gives $9.6\times10^{4}$ in these units at converged
resolution ($m_\theta=18$, $n_\phi=24$).}
\label{tab:qh_cmp}
\begin{ruledtabular}
\footnotesize
\begin{tabular}{lccccc}
 & constraint (g)/(h) & (g) QH & (h) QH & LP QH, Wiedman & HSX \\
\hline
design & & cold-start QSS & cold-start QSS & two-stage & two-stage \\
$N_{\mathrm{fp}}$ & & 3 & 4 & 4 & 4 \\
$A$ / $\bar\iota$ / $E$ & 8 / 0.93, 1.24 / $\le3.42$ & 8.0 / 0.93 / 3.42 &
  8.0 / 1.24 / 3.39 & 8.0 / 1.24 / 3.42 & 9.6 / 1.06 / 3.60 \\
$\iota(0)$ / $\iota(1)$ & & 0.92 / 0.94 & 1.24 / 1.24 & 1.24 / 1.24 & 1.05 / 1.09 \\
$W$ [\%] & $\ge-1.4$ & $-1.4$ & $-1.3$ & $-1.4$ & \best{0.5} \\
\hline
$N_c$ / total coils / Fourier order $M$ & & 6 / \best{36} / 5 & 6 / 48 / 5 & \best{5} / 40 / 7 &
  6 / 48 / 16 \\
$L_i$ [m] & & 28.7--35.2 & 25.8--27.8 & 35.1--36.3 & 27.7--31.4 \\
$L_\mathrm{total}$ [m] & $\le196.0$/162.0 & 196.0 & \best{162.0} & 177.7 & 181.2 \\
$\kappa_{\max}$ / $\kappa_\mathrm{MSC}$ [m$^{-1}$ / m$^{-2}$] &
  $\le$0.95/0.12, 0.94/0.18 & 0.95 / 0.12 & \best{0.79} / 0.098 & 0.82 / \best{0.079} &
  0.91 / 0.25 \\
$\bar\tau$ / $\gamma_c$ [m$^{-1}$ / $10^{-3}$~m$^{-2}$] & -- / 0 for (h) & 1.1 / 1.4 &
  \best{0.56} / \best{0.00} & 1.2 / 4.3 & 5.3 / 2.3 \\
$d_{cc}$ / $d_{cs}$ [m] & $\ge$0.68/1.37, 0.69/1.38 & \best{1.41} / 1.87 & 0.85 / 1.57 &
  1.08 / 1.61 & 1.25 / 1.90 \\
$\bn$ [\%] & & 0.15 & 0.12 & \best{0.056} & 0.56 \\
\hline
fixed-boundary $f_{\mathrm{QS}}$ [$10^{-5}$] & & 33 & 5.3 & \best{2.7} &
  $3.6\times10^{4}{}^\ddagger$\\
$f_{\mathrm{QS}}$ [$10^{-3}$] & & 33 & 27 & \best{3.0}${}^\dagger$ & 280 \\
$\delta_\mathrm{LCFS}$ mean / max [mm] & & 2.6 / 11 & \best{2.0} / \best{5.8} &
  2.4 / 6.1 & 20 / 83 \\
$\alpha$-loss [\%] & & 4.0 & 0.09 & 0.01 & 17 \\
\end{tabular}
\end{ruledtabular}
\end{table*}

\begin{figure*}[!t]
\includegraphics[width=0.9\textwidth]{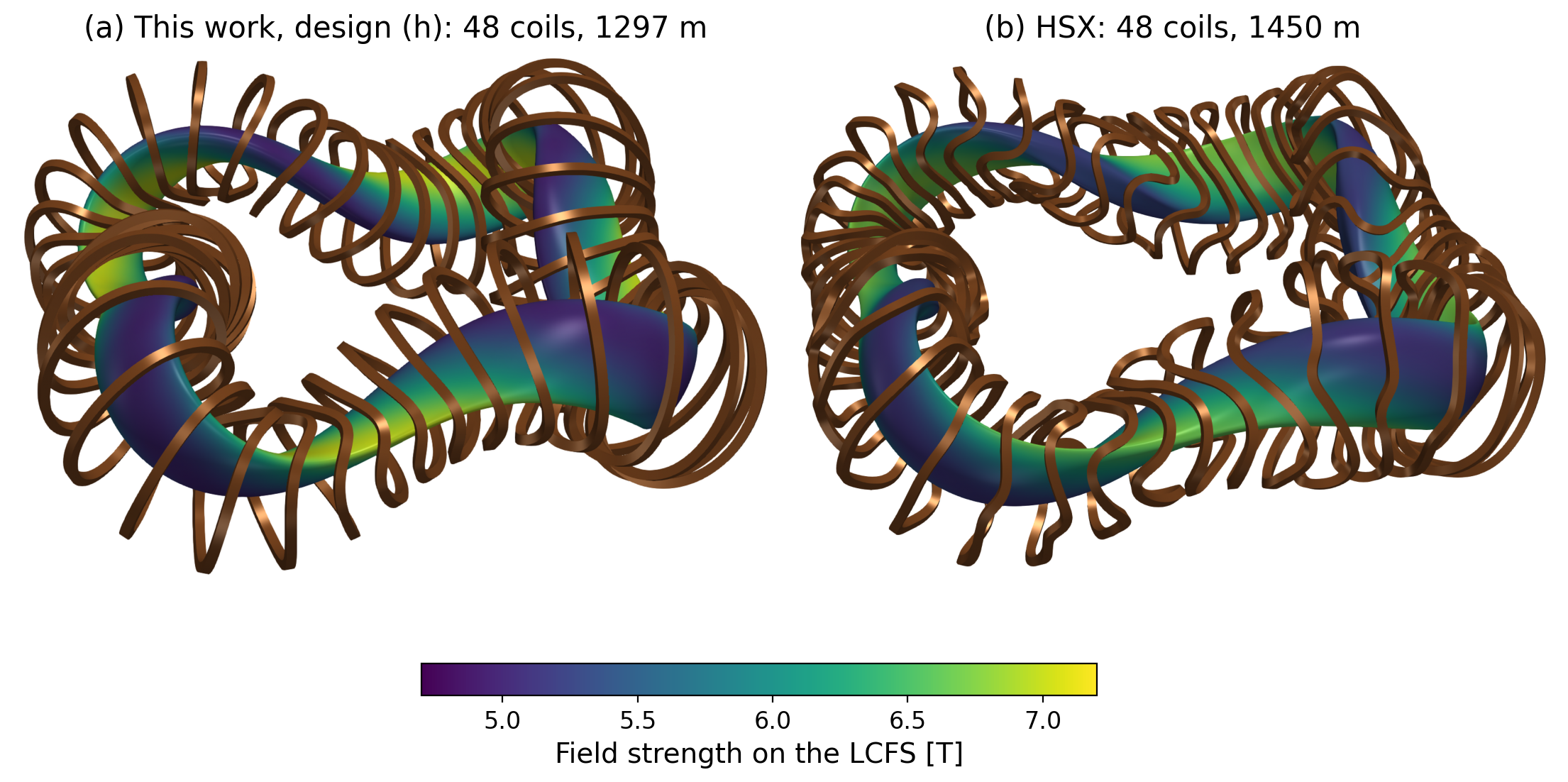}
\caption{(a) The quasi-helically symmetric design (h) of Sec. IV B, with its 48 certified, fully convex modular coils as finite-build winding packs of 0.47 × 0.47 m². (b) HSX with its 48 modular coils \cite{anderson_almagri_anderson_matthews_talmadge_shohet_1995}. Both are scaled to the minor radius and volume-averaged field of ARIES-CS. The coil lengths in the titles are totals over all coils, eight times the values per half field period in Table IV.}
\label{fig:h_vs_hsx}
\end{figure*}

\subsection{Quasi-helically symmetric stellarators}
\label{sec:qh_cold}
Designs (g) and (h) are quasi-helically symmetric, with three and four field periods, $A=8$ and $N_c=6$. Both reach low fixed-boundary quasisymmetry errors with coils that look remarkably simple for quasi-helical symmetry (Fig.~\ref{fig:cold}). Their free-boundary quasisymmetry errors, given in Table~\ref{tab:qh_cmp} with their normal fields, are far above the fixed-boundary values, even though the free-boundary LCFS stays close to its target. Yet design (h) exhibits practically perfect alpha confinement, and was obtained from a cold-start. Extremely low normal fields and free-boundary quasisymmetry errors are therefore not necessary for good fast-particle confinement. Coil ripple raises the alpha loss of design (g) to a few percent. This design, however, uses fewer coils than (h), given its three field periods.

Table~\ref{tab:qh_cmp} compares the two designs with the precise quasi-helically symmetric configuration of Landreman and Paul~\cite{LandremanPaul}, here called LP QH, with the coils of Wiedman \emph{et al.}~\cite{Wiedman_Buller_Landreman_2024}, and with HSX~\cite{anderson_almagri_anderson_matthews_talmadge_shohet_1995}. Design (h) was optimized with the aspect ratio, mean rotational transform, elongation and magnetic well of LP QH, but initializing from a cold-start. It also asks for convex coil shapes, it is shown in Fig. \ref{fig:h_vs_hsx} next to HSX. It reaches a similar fixed-boundary quasisymmetry error using one more coil per half field period but exhibiting shorter coils and less total coil length. Its coils lie closer to each other and to the plasma, and its $\bn$ is about twice that of the Wiedman coils.
In the free-boundary equilibria of the coils, which resolve the ripple, the quasisymmetry error of the Wiedman coils is an order of magnitude smaller than that of (h). Both coil sets of Wiedman and (h) nevertheless confine nearly all alphas. While (h) exhibits a one-coil difference and improved coil metrics compared with the Wiedman solution, discussed below, the plasma design is very similar to the original LP QH stellarator. 

Compared with HSX at the same minor radius, both (g) and (h) designs have a lower $\bn$, about half the mean-squared curvature, less torsion, and exhibit a similar maximum curvature. HSX keeps its coils farther from the plasma. In the free-boundary equilibrium of its coils, HSX has a quasisymmetry error an order of magnitude above those of (g) and (h), its LCFS is displaced several times further, and it loses far more alphas. HSX is an experiment designed in the 1990s at small scale, with access requirements and without alpha confinement at reactor scale as a goal, so this comparison shows what the QSS method reaches and is no criticism of the HSX design. Lastly, we want to emphasize that the coils of (h) are fully convex. It is noteworthy that there is a $N_\text{fp} = 4$ QH coil solution at moderate aspect ratio that can exhibit convex coils. The other four-coil QH designs we compared with here exhibit substantial nonconvexity and at least twice as much torsion. 

\subsection{Quasi-isodynamic stellarators}
\label{sec:qi_cold}
Designs (d)--(f) are quasi-isodynamic, with two, three and five field periods and aspect ratios of 8, 9 and 10, all with $N_c=6$. With two field periods, design (d) is optimized at $\langle\beta\rangle=2\%$ from a cold-start, at $A=8$, with the pressure raised along the continuation. From the third continuation step on, it also carries the maximum-$J$ constraint~\cite{Proll2012} of Appendix~\ref{app:errors}, with a ramped bound, combined with the closure of the $J$ contours at mid-well depths\footnote{We add it later in the optimization because activating maximum-$J$ from the first continuation step can produce a gradient that points somewhere unhelpful or strongly violates the constraints given that a circular tokamak with a few Fourier modes available has little room to satisfy it. We usually do the same for similar terms like $J$-closure and Mercier stability.}.
Design (d) stands out because its free-boundary equilibrium has a magnetic well and exhibits the max-$J$ property, although it exhibits about $9\%$ $\alpha$-loss.
The loss became moderate only with the larger aspect ratio and the lower mirror ratio, which reduce the fraction of deeply trapped alphas, together with the ramped maximum-$J$ constraint, which at $A=8$ also lowers the fixed-boundary QI error by more than half. Design (e), with three field periods at $A=9$, reaches a low free-boundary QI error. Like design (h), it confines practically every alpha particle. Table~\ref{tab:coils_all} gives the QI errors and alpha losses of designs (d)--(f).

A prime example of why the coupled method is needed is design (f), with five field periods. Its physics constraints are those of the precise quasi-isodynamic configurations of Ref.~\cite{goodman_prx}, here with its own bound on the mirror ratio. 
At five field periods, the circular seed of the other designs happens to steer the optimization towards a suboptimal branch for QI, a symptom of local optimization. A seed with a small rotating ellipse and helical axis shift avoids it; the change is nearly imperceptible. With this seed, the method finds a quasi-isodynamic design whose fixed boundary has a low QI error and a magnetic hill, with certified coils. In free-boundary, its QI error rises, and it loses few of its alphas.

With the same seed, schedule and physics constraints but without the coil terms, the boundary reaches a lower fixed-boundary QI error. However,
the two-stage alternative fails at this length, finding no certified coils, and its free-boundary equilibrium does not converge at the resolution of our other designs. In a two-stage design, nothing prevents the plasma optimization from reaching a boundary that no coils within the engineering limits can produce, and this is found only afterwards, in the coil stage. 

\begin{figure*}[!t]
\includegraphics[width=0.9\textwidth]{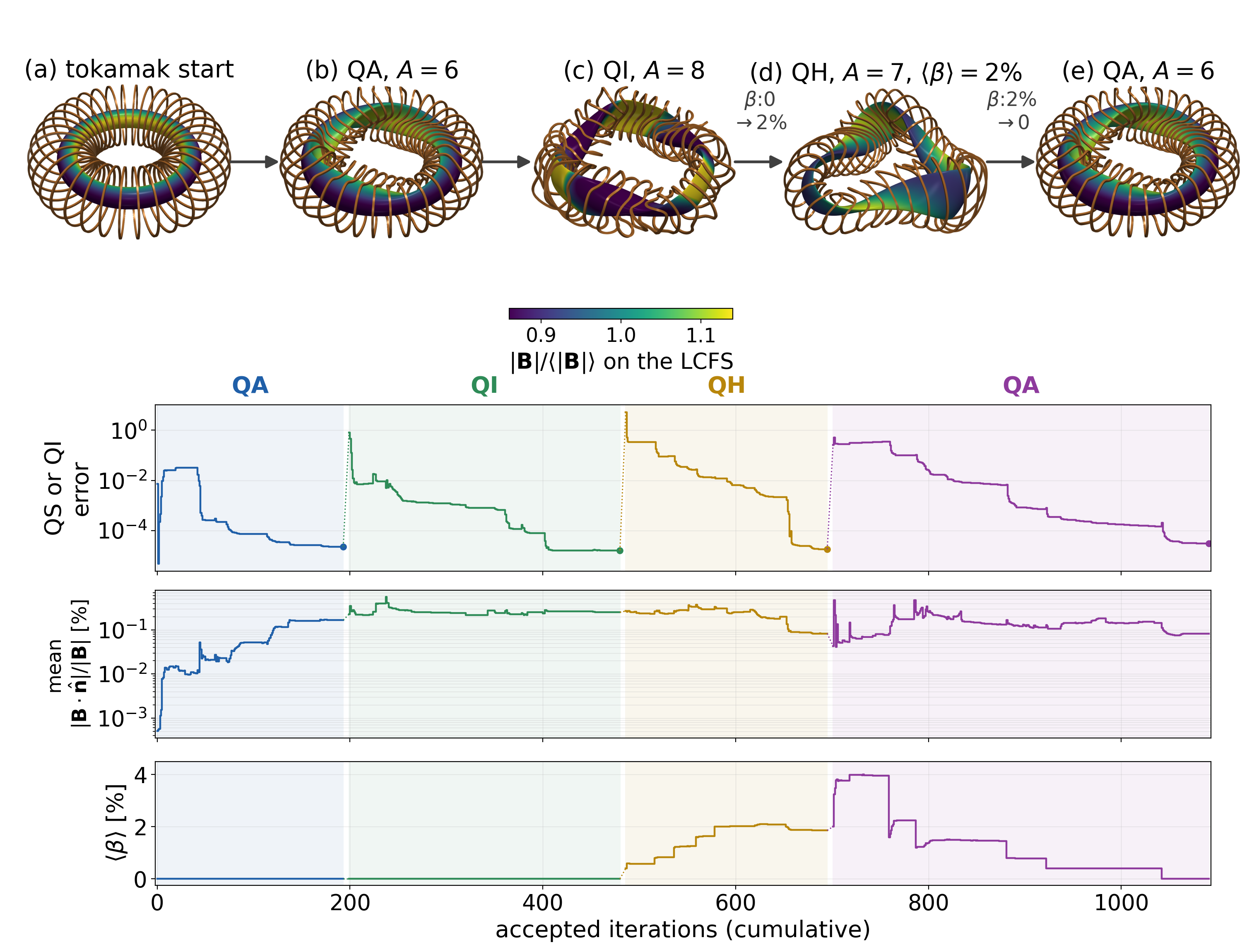}
\caption{Closed symmetry loop of Sec.~\ref{sec:loop}, three field periods, $N_c=6$. The top row shows (a) the starting tokamak and the final designs of the four legs, (b) quasi-axisymmetric with $A=6$, (c) quasi-isodynamic with $A=8$, (d) quasi-helically symmetric with $A=7$ and $\langle\beta\rangle=2\%$, and (e) quasi-axisymmetric with $A=6$ in vacuum. The bottom row shows the fixed-boundary QS or QI error, the mean $\mathbf B\cdot\hat{\mathbf n}/|\mathbf B|$ of the in-loop coils and $\langle\beta\rangle$ at every accepted iterate, against the cumulative number of accepted iterations. Dotted lines join the end of each leg to the start of the next.}
\label{fig:loop}
\end{figure*}

\subsection{Robustness to the starting point}
\label{sec:loop}
All the designs of Sec.~\ref{sec:cold} start from a cold-start. To further investigate the dependence on the starting condition, we chain four optimizations with three field periods into a closed loop. A circular tokamak is optimized into a quasi-axisymmetric stellarator, which seeds a quasi-isodynamic one, which seeds a quasi-helically symmetric one, which in turn seeds a quasi-axisymmetric design again (Fig.~\ref{fig:loop}). Each leg starts from the final boundary and the certified coils of the previous leg, with $N_c=6$, and changes only the quasisymmetry target and the aspect ratio, $A=6$ for the quasi-axisymmetric legs, 8 for the quasi-isodynamic one and 7 for the quasi-helically symmetric one. The coil limits are those of design (e), scaled as always with the minor radius. As an addition to the robustness test, the quasi-helically symmetric leg also raises the pressure from vacuum to $\langle\beta\rangle=2\%$, with $p\propto1-s$, and the last leg lowers it back to vacuum, one step per boundary-mode stage.

All four legs converge, and every final coil set is certified. The errors quoted for this loop are fixed-boundary values. The quasi-isodynamic design reaches a QI error comparable to that of design (e) at $A=9$, and the quasi-helically symmetric design reaches a much lower error at $\langle\beta\rangle=2\%$ than the similar run in vacuum. In the last leg, the QS error falls by several orders of magnitude as the pressure is lowered to vacuum. Figure~\ref{fig:loop} shows these errors along the loop. Note that $\bn$ increases in the beginning and this is benign since we start with a tokamak configuration that naturally has an extremely low field error. In the rest of the optimization $\bn$ is kept below its upper bound. Since the pressure is fixed at the start of each stage while the boundary moves, $\langle\beta\rangle$ rises within the first stages of this leg, up to 4\% as the shape leaves the quasi-helically symmetric family, before it follows the ramp down. After going around the whole loop, the design returns to essentially the quasi-axisymmetric configuration it started from, exhibiting the essentially the same aspect ratio and rotational transform profile, and a similar QS error. The method therefore shows strong indications that it does not require a good initial guess, or even a neutral initial guess. It moves between very different stellarators with the three types of symmetry, and between vacuum and finite pressure, while keeping coils within all engineering limits at every step. 

\section{NCSX revisited at reactor-scale}
\label{sec:ncsx78}
NCSX shows how strongly coils can weigh on a stellarator project. Its plasma was optimized first and its coils afterwards, and the result needed 78 coils: 18 modular coils of three shapes, 18 toroidal-field coils and 42 poloidal-field (PF) coils. Their complexity contributed to the cost growth that led to its cancellation~\cite{ncsx}. We ask whether a single cold-start run can find a comparable design, using only 18 modular coils.

The design is meant to replicate the NCSX configuration~\cite{Zarnstorff2001} at reactor scale. It has three field periods and $A=4.4$, at $\langle\beta\rangle=4\%$ with $p\propto1-s$ and the plasma current of NCSX, with the constraints on the rotational transform, the elongation, the magnetic well and the Mercier criterion given in Table~\ref{tab:ncsx78}. An axisymmetric plasma has no equilibrium at this pressure with this current, so the pressure is raised together with the shaping along the continuation. We allow only 18 modular coils, that is $N_c=3$, each shorter, less curved and more widely spaced than the modular coils of NCSX. NCSX has the same number of modular coils, but with its modular coils only, the free-boundary equilibrium does not converge. Its extra toroidal-field and poloidal-field coils are part of the configuration itself, beyond providing experimental flexibility. 

\begin{figure*}[!t]
\includegraphics[width=0.9\textwidth]{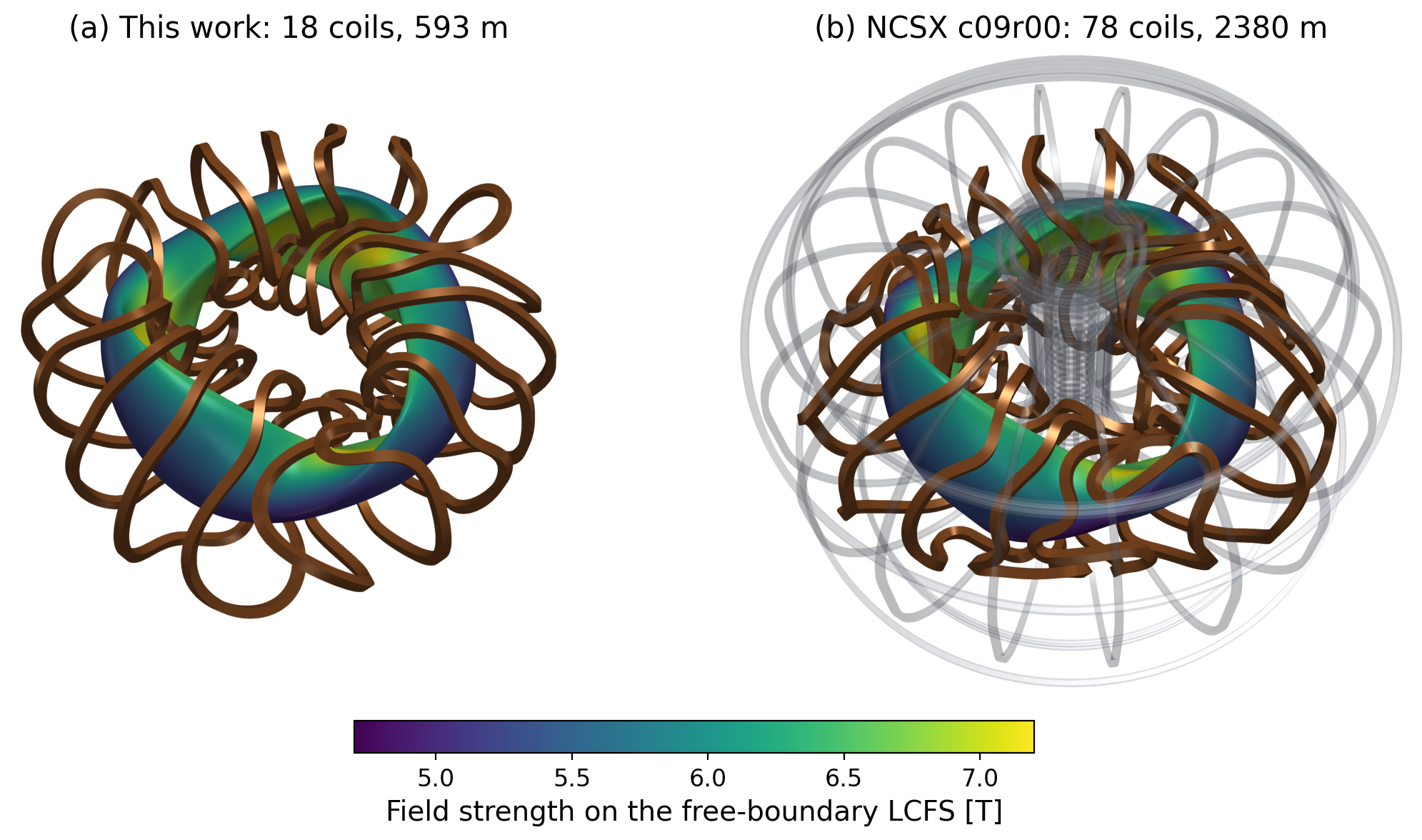}
\includegraphics[width=0.9\textwidth,trim=0 0 0 51.9bp,clip]{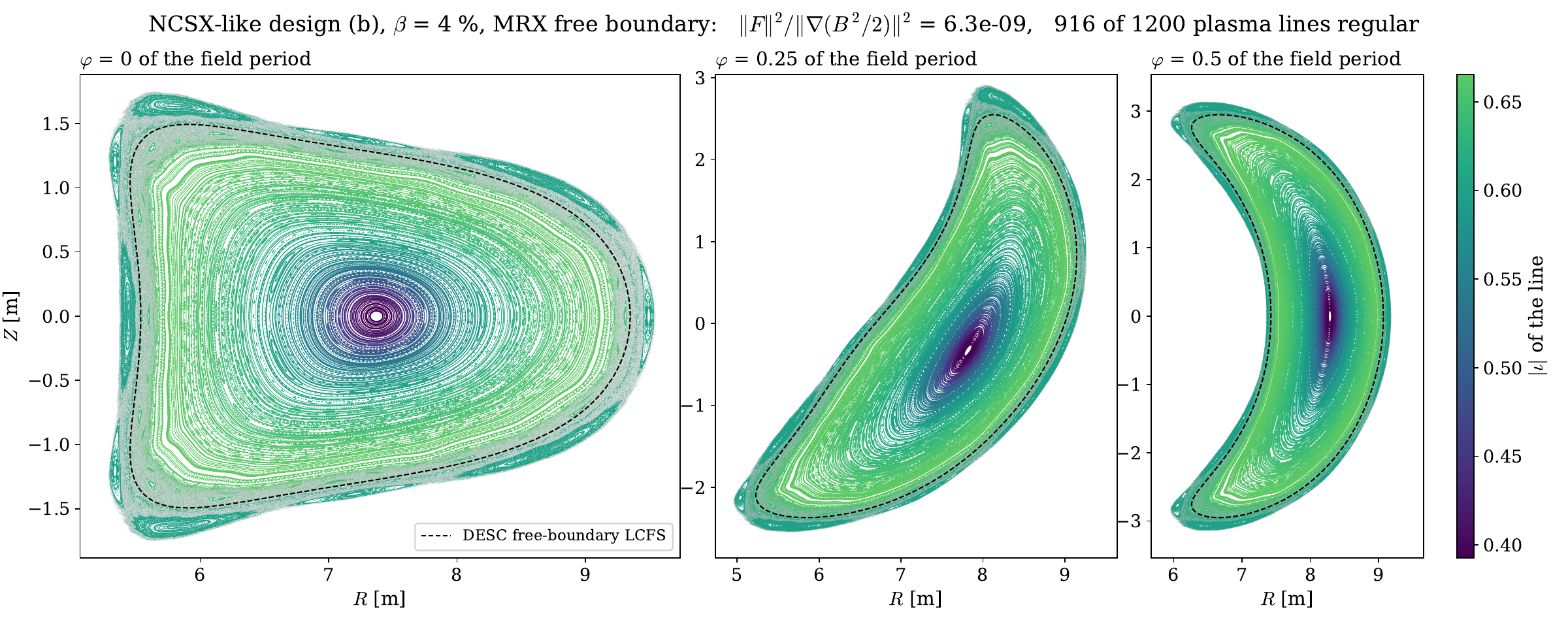}
\caption{(a) The NCSX-like design of Sec.~\ref{sec:ncsx78}, optimized in a single run from a cold-start, with its 18 certified modular coils as finite-build winding packs of $0.47\times0.47$~m$^2$. (b) The NCSX reference configuration c09r00 with its full coil set of 78 coils: 18 modular coils (copper), 18 toroidal-field coils and 42 poloidal-field coils (grey). 
(c) Poincar\'e sections of design (a) at the toroidal angles $\phi=0$, $1/4$ and $1/2$ of a field period, from free-boundary MRX~\cite{MRX} initialized from the DESC free-boundary equilibrium.}
\label{fig:ncsx78}
\end{figure*}

\begin{table*}[!t]
\caption{The reactor-scale NCSX-like design of Sec.~\ref{sec:ncsx78} and the NCSX reference configuration c09r00 with its full set of 78 coils. The constraint column gives the limits of this work. The coil metrics of c09r00 are those of its modular coils, given as point filaments. $^\dagger$Order of the Fourier fit of the filaments used for its free-boundary equilibrium. $^\ast$In parentheses, the longest PF coil.}
\label{tab:ncsx78}
\begin{ruledtabular}
\begin{tabular}{lccc}
 & constraint & this work & NCSX c09r00 \\
\hline
$N_c$ / total coils / Fourier order $M$ & & 3 / \best{18} / 5 & 3 / 78 / 40$^\dagger$ \\
total length, all coils [m] & & \best{593} & 2{,}380 \\
$L_i$ [m] & $\le34.0$ & \best{31.9--34.0} & 34.7--38.5 (90.3)$^\ast$ \\
$\kappa_{\max}$ / $\kappa_\mathrm{MSC}$ [m$^{-1}$ / m$^{-2}$] & $\le0.69$ / $\le0.15$ & \best{0.69} / \best{0.15} & 1.70 / 0.53 \\
$\bar\tau$ / $\gamma_c$ [m$^{-1}$ / $10^{-3}$~m$^{-2}$] & & \best{0.68} / \best{13} & 1.02 / 32 \\
$d_{cc}$ / $d_{cs}$ [m] & $\ge0.90$ / $\ge1.13$ & \best{0.91} / \best{1.13} & 0.80 / 1.11 \\
\hline
$\langle\beta\rangle$ [\%] & 4 & 4.0 & 4.1 \\
$A$ / $E$ & 4.4 / $\le4.5$ & 4.41 / \best{4.49} & 4.47 / 5.56 \\
fixed-boundary $f_{\mathrm{QS}}$ [$10^{-5}$] & & \best{4{,}600} & 9{,}400 \\
$W$ [\%] & $\ge1$ & 10.1 & 9.9 \\
$\iota(0)$ / $\iota(1)$ & 0.40 / 0.65 & 0.41 / 0.63 & 0.36 / 0.65 \\
$f_{\mathrm{QS}}$ [$10^{-3}$] & & \best{58} & 96 \\
$\hat D_\mathrm{Merc}$ [$10^{-3}$] & $\ge0$ & 4.3 & \best{13} \\
$\delta_\mathrm{LCFS}$ mean / max [mm] & & \best{11} / \best{54} & 14 / 74 \\
$\alpha$-loss [\%] & & $23.6\pm0.5$ & \best{$22.5\pm0.5$} \\
\end{tabular}
\end{ruledtabular}
\end{table*}

The result is shown in Fig.~\ref{fig:ncsx78} and Table~\ref{tab:ncsx78}. NCSX needs several times as many coils and several times the total coil length of our design, whose coils are even shorter in total than the 18 modular coils of NCSX alone. These coils are also simpler, with a much lower curvature, and they sit farther from each other and at least as far from the plasma. Even though our coils are simpler with respect to every metric, the plasma is still at least as good as the original design. Its quasisymmetry error is lower in both fixed and free boundary, and its free-boundary LCFS stays closer to its target. The alpha losses are very similar, within sampling error, and its Mercier margin is smaller. With shorter and more regularized coils, adding Mercier stability costs more than an order of magnitude in quasisymmetry in these runs, so we did not strongly optimize the Mercier criterion.

The NCSX rotational transform passes through the 3/5 and 3/6 rational values in the plasma. Therefore, we diagnose and illustrate the solution with a non-nested equilibrium code. 
A free-boundary MRX run~\cite{MRX},
which does not assume nested surfaces, shows in Fig.~\ref{fig:ncsx78}c significant internal and exterior island chains. The 3/6 resonance near the core stays essentially nested (a 3/6 island chain can be seen at the numerical grid scale), potentially because of the high shear. There is also an internal 3/5 resonance, and externally the solution actually exhibits two 3/5 resonances that overlap. The two overlapping 3/5 chains form a layer of islands and chaotic field lines around the confined plasma. Such a layer resembles the edge of an island divertor~\cite{Feng2006, SunnPedersen2019}. 

The design of NCSX required years of plasma and coil design; the method here reaches a comparable quality solution in a single run from a cold-start, with a quarter of the coils.

\section{Towards a Reactor Design}
\label{sec:reactor}
In this section, we ask whether the method can produce a configuration that meets multiple requirements of a reactor at once: good fast-particle confinement in free-boundary, Trapped-Electron Mode (TEM) stability, Mercier stability, a magnetic well at finite pressure, and certified coils with sufficient spacing to accommodate inter-coil support structures, alongside a blanket and neutron shield positioned between the coils and the plasma. Recent reactor studies of quasi-isodynamic stellarators~\cite{LION2025114868, Goodman2025reactor} show that such a combination exists, but they typically reach it through a sequence of separate plasma and coil optimizations. Here, plasma and coils are designed together in a single run, warm-started from a four-period boundary derived from W7-X. 

\subsection{Problem and optimization}
\label{sec:reactor_problem}
The configuration has four field periods, $A=11$, $\langle\beta\rangle=2.2\%$ with $p\propto1-s$ and no net toroidal current, and five coils per half field period. The physics constraints bound the mirror ratio, the elongation and the rotational transform on axis and at the edge, and add the maximum-$J$ constraint and Mercier stability. Their values are given in Table~\ref{tab:reactor}. The limits on the length and curvature of each coil and on the coil--coil and coil--surface distances are given in Table~\ref{tab:reactor_coils}. An additional coil constraint keeps adjacent coils apart by a minimum toroidal angle (Appendix~\ref{app:slot}). The optimization starts from a four-period boundary derived from W7-X and rescaled to $A=11$, and runs the nested method in three continuation stages, with Mercier stability added in the second.

In the next few subsections, we will compare the design with W7-X in its standard configuration. Its coil set is the 50 non-planar coils at the design current, with the geometry distributed with SIMSOPT~\cite{landreman_simsopt_2021} and excluding the planar coils. Its free-boundary equilibrium is computed as in Sec.~\ref{sec:validation} at $\langle\beta\rangle=2\%$, with $p\propto1-s$ and no net toroidal current, starting from the published vacuum boundary of this configuration at its full resolution.

For our four-field-period final design, we also plot the field outside the plasma, which determines where the edge of the confined region lies and how field lines reach the wall. 
Figure~\ref{fig:reactor_render} shows Poincar\'e sections of field lines inside and outside the LCFS. Inside the LCFS, the rotational transform of this design is nearly flat and stays away from the low-order resonances inside the plasma. Outside the LCFS, the rotational transform reaches one, and a chain of four islands with $\iota=1$ lies close to it. They are surrounded by open field lines, which reach the coils within a few toroidal transits. With four field periods, this $\iota=1$ chain is the analogue of the 5/5 island chain of the island divertor of W7-X~\cite{Feng2006, SunnPedersen2019}. 

\begin{figure*}[t]
\centering
\includegraphics[width=0.74\textwidth]{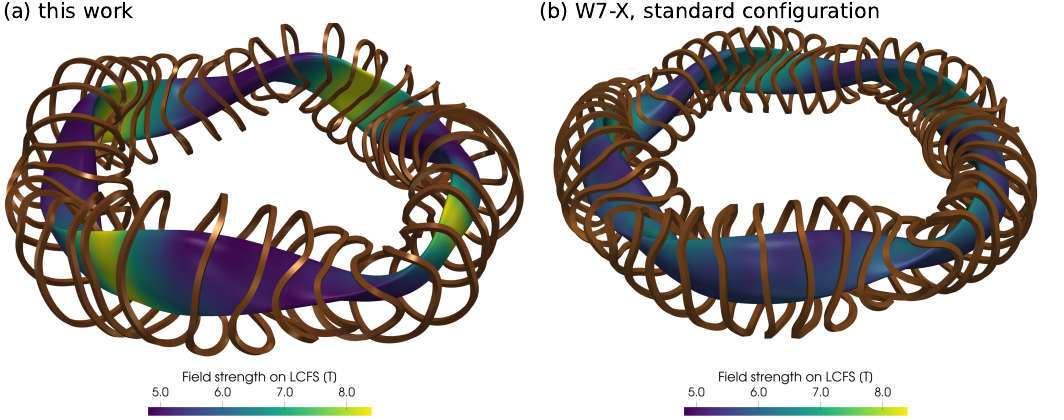}
\begin{tikzpicture}
\node[anchor=south west, inner sep=0] (im) {\includegraphics[width=\textwidth]{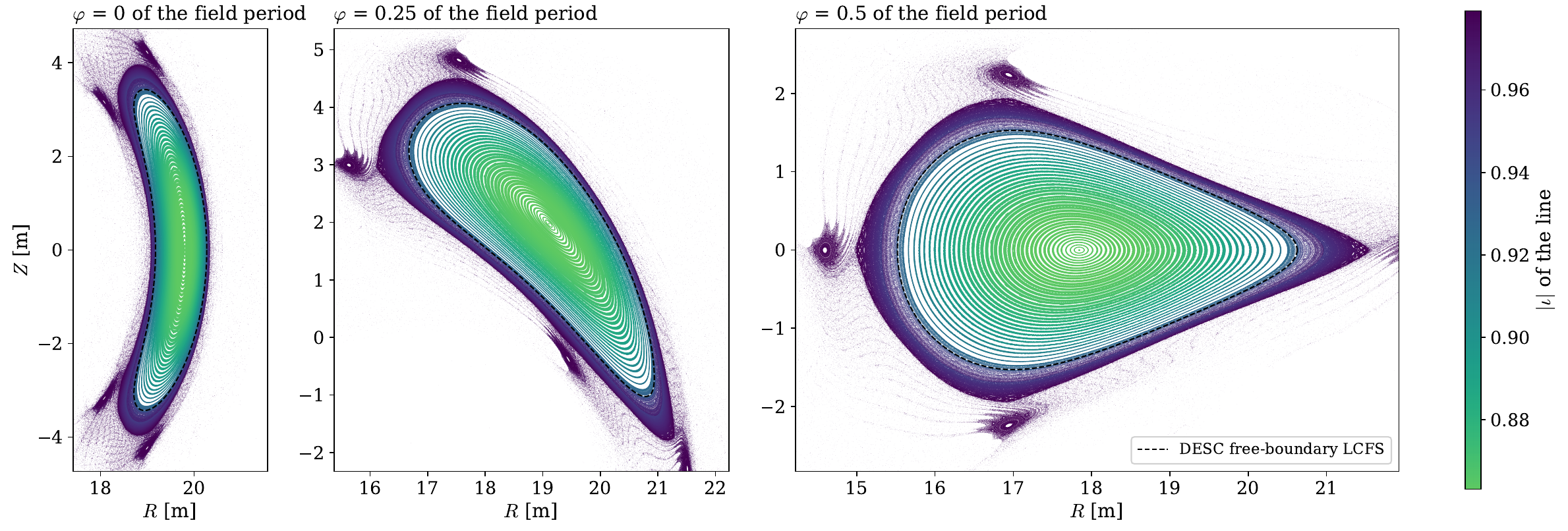}};
\begin{scope}[x={(im.south east)}, y={(im.north west)}]
\fill[white] (0.03,0.9515) rectangle (0.70,1.0);
\end{scope}
\end{tikzpicture}
\caption{Reactor-like quasi-isodynamic design. (a) The LCFS of the free-boundary equilibrium of its certified coils, colored by $|\mathbf B|$, and the 40 coils as finite-build winding packs with a square cross-section of $0.47\times0.47$~m$^2$. (b) W7-X in its standard configuration with its 50 non-planar coils, at the same size and volume-averaged field strength, drawn in the same way from the same viewpoint and with the same color scale. Its planar coils carry no current and are not drawn. The bottom row shows Poincar\'e sections of the design at the toroidal angles $\phi=0$, $1/4$ and $1/2$ of a field period.}
\label{fig:reactor_render}
\end{figure*}

\begin{table*}[!t]
\caption{Reactor-like quasi-isodynamic design and W7-X. The constraint column gives the limits of the design. W7-X is the free-boundary equilibrium of its standard-configuration coils at $\langle\beta\rangle=2\%$, with $p\propto1-s$ and no net toroidal current. The maximum-$J$ row gives the share of the cross-section $0.05<s<0.9$ on which $J$ increases outwards, at the trapping depths $K=0.02$, 0.15 and 0.3 of Appendix~\ref{app:errors}. W7-X has no target boundary at this pressure, so no $\delta_\mathrm{LCFS}$ is given.}
\label{tab:reactor}
\begin{ruledtabular}
\begin{tabular}{lccc}
 & constraint & this work & W7-X\\
\hline
$A$ / $E$ & 11 / $\le6$ & 10.95 / 5.79 & 10.79 / \best{3.80} \\
$\langle\beta\rangle$ [\%] & 2.2 & 2.22 & 2.02 \\
$\iota(0)$ / $\iota(1)$ & 0.86 / 0.95 & 0.86 / 0.93 & 0.86 / 0.96 \\
$\Delta$ & $\le0.25$ & 0.242 & 0.042 \\
$W$ [\%] & & 4.0 & 4.1 \\
$\hat D_\mathrm{Merc}$ at $s=0.5$/0.9 [$10^{-3}$] & $s\hat D_\mathrm{Merc}\ge20$ & 42 / 11 & \best{68} / 11 \\
$f_{\mathrm{QI}}$ [$10^{-3}$] & & \best{0.64} & 110 \\
$\delta_\mathrm{LCFS}$ mean / max [mm] & & 25 / 100 & -- \\
$\epsilon_{\mathrm{eff}}$, $s=0.25$/0.5/0.9 [\%] & & 2.0/2.7/3.6 & \best{0.46}/\best{0.52}/\best{0.95} \\
maximum-$J$ violation [\%] & 0 & \best{8.1}/\best{1.9}/\best{0} & 100/93/66 \\
$\alpha$-loss [\%] & & \best{$4.7\pm0.2$} & $27.1\pm0.5$ \\
\end{tabular}
\end{ruledtabular}
\end{table*}

\subsection{Trapped particles, stability and $\alpha$-particle confinement}
\label{sec:reactor_physics}
A reactor plasma must stay stable as its pressure rises, and it must confine its particles and heat well enough to reach and sustain ignition. In a quasi-isodynamic stellarator, much of both rests on the trapped particles. When their second adiabatic invariant $J$ decreases outwards, the maximum-$J$ property (Appendix~\ref{app:errors}), trapped particles precess in the direction that stabilizes the trapped-electron mode and the ion-temperature-gradient mode driven by the density gradient, which lowers turbulent transport~\cite{goodman_prx, Helander_2012}. A magnetic well and Mercier stability, in turn, protect the plasma against pressure-driven interchange modes, and a low effective ripple keeps the neoclassical transport of the thermal plasma small. None of these properties is guaranteed once the target boundary is replaced by the field of discrete coils, so we evaluate all of them on the free-boundary equilibrium.

The second adiabatic invariant $J$ of trapped particles, from the most deeply trapped ($K=0.02$) to the barely trapped ($K=0.9$), is shown in Fig.~\ref{fig:reactor_maxj} of Appendix~\ref{app:reactor}. For all but the most deeply trapped particles, $J$ decreases outwards on the whole cross-section or on nearly all of it. Even the most deeply trapped particles, whose bounce points lie close to the minimum of $|\mathbf B|$ and are therefore the most sensitive to the coil ripple, keep the property on most of the cross-section, with the exceptions confined to small pockets. In W7-X at $\langle\beta\rangle=2\%$, which does not have this property at this pressure, $J$ increases outwards on most of the cross-section for the deeply and moderately trapped particles. Table~\ref{tab:reactor} gives these shares, which vanish when the maximum-$J$ property holds everywhere. Table~\ref{tab:startup} reports the value of the maximum-$J$ constraint itself, which is met when it is not positive. The configuration has a magnetic well in free-boundary, although the optimization does not ask for one, and it is Mercier stable on all but the last surface, $s=0.99$. The effective ripple $\epsilon_{\mathrm{eff}}$~\cite{Nemov1999}, computed with NEO\_JAX~\cite{NEOJAX}, a JAX implementation of NEO~\cite{Nemov1999}, is almost unchanged by the coils.

In free-boundary, the design basically keeps the alpha losses of the fixed-boundary equilibrium. Table~\ref{tab:reactor} gives the total loss, and Table~\ref{tab:alpha} of Appendix~\ref{app:alpha} the losses by birth radius. As for the other designs, the core is well confined. For comparison, at $\langle\beta\rangle=2\%$, the reactor-scale QI design Stellaris~\cite{LION2025114868} loses a similar fraction of its alphas with a similar birth profile, although Stellaris operates at 9~T on axis with minor radius of $1.3$~m, which gives an alpha gyroradius about 15\% smaller relative to the minor radius. 

\subsection{Coils}
\label{sec:reactor_coils}
Coils that a reactor can use are hard to obtain. Between the coils and the plasma there must be room for the first wall, the blanket and the neutron shield, about a metre in recent reactor studies~\cite{LION2025114868,Anderson_Canik_Hegna_Mowry_2025}. Between neighbouring coils there must be room for winding packs tens of centimetres wide, their steel casings and a support structure that carries net forces of 100--350~MN per coil~\cite{LION2025114868}. The coils must also stay gently curved, so that the superconducting tape can be wound without damage. Each of these requirements pushes the coils away from the plasma and from each other, where their field is smoother and can reproduce the strong shaping of a quasi-isodynamic boundary only with difficulty. 

\begin{table*}[!t]
\caption{Certified coils of the reactor-scale QSS design, compared with other reactor-scale QI designs, all at ARIES-CS scale. Each cell gives the range over the coils of a half field period. $\bar\tau$ and $\gamma_c$ were not reported for any of the comparison designs, so we omit them here. Blank cells are not reported and we exclude the W7-X planar coils from the metrics.}
\label{tab:reactor_coils}
\begin{ruledtabular}
\begin{tabular}{lccccc}
 & $N_c$ / Fourier order $M$ & $L_i$ / $L_\mathrm{total}$ [m] & $d_{cc}$ [m] & $d_{cs}$ [m] & $\kappa_{\max}$ [m$^{-1}$] / $\kappa_\mathrm{MSC}$ [m$^{-2}$] \\
\hline
this work & 5 / 8 & 33.0--33.1 / 165.4 & \best{1.73--1.74} & 1.68--1.97 & \best{0.475} / \best{0.10}\\
Ref.~\cite{Goodman2025reactor} & 5 & 30.8--38.9 / 180.5 & 0.83--1.01 & 1.48--1.69 & 0.98--1.19 \\
Stellaris~\cite{LION2025114868} & 6 & $\sim33$ / $\sim197$ & 0.88 & \best{1.80} & 1.19 / 0.20 \\
Infinity Two~\cite{Hegna2025InfinityTwo} & 6 & & 0.90 & 1.65 & 1.41 / 0.43 \\
W7-X~\cite{Goodman2025reactor} & 5 & $\sim27$ / \best{$\sim136$} & $\sim1.2$ & $\sim0.94$ & $\sim0.76$ / 0.19 \\
\hline
\textit{limits of this work} & & $\le33.1$ & $\ge1.66$ & $\ge1.65$ & $\le0.475$ / $\le0.10$ \\
\multicolumn{6}{l}{\textit{and adjacent coils of this work at least $1.5^\circ$ apart in toroidal angle}} \\
\end{tabular}
\end{ruledtabular}
\end{table*}

Table~\ref{tab:reactor_coils} compares our coils with those of three recent reactor studies~\cite{Goodman2025reactor, LION2025114868, Hegna2025InfinityTwo} and of W7-X, all at ARIES-CS scale, with a minor radius of 1.704~m. We emphasize that the reactor studies we compared with are integrated designs. They also satisfy requirements that are outside the scope of this work, such as finite-build winding packs and conductor limits, structural loads, maintenance access, kinetic and turbulent transport and time-dependent magnetohydrodynamic stability. The comparison here concerns the centre-line coil geometry only. 

Within that narrow scope, we now compare the filamentary coil solutions. Of the reactor study designs, only Infinity Two reports a mean-squared curvature, and it is roughly four times ours. Compared with the five coils of Goodman \emph{et al.}~\cite{Goodman2025reactor} or the six coils of Stellaris~\cite{LION2025114868},
our coils exhibit roughly double the minimum coil-coil distance (leaving about twice the space for the winding packs and their support structure), with less than half the curvature. Part of this advantage just reflects the aspect ratio, 11 against about 10.5 and 9.8, since a larger aspect ratio leaves more room between the coils. 

\begin{figure*}[!t]
\centering
\includegraphics[width=0.84\textwidth]{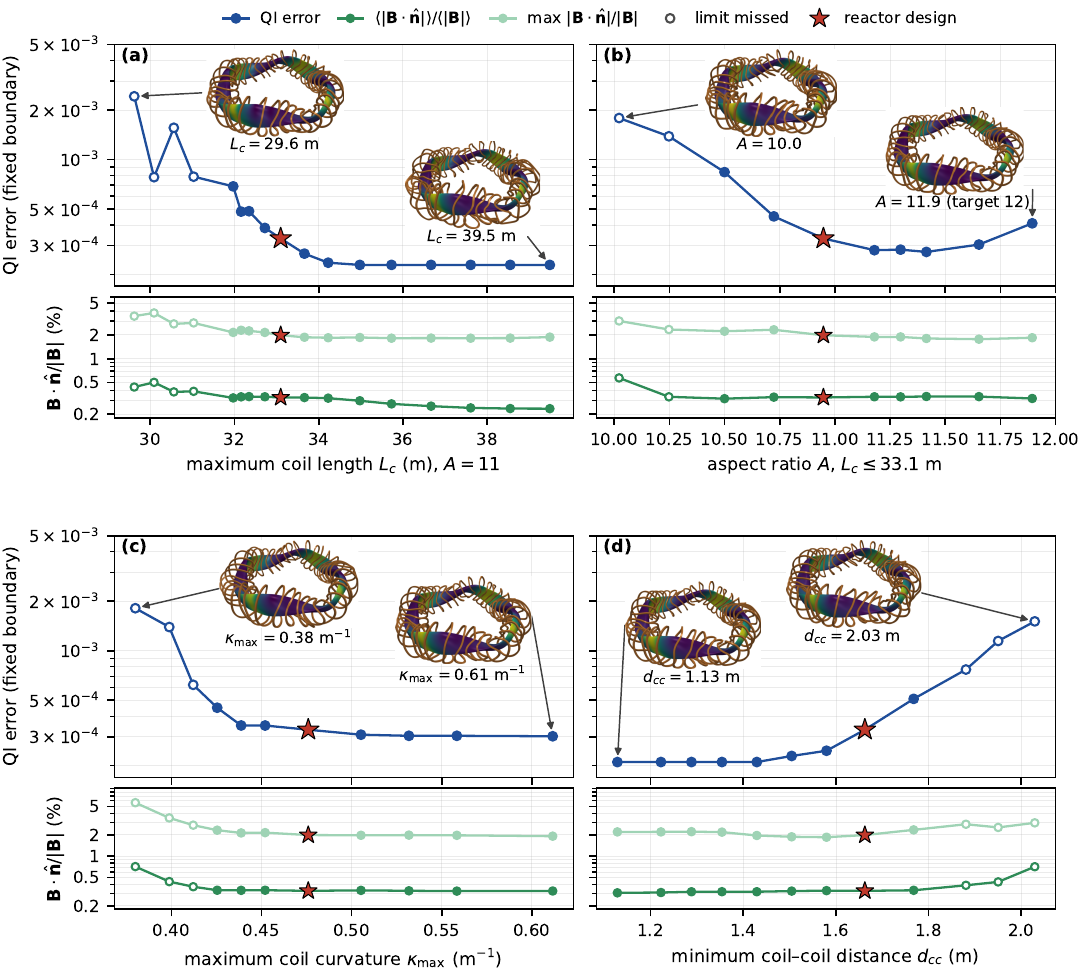}
\caption{Continuations of the reactor-like design, marked by the star, in (a) the maximum coil length at $A=11$, (b) the aspect ratio at a coil length of at most 33.1~m, (c) the maximum coil curvature and (d) the minimum coil--coil distance $d_{cc}$. The upper panels show the fixed-boundary QI error and the lower panels the normal field $\bn$ of the certified coils on the target boundary. Open symbols mark points that miss the bound on the squared flux by more than 25\% or miss a physics or coil limit. The renders show the LCFS colored by $|\mathbf B|$ and the certified coils at the ends of each continuation.}
\label{fig:reactor_homotopy}
\end{figure*}
\subsection{Trade-offs around the design}
\label{sec:reactor_homotopy}
The design is one point of a family, and the nested method can move along that family. Each new point is warm-started from its certified neighbour and re-solved with the full method, so a converged design can be continued in any single limit without restarting the optimization or retuning weights. This turns the design into a map of the nearby trade-offs between engineering limits and plasma quality. We continue the design in the four limits that most affect the size and cost of a reactor, the coil length, the aspect ratio, the coil curvature and the coil--coil distance, each time with all other limits fixed (Fig.~\ref{fig:reactor_homotopy}). In the curvature continuation, the limit on the mean-squared curvature scales with $\kappa_{\max}^2$.

The four fronts answer different questions. Longer coils lower the QI error substantially up to a length slightly above that of the design, after which it levels off and extra length only lowers $\bn$. Slightly shorter coils can no longer reproduce the target within the bound on the squared flux. The design thus sits where each metre of coil still buys QI quality. In aspect ratio, the QI error is lowest slightly above the design value and rises steeply below it, so a more compact plasma costs QI quality well before it costs coil quality. Of the two clearance limits, the coil--coil distance is the one that constrains the design. Reducing it lowers the QI error noticeably at first, after which it stops improving, while widening it raises the QI error several-fold. The curvature has little room left, since loosening it gains little, while tightening it degrades the QI error quickly, until the coils can no longer reproduce the target within the bound on the squared flux and the physics limits. Along all four fronts, every certified design remains Mercier stable in fixed boundary, keeps the maximum-$J$ property and has a magnetic well. A designer can therefore read off directly, for example, how much QI quality wider coil spacing or a more compact plasma would cost.

\subsection{One coil set for the pressure ramp}
\label{sec:startup}
\begin{figure*}[!t]
\centering
\includegraphics[width=0.9\textwidth]{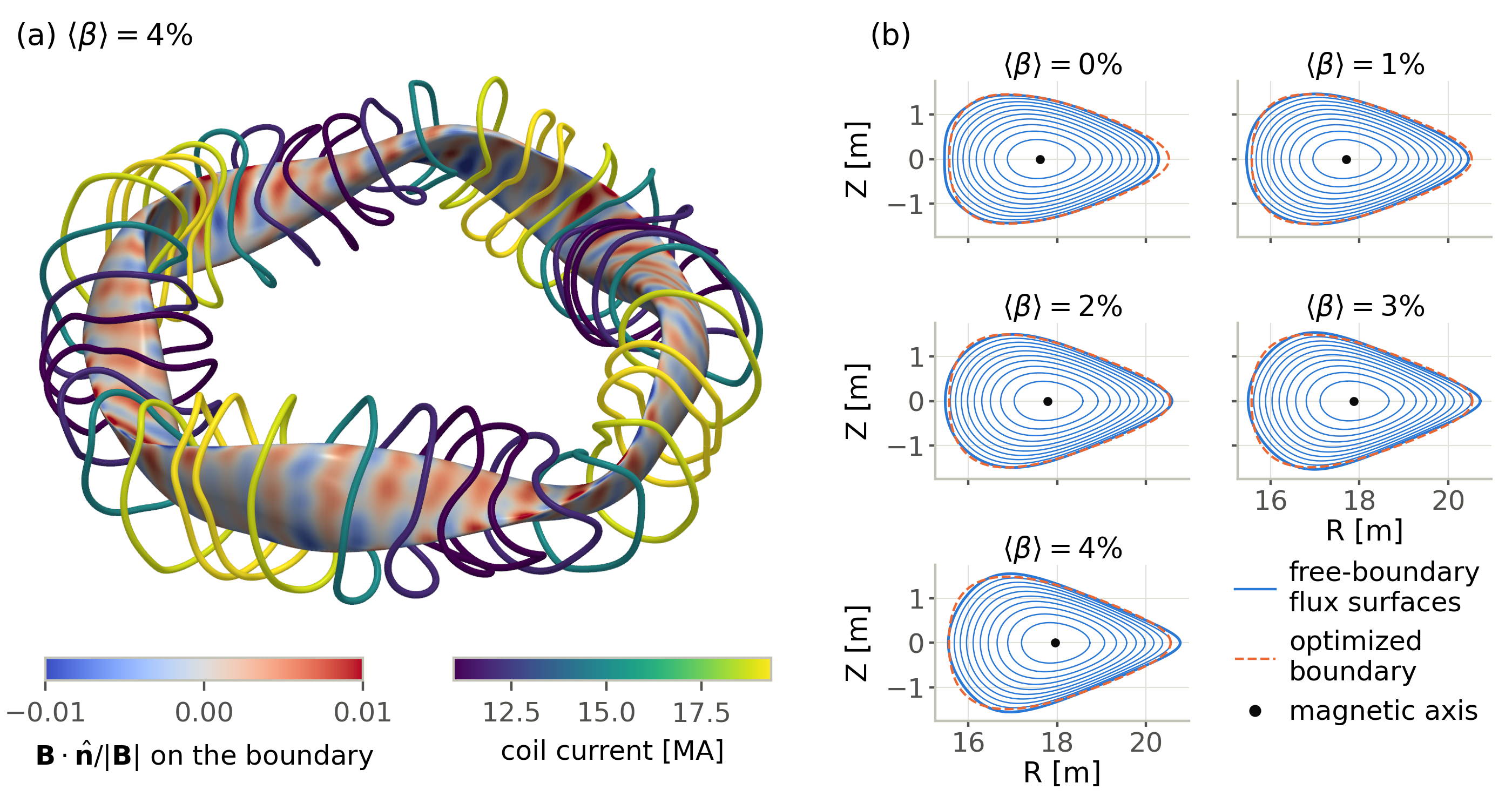}
\caption{The warm-started reactor-scale QI design with one set of coil shapes for $\langle\beta\rangle$ from 0 to 4\%. (a) The 4\% state, with the normal field of the shared coils and of the plasma on the LCFS. (b) Flux surfaces of the DESC free-boundary equilibria of the five states, with the shared coils and the currents of each state, at the toroidal angle $\phi=1/2$ of a field period. The dashed curves are the optimized boundaries.}
\label{fig:startup_render}
\end{figure*}

A reactor does not operate at a single pressure. It starts from a vacuum field and raises the plasma pressure to its operating point. A configuration optimized at one value of $\beta$, such as the design of Section~\ref{sec:reactor_problem}, therefore says little about the plasmas its coils confine at lower pressure. This section extends the nested method to several pressures at once. One set of coil shapes is designed together with the plasmas of five states, from vacuum to $\langle\beta\rangle=4\%$, each with its own boundary and its own coil currents.

\subsubsection{Formulation}
The states $j=0,\dots,4$ have the pressure profiles $p_j(s)=P_j\,(1-s)$, where $P_j$ is chosen so that the volume-averaged $\beta$ of state $j$ is $j\%$. All states have the toroidal flux of the design and no net toroidal current, so the field strength stays fixed while the pressure rises. Each state has its own boundary $\mathbf s_j$, which may shift and deform with pressure as a free-boundary plasma does. All states share the coil geometry $\mathbf c$, and each state has its own currents $\mathbf I_j$, one per base coil. The total current of each state is set by its own equilibrium through the factor $\lambda_G$ of Section~\ref{sec:methods_coils}, so each state has $N_c-1$ free current ratios. The coil problem of Equation~\eqref{eq:inner} becomes
\begin{align}
\label{eq:startup_inner}
J^\star_{\mathrm{tot}}(\mathbf s_0,\dots,\mathbf s_4)=\min_{\mathbf c,\,\mathbf I_0,\dots,\mathbf I_4}\;&\sum_{j=0}^{4}w_j\,J_f(\mathbf c,\mathbf I_j;\mathbf s_j)\\ \notag
\text{subject to}\;&\mathbf g(\mathbf c,\mathbf s_j)\le0,\quad j=0,\dots,4,
\end{align}
where $J_f(\mathbf c,\mathbf I_j;\mathbf s_j)$ is the squared flux of Equation~\eqref{eq:squared_flux_continuous_def} on the boundary of state $j$, with the plasma field of that state, and $w_j$ is the weight of state $j$. The coil--surface distance is imposed with respect to every boundary. The plasma problem of Equation~\eqref{eq:outer} becomes
\begin{equation}
\begin{aligned}
\min_{\mathbf s_0,\dots,\mathbf s_4}\;&\sum_{j=0}^{4}f_{\mathrm{QI}}(\mathbf s_j;p_j)\\
\text{subject to}\;&\mathbf h_j(\mathbf s_j;p_j)\le0,\quad j=0,\dots,4,\\
&J^\star_{\mathrm{tot}}(\mathbf s_0,\dots,\mathbf s_4)\le\epsilon,
\end{aligned}
\label{eq:startup_outer}
\end{equation}
with one flux constraint for the whole ramp. The envelope theorem of Appendix~\ref{app:envelope} applies to the summed problem. 
A separate bound on the squared flux of each state would need the response of the shared coils to every boundary, so the formulation keeps a single bound and monitors the squared flux of each state. 

The physics constraints $\mathbf h_j$ and the coil limits are those of the design of Section~\ref{sec:reactor} in every state, with two exceptions. Separate optimizations of the vacuum and 1\% states with the full set of constraints showed that the Mercier bound of the design and the maximum-$J$ constraint cannot be met at these pressures. These two states therefore require only marginal Mercier stability and carry no maximum-$J$ constraint. We considered it reasonable to optimize less aggressively for the low-pressure states, which are not the final operating state and where good alpha losses are not particularly relevant.

\begin{table*}[!t]
\caption{Five states of the warm-started QI reactor design with one set of coil shapes, each with the coil currents of that state.}
\label{tab:startup}
\begin{ruledtabular}
\begin{tabular}{lccccc}
 & 0\% & 1\% & 2\% & 3\% & 4\% \\
\hline
$A$ / $E$ & 10.93 / 5.82 & 10.97 / 5.74 & 10.96 / 5.76 & 10.97 / 5.80 & 10.99 / 5.96 \\
$f_{\mathrm{QI}}$ [$10^{-3}$] & 0.84 & 0.65 & 0.57 & 0.68 & 0.75 \\
$\Delta$ & 0.218 & 0.227 & 0.236 & 0.245 & 0.253 \\
maximum-$J$ & 0.091 & 0.036 & $-0.004$ & $-0.024$ & $-0.048$ \\
$\iota(0)$ / $\iota(1)$ & 0.90 / 0.94 & 0.89 / 0.94 & 0.88 / 0.94 & 0.87 / 0.94 & 0.85 / 0.93 \\
$W$ [\%] & 0.43 & 1.95 & 3.49 & 4.98 & 6.44 \\
$\hat D_\mathrm{Merc}$ [$10^{-3}$] & $-0.3$ & 5.0 & 8.8 & 11.8 & 6.8 \\
$\delta_\mathrm{LCFS}$ mean / max [mm] & 47 / 231 & 33 / 138 & 28 / 104 & 37 / 178 & 38 / 216 \\
$\alpha$-loss [\%] & $17.0\pm0.4$ & $16.8\pm0.4$ & $8.2\pm0.3$ & $2.0\pm0.2$ & $1.8\pm0.1$ \\
\hline
$\bn$ [\%] & 0.34 & 0.28 & 0.29 & 0.27 & 0.29 \\
current of coil 1 [MA] & 18.27 & 18.62 & 18.81 & 19.05 & 19.32 \\
current of coil 2 [MA] & 17.73 & 18.05 & 18.24 & 18.53 & 18.80 \\
current of coil 3 [MA] & 14.72 & 14.85 & 14.83 & 14.89 & 14.96 \\
current of coil 4 [MA] & 12.11 & 12.10 & 12.04 & 12.03 & 12.03 \\
current of coil 5 [MA] & 11.02 & 11.08 & 10.99 & 10.99 & 11.01 \\
\hline
\multicolumn{6}{l}{shared coil shapes: $N_c=5$, $M=5$, $L_i=33.3$--34.1~m, $d_{cc}=1.71$~m, $d_{cs}=1.68$--1.69~m,} \\
\multicolumn{6}{l}{$\kappa_{\max}=0.479$~m$^{-1}$, $\kappa_\mathrm{MSC}=0.10$~m$^{-2}$, $\bar\tau=0.55$~m$^{-1}$, $\gamma_c=6.7\times10^{-3}$~m$^{-2}$} \\
\end{tabular}
\end{ruledtabular}
\end{table*}

\subsubsection{Optimization}
The run is warm-started from the reactor-scale QI solution presented throughout Sec.~\ref{sec:reactor}. 
We found it useful to first optimize for a solution at each $\langle\beta\rangle$ with separate coils, producing five plasma solutions with five coil solutions. Afterwards, the coil solution obtained for $\langle\beta\rangle=2\%$ are used to warm-start the one-coil-set run for all five plasmas. This run first uses equal weights $w_j$ and then weight is doubled for the 3\% state and quadrupled for the 4\% state, to prioritize the two highest pressure design points, which ends up lowering the squared flux of every state.

\subsubsection{Results}
Table~\ref{tab:startup} lists the free-boundary properties of the five states and the coils that serve them. Interestingly, the five boundaries differ from each other only by a few centimetres, and the currents change by only a few percent along the ramp-up. The two largest currents rise marginally with pressure, while the other three stay nearly constant. Figure~\ref{fig:startup_render}a shows the 4\% state with the shared coils.

Every state is validated in free boundary with DESC, as in Section~\ref{sec:validation}.
Figure~\ref{fig:startup_render}b shows the flux surfaces of the five free-boundary equilibria. In every state, the mean deviation of the LCFS from the optimized boundary is somewhat larger than for the design of Section~\ref{sec:reactor_problem} but the alpha losses remain very low at the higher $\langle\beta\rangle$ design points. The vacuum and 1\% states lose many more alphas. Since a reactor holds few alphas during the ramp-up, these losses matter less than those at the operating point. These are the states without the maximum-$J$ constraint, and their maximum-$J$ value is positive. Moreover, the magnetic well deepens with pressure, the plasma is Mercier stable at 1 to 4\% on $0.15\le s\le0.95$, and the vacuum state is marginally Mercier stable near $s=0.15$.

These results show that one set of coil shapes can serve a reactor-relevant quasi-isodynamic plasma from vacuum to $\langle\beta\rangle=4\%$ with modest current changes, at a small cost in the QI error of each state.

\FloatBarrier
\begin{table*}[!t]
\caption{Every coil set of this paper, per half field period, with $N_c$ the number of base coils. Rows in italics give the coil limits imposed on the coil sets listed below them. Letters (a)--(h) refer to the designs of Fig.~\ref{fig:cold}. $^\ddagger$Limit on $L_\mathrm{total}$, in place of a limit per coil. $^\S$One set of coil shapes for five states from vacuum to $\langle\beta\rangle=4\%$, each with its own currents. The ranges are over the five states. $^\P$Modular coils, plus 18 toroidal-field and 42 poloidal-field coils over the device. Its $M$ is the order of the Fourier fit of its filaments, and its $\bn$ is that of all 78 coils. The c09r00 limits are not shown, since its coils were not designed under the limits of our problem.}
\label{tab:coils_all}
\begin{ruledtabular}
\footnotesize
\begin{tabular}{lccccccccc}
coil set & Sec. & $N_{\mathrm{fp}}$ & $N_c$ / $M$ & $L_i$ [m] & $\kappa_{\max}$ / $\kappa_\mathrm{MSC}$ & $d_{cc}$ / $d_{cs}$ [m] & $f_{\mathrm{QS}}$, $f_{\mathrm{QI}}$ [$10^{-3}$] & $\bn$ [\%] & $\alpha$-loss [\%] \\
\hline
\textit{LP QA coil limits} &  &  &  & $\le40.6$ & $\le0.30$ / $\le0.039$ & $\ge1.01$ / $\ge1.52$ &  &  &  \\
LP QA vac., two-stage & \ref{sec:lpqa} & 2 & 5 / 5 & 40.6 & 0.30 / 0.039 & 1.47 / 1.90 & 7.9 & 0.22 & 20.6 \\
LP QA vac., QSS & \ref{sec:lpqa} & 2 & 5 / 5 & 40.6 & 0.28 / 0.039 & 1.12 / 2.38 & 0.62 & 0.053 & 6.1 \\
LP QA $\beta$, two-stage & \ref{sec:lpqa} & 2 & 5 / 5 & 40.6 & 0.30 / 0.039 & 1.42 / 2.06 & 3.4 & 0.13 & 16.3 \\
LP QA $\beta$, QSS & \ref{sec:lpqa} & 2 & 5 / 5 & 40.6 & 0.30 / 0.039 & 1.19 / 2.43 & 0.68 & 0.054 & 3.3 \\
(c) QA, cold-start & \ref{sec:cold_start_lpqa} & 2 & 5 / 5 & 40.6 & 0.28 / 0.039 & 1.16 / 2.32 & 0.74 & 0.056 & 8.0 \\
\hline
\textit{(f) coil limits} &  &  &  & $\le22.2$ & $\le0.76$ / $\le0.12$ & $\ge0.68$ / $\ge1.03$ &  &  &  \\
(f) QI, cold-start & \ref{sec:qi_cold} & 5 & 6 / 5 & 22.2 & 0.56 / 0.12 & 0.74 / 1.11 & 1.06 & 0.49 & 2.1 \\
\hline
\textit{(a) coil limits} &  &  &  & $\le25.7$ & $\le1.2$ / $\le0.46$ & $\ge0.21$ / $\ge0.63$ &  &  &  \\
(a) QA, $A=2.5$ & \ref{sec:lowA_tokamak} & 2 & 4 / 5 & 25.7 & 1.2 / 0.35 & 0.23 / 0.63 & 10 & 0.27 & 8.3 \\
\hline
\textit{(b) coil limits} &  &  &  & $\le35.7$ & $\le0.89$ / $\le0.51$ & $\ge0.49$ / $\ge1.28$ &  &  &  \\
(b) QA, bootstrap & \ref{sec:qa_nfp1} & 1 & 6 / 5 & 32.2--35.6 & 0.89 / 0.16 & 0.49 / 1.29 & 1.3 & 0.17 & 7.1 \\
\hline
\textit{(d) coil limits} &  &  &  & $\le45.6$ & $\le0.57$ / $\le0.067$ & $\ge0.70$ / $\ge2.04$ &  &  &  \\
(d) QI, $\langle\beta\rangle=2\%$ & \ref{sec:qi_cold} & 2 & 6 / 5 & 42.0--45.6 & 0.57 / 0.067 & 0.86 / 2.49 & 0.34 & 0.17 & 8.8 \\
\hline
\textit{(e) coil limits} &  &  &  & $\le34.4$ & $\le0.83$ / $\le0.14$ & $\ge0.78$ / $\ge1.39$ &  &  &  \\
(e) QI & \ref{sec:qi_cold} & 3 & 6 / 5 & 31.6--34.4 & 0.83 / 0.14 & 0.87 / 1.61 & 0.068 & 0.25 & 0.02 \\
\hline
\textit{(g) coil limits}$^\ddagger$ &  &  &  & $\le196.0$ & $\le0.95$ / $\le0.12$ & $\ge0.68$ / $\ge1.37$ &  &  &  \\
(g) QH & \ref{sec:qh_cold} & 3 & 6 / 5 & 28.7--35.2 & 0.95 / 0.12 & 1.41 / 1.87 & 33 & 0.15 & 4.0 \\
\hline
\textit{(h) coil limits}$^\ddagger$ &  &  &  & $\le162.0$ & $\le0.94$ / $\le0.18$ & $\ge0.69$ / $\ge1.38$ &  &  &  \\
(h) QH & \ref{sec:qh_cold} & 4 & 6 / 5 & 25.8--27.8 & 0.79 / 0.098 & 0.85 / 1.57 & 27 & 0.12 & 0.09 \\
\hline
\textit{reactor coil limits} &  &  &  & $\le33.1$ & $\le0.475$ / $\le0.10$ & $\ge1.66$ / $\ge1.65$ &  &  &  \\
reactor-like QI & \ref{sec:reactor} & 4 & 5 / 8 & 33.0--33.1 & 0.475 / 0.10 & 1.73 / 1.68 & 0.64 & 0.33 & 4.7 \\
\hline
\textit{start-up coil limits} &  &  &  & $\le34.1$ & $\le0.475$ / $\le0.10$ & $\ge1.66$ / $\ge1.65$ &  &  &  \\
start-up, 0--4\%$^\S$ & \ref{sec:startup} & 4 & 5 / 5 & 33.3--34.1 & 0.479 / 0.10 & 1.71 / 1.68 & 0.57--0.84 & 0.27--0.34 & 1.8--17.0 \\
\hline
\textit{NCSX-like coil limits} &  &  &  & $\le34.0$ & $\le0.69$ / $\le0.15$ & $\ge0.90$ / $\ge1.13$ &  &  &  \\
NCSX-like & \ref{sec:ncsx78} & 3 & 3 / 5 & 31.9--34.0 & 0.69 / 0.15 & 0.91 / 1.13 & 58 & 0.21 & 23.6 \\
NCSX c09r00 & \ref{sec:ncsx78} & 3 & 3$^\P$ / 40 & 34.7--38.5 & 1.70 / 0.53 & 0.80 / 1.11 & 96 & 0.22 & 22.5 \\
\end{tabular}
\end{ruledtabular}
\end{table*}

\section{Conclusion}
\label{sec:conclusion}
Table~\ref{tab:coils_all} lists the limits, the certified coils, the quasisymmetry or QI errors, the normal fields and the alpha losses of every design of this paper, with the section in which each design is discussed.

In this work, we have presented a single-stage method for stellarator design, in which the plasma boundary is optimized while filamentary coils are kept at an optimum, under hard engineering limits, for every candidate boundary. The squared flux of these coils enters the plasma problem as a constraint whose exact gradient follows from the envelope theorem, in vacuum and at finite pressure, so that the method needs no weights between physics and coils and always ends with coils that satisfy the engineering limits.

Starting from a published quasi-axisymmetric design, the method finds configurations whose coils fit them better and which are more quasisymmetric and confine alpha particles much better in free-boundary equilibria. With a cold-start in both the plasma and coils, it reaches all three classes of omnigenous stellarators over a wide range of aspect ratios, field periods and pressures, compares favourably with recent single-stage and coil-design methods in the benchmarks of Appendix~\ref{sec:benchmarks}, and returns to its starting design after a closed loop through all three symmetries, which shows that it is robust to its starting point. It also reaches a new class of very compact quasi-axisymmetric stellarator, with one field period and self-consistent bootstrap current. 

We then applied the method at reactor scale. A single run from a cold-start gives a quasi-axisymmetric design at $\langle\beta\rangle=4\%$ that is comparable to NCSX with 18 modular coils in place of 78 coils. A second run gives a reactor-like quasi-isodynamic design, which shows that stability, fast-particle confinement and well-separated, gently curved coils can be obtained together, and continuations around it map the trade-offs between plasma quality and engineering limits. A third run designs one set of coil shapes that serves its plasma from vacuum to $\langle\beta\rangle=4\%$, with only the currents changing along the ramp. The plasma and coil properties are comparable with published reactor-scale QI designs.

Several limitations and open questions remain. The optimization is local, and the coil problem is nonconvex, so a result can depend on the branch of coils that the continuation follows, as discussed in Appendix~\ref{app:envelope}. The coils are filaments, without finite build, forces or tolerances, except where stated. None of the optimizations here directly targeted the effective ripple. Mercier stability and quasi-axisymmetry together appear very hard to satisfy at low aspect ratio. Design (b) is not Mercier stable, and in the NCSX-like runs Mercier stability cost more than an order of magnitude in quasisymmetry. The bootstrap current is self-consistent only in design (b). Turbulence and other microinstabilities must enter as further constraints for reactor designs and the alpha-particle calculations are collisionless. All of the equilibria computed here, except the MRX run for the NCSX-like QSS solution, assume nested flux surfaces. Finally, the method cannot reconcile incompatible requirements. If the prescribed physics and engineering limits admit no common solution, no algorithm can supply one. The optimization then returns the point of least violation, and the limits themselves must be revisited.

\section*{Acknowledgments}
We thank Tobias Blickhan, Matt Landreman, and Joshua Burby for useful discussions.
We acknowledge funding support through the Simons Foundation under award 560651 and through a Google Award for Machine Learning Research and Education
with TPUs. This work was also supported by a Google.org Impact Challenge: AI for Science award for the project ``Transport-consistent optimization of high-beta stellarators,'' led by J.~W.~Burby at the University of Texas at Austin. 
This work was supported in part through the University of Washington’s Tillicum GPU-Accelerated Research Computing Platform, operated by UW-IT Research Computing. This work used the DeltaAI system at the National Center for
Supercomputing Applications, award OAC 2320345, through allocation PHY260492 from the NSF-funded Advanced Cyber-infrastructure Coordination Ecosystem: Services \& Support program (ACCESS). 

\section*{AI Use Statement}
\label{sec:AI_use}
Claude Opus 5.5 (Anthropic), a large language model, was used to polish the code and to prepare the manuscript, including drafting and revising its text. The authors directed all of its use, checked every change to the code against the gradient tests and the results reported here, and reviewed and revised all text. The authors take full responsibility for the content of this paper.

\section*{Data Availability}
\label{sec:data}
The XCoil coil solver and the QSS optimization code are available on GitHub at the tagged releases used for this paper: qss-filaments, release v1.0.0 (\url{https://github.com/PedroFranciscoGil/qss-filaments/releases/tag/v1.0.0}),
and XCoil, release v2.0.0 (\url{https://github.com/PedroFranciscoGil/XCoil/releases/tag/v2.0.0}), on which qss-filaments v1.0.0 depends. The fixed-boundary target boundaries and equilibria, the DESC free-boundary equilibria, the certified coil sets with their currents and the alpha-loss tables of every design in Table~\ref{tab:coils_all} are archived on Zenodo at \url{https://doi.org/10.5281/zenodo.24418301}.

\appendix
\section{Fast-particle and equilibrium constraints}
\label{app:errors}
The quasisymmetry error is the two-term residual of Refs.~\cite{Rodríguez_Paul_Bhattacharjee_2022, LandremanPaul}. A field is quasisymmetric with helicity $(M_B,N_B)$ when $|\mathbf B|$ depends on the Boozer angles of Appendix~\ref{app:alpha} only through $M_B\theta_B-N_B\phi_B$. The QS error measures the departure from this condition on a set of surfaces $s_i$,
\begin{equation}
\begin{aligned}
f_{\mathrm{QS}}=\sum_i\bigg\langle\frac{1}{B^6}\Big[&(N_B-\iota M_B)\,\mathbf B\times\nabla B\cdot\nabla\psi\\
&-(M_BG+N_BI)\,\mathbf B\cdot\nabla B\Big]^2\bigg\rangle_{s_i},
\end{aligned}
\label{eq:qs}
\end{equation}
where $\langle\cdot\rangle_{s_i}$ is the flux-surface average on the surface $s_i$, $\psi$ the toroidal flux function of Sec.~\ref{sec:equilibrium}, $\iota$ the rotational transform, and $G$ and $I$ the poloidal and toroidal current functions of Appendix~\ref{app:alpha}. The expression in brackets vanishes on every surface of a quasisymmetric field. The residual vector $\mathbf q$ of Sec.~\ref{sec:outer} collects the values of this expression divided by $B^3$ on an angular grid on each surface, weighted so that the sum of their squares gives the flux-surface averages, so that $f_{\mathrm{QS}}=\|\mathbf q\|^2$. It does not require a Boozer transform, since the expression is computed directly from the VMEC representation of the equilibrium. For the quasi-axisymmetric configurations, the helicity is $(M_B,N_B)=(1,0)$ and $f_{\mathrm{QS}}$ is evaluated on the 11 surfaces $s=0,0.1,\dots,1$. For the quasi-helically symmetric configurations, the helicity is $(1,-N_{\mathrm{fp}})$.

The aspect ratio and the elongation describe the shape of the plasma. The aspect ratio is that of VMEC,
\begin{equation}
A=\frac{R_0}{a},\qquad a=\sqrt{\bar{\mathcal S}/\pi},\qquad R_0=\frac{V}{2\pi^2a^2},
\label{eq:aspect}
\end{equation}
where $V$ is the plasma volume and $\bar{\mathcal S}$ the toroidal average of the area of its cross-sections. The elongation follows Ref.~\cite{Goodman_Camacho}. The cross-section of the boundary by the plane normal to the magnetic axis at the toroidal angle $\phi$ has an area $\mathcal S(\phi)$ and a perimeter $\mathcal P(\phi)$. The ellipse with the same area and perimeter has the semi-axes $a_e\ge b_e$, and the elongation is the largest ratio of the two,
\begin{equation}
E=\max_\phi\frac{a_e(\phi)}{b_e(\phi)}.
\label{eq:elongation}
\end{equation}

Among the constraints $\mathbf h$, the mirror ratio follows Ref.~\cite{Goodman_Camacho} and the magnetic well Ref.~\cite{landreman2021stellarator}. The mirror ratio of a surface is
\begin{equation}
\Delta=\frac{B_{\max}-B_{\min}}{B_{\max}+B_{\min}},
\label{eq:mirror}
\end{equation}
with the extrema of $|\mathbf B|$ on that surface. We evaluate and bound it on the innermost surface. The magnetic well is
\begin{equation}
W=\frac{V'(0)-V'(1)}{V'(0)},
\label{eq:well}
\end{equation}
where $V(s)$ is the volume enclosed by the surface $s$ and the prime denotes $d/ds$. A positive $W$ means that $V'$ decreases outwards, a magnetic well, which is favourable for interchange stability. A negative $W$ is a magnetic hill. The Mercier criterion is reported as
\begin{equation}
\hat D_\mathrm{Merc}=D_\mathrm{Merc}\,\Psi_\mathrm{edge}^2,
\label{eq:dmerc}
\end{equation}
where $D_\mathrm{Merc}$ is the Mercier coefficient evaluated by VMEC, the sum of its shear, current, well and geodesic-curvature terms, and $\Psi_\mathrm{edge}=2\pi\psi_{\mathrm{LCFS}}$ is the total toroidal flux. It is positive for stability. Unless a radius is given, the tables report its minimum on $0.15\le s\le0.95$.

The distance between the free-boundary LCFS $\Gamma_\mathrm{fb}$ and the target boundary $\Gamma$ is measured in 32 poloidal planes per field period. For a point $\mathbf x$ of $\Gamma_\mathrm{fb}$ in one of these planes,
\begin{equation}
d(\mathbf x)=\min_{\mathbf x'\in\Gamma}|\mathbf x-\mathbf x'|,
\label{eq:lcfs_dev}
\end{equation}
with $\mathbf x'$ in the same plane, and $\delta_\mathrm{LCFS}$ denotes the mean and the maximum of $d$ over the points of all planes.

For the quasi-isodynamic configurations, the objective is the QI error. 
On each surface $s_i$, the field strength is evaluated in Boozer coordinates along $N_\alpha$ field lines $\alpha_j=\theta_B-\iota\phi_B$ at $N_\phi$ points $\phi_l$ over one field period, and normalized as $b=(B-B_{\min})/(B_{\max}-B_{\min})$ with the extrema of the surface. From this field, a perfectly quasi-isodynamic target $b^{\mathrm{QI}}$ is constructed by the squash, stretch and shuffle operations of Ref.~\cite{Goodman_Camacho}. Every magnetic well is stretched to the full range of $b$, and its bounce distances are replaced by their average over the field lines, placed symmetrically about the centre of the well, so that the maxima of $b$ lie at the ends of the field period. The QI error is the distance between the two fields,
\medmuskip=0mu
\begin{equation}
f_{\mathrm{QI}}=\frac{1}{N_\alpha N_\phi}\sum_{i,j,l}\Big[b(s_i,\alpha_j,\phi_l)-b^{\mathrm{QI}}(s_i,\alpha_j,\phi_l)\Big]^2 ,
\label{eq:qi}
\end{equation}
\medmuskip=4mu
which vanishes for an exactly quasi-isodynamic field. The factor $1/(N_\alpha N_\phi)$ makes every surface contribute the mean squared difference over its grid, so that the value does not depend on the angular resolution. All QI errors quoted in this paper use this normalization. The QI errors in the tables are those of the free-boundary equilibria, summed over the five surfaces $s=0.1$, 0.3, 0.5, 0.7 and 0.9. Unlike the original construction, which splits each field line at the grid point of its minimum, we obtain the bounce points of every well from the exact measure of its sublevel sets on the piecewise-linear interpolant of $b$, so that $f_{\mathrm{QI}}$ is a continuous and differentiable function of the boundary. Like $f_{\mathrm{QS}}$, $f_{\mathrm{QI}}$ is a sum of squares, and its residuals are used for the least-squares subproblems of Sec.~\ref{sec:outer} in the quasi-isodynamic cases.

Two fast-particle constraints act on the second adiabatic invariant. A trapped particle conserves $B^*=\mathcal E/\mu$, the ratio of its energy to its magnetic moment, and
\begin{equation}
J(s,\alpha,B^*)\propto|G+\iota I|\int\sqrt{1-B/B^*}\,\frac{d\phi_B}{B},
\label{eq:maxj_J}
\end{equation}
where the integral runs along the field line $\alpha$ between consecutive bounce points of one magnetic well, and the omitted factor $\sqrt{2\mu B^*/m}$ is constant at fixed $B^*$. We label trapped particles by the depth $K$ of their bounce points, $B^*=B_{\min}+K\,(B_{\max}-B_{\min})$, with the extrema of $|\mathbf B|$ on the surface, so that $K\to0$ is the most deeply trapped particle and $K\to1$ the barely trapped one. The particle therefore drifts on a contour of $J$ at fixed $B^*$, and it is confined if this contour is closed, so that the contours that start on an inner surface cannot reach the edge.

The \emph{$J$-closure constraint}, used at mid-well depths for design (d) of Sec.~\ref{sec:qi_cold}, asks for closed contours of $J$ on the field of the equilibrium. A particle born in the core stays confined if the contour of $J$ through its birth point closes before it reaches the edge. The maximum-$J$ property guarantees this but is a stronger requirement, and a low QI error guarantees it only in the limit of an exactly quasi-isodynamic field. The closure constraint asks for it directly, for the particles born on the innermost surfaces, where most of the fusion reactions take place. We evaluate $J$ on 15 surfaces between $s=0.02$ and $0.9$, along 36 field lines, and at several values of $B^*$, at the depths $K=0.03$, 0.1, 0.2, 0.35, 0.5, 0.65 and 0.8 of the range of $|\mathbf B|$ that is trapped on every field line. A starting point $(s_i,\alpha_j)$ on one of the eight innermost surfaces is enclosed if some outer surface $s_c$ has its whole range of $J$ on one side of $J(s_i,\alpha_j)$, that is, if
\begin{multline}
o_{ij}=\min_{c>i}\,\min\Bigg[\frac{\max_\alpha J(s_c)-J(s_i,\alpha_j)}{\bar J_{ic}},\\
\frac{J(s_i,\alpha_j)-\min_\alpha J(s_c)}{\bar J_{ic}}\Bigg]<0,
\end{multline}
where $\bar J_{ic}$ is the mean of $J$ on the two surfaces. The constraint is then the mean opening $\delta=\langle\ln(1+e^{o_{ij}/w})\rangle_{i,j}$, with $w=10^{-3}$. This quantity vanishes when every start is enclosed and is about $\ln2$ at the edge of enclosure. Therefore, the enforced inequality is $\delta\le0.1$ for every $B^*$.

The \emph{maximum-$J$ constraint}, used for design (d) of Sec.~\ref{sec:qi_cold} and for the reactor-like configuration, follows Ref.~\cite{goodman_prx} and asks for $J$ to decrease outwards, $\partial J/\partial s<0$, which closes the contours, is favourable for the stability of trapped-particle modes, and provides stability against Trapped-Electron Modes (TEM), a type of drift-wave microinstability. It is evaluated on the constructed field $b^{\mathrm{QI}}$ of the QI error, on which $J$ does not depend on $\alpha$. On every surface, the constructed field is fixed by the mean bounce distances of its wells, which are piecewise linear in the level of $b$, so Eq.~\eqref{eq:maxj_J} has a closed form, smooth in the Boozer spectrum. For each pair of adjacent surfaces of the QI error, $s=0.1,0.3,\dots,0.9$, the values of $B^*$ lie at the depths 0.35, 0.55 and 0.65 of the range of $|\mathbf B|$ trapped on both surfaces, measured from its top, that is, at $K=0.65$, 0.45 and 0.35, and the constraint requires
\begin{equation}
\frac{1}{\bar J}\,\frac{J(s_{i+1},B^*)-J(s_i,B^*)}{s_{i+1}-s_i}\le-0.06,
\end{equation}
for every pair and every $B^*$, where $\bar J$ is the mean of the two values.

In both constraints, all maxima and minima are replaced by log-sum-exp soft maxima and minima, so that each constraint is a single differentiable function of the boundary. The soft minima over the outer surfaces and over the two sides of the enclosure test make it slightly less conservative than the hard test, while the soft extrema over the field lines make it more conservative. At the sharpness used, the difference between the soft and the hard values is small compared with the tolerance on $\delta$.

\subsection{Equilibrium settings}
\label{app:vmex_settings}
The fixed-boundary equilibria of the optimization use the following VMEX settings. These are a radial ladder of 16 and 50 surfaces with force tolerances of $10^{-16}$ and $10^{-11}$, and maximum poloidal and toroidal mode numbers $m_\theta=n_\phi=m_{\max}+2$ in the cold-starts when the boundary modes up to $m,|n|\le m_{\max}$ are free. The angular grid is $16\times14$ for $m_{\max}\le4$ and follows the rule of VMEC ($2\,m_\theta+6$ by $2\,n_\phi+4$, i.e., $20\times18$) for $m_{\max}=5$. On these settings, VMEX reproduces the quasisymmetry, aspect ratio, rotational transform and elongation of VMEC2000 to $10^{-8}$.

\section{Warm-started quasi-axisymmetric designs}
\label{app:warm}
This appendix applies the method to a published configuration, the precise quasi-axisymmetric design of Landreman and Paul~\cite{LandremanPaul}, in vacuum and at finite pressure. These warm-starts complement the cold-starts of Sec.~\ref{sec:cold}. They show how far the method improves a design that is already highly optimized, when the coils are designed together with the plasma.

\subsection{Comparison methodology}
\label{sec:protocol}
In the comparisons of this appendix, the two-stage baseline and the QSS design start from the same published configuration. The two-stage reference keeps the published boundary and fits coils to it with the final coil stage of Sec.~\ref{sec:methods_coils}. For the LP QA designs, the QSS design is obtained by the flux-bound continuation of Sec.~\ref{sec:homotopy}, started from the same boundary, with the boundary modes up to $m,|n|\le5$ free (120 coefficients) and the higher modes held at their published values. The goal of the QSS design is to change the plasma shape in order to obtain more coil-compatible equilibrium shapes, while the two-stage reference keeps the published shape. For LP QA, we choose as the QSS design the certified point of the continuation with the lowest optimal squared flux whose fixed-boundary quasisymmetry error does not exceed $5\times10^{-4}$, the relaxation of quasisymmetry we accept for this comparison. 
The free-boundary equilibria of all the final designs and their two-stage references are verified independently with DESC in Appendix~\ref{app:desc}. 

\subsection{Quasi-axisymmetric stellarators in vacuum and at finite pressure}
\label{sec:lpqa}
\label{sec:fbqa}
We consider the precise quasi-axisymmetric design of Landreman and Paul~\cite{LandremanPaul}, with two field periods, in vacuum and in the finite-pressure variant, with $\beta=1.05\%$, $p\propto1-s$ and a self-consistent Redl bootstrap current of $-1.15$~MA. The fixed-boundary target of this variant has $\langle\beta\rangle=0.91\%$. Its free-boundary equilibria are computed with the design currents of the coils, and their toroidal flux is the flux of the coil field through the target cross-section, which is 7\% lower than that of the target. The field strength is lower by the same fraction at the same pressure, and the volume-averaged value becomes $\langle\beta\rangle=1.05\%$. We quote this free-boundary value throughout. 
The physics constraints are the values of each design itself and are given in Table~\ref{tab:qa_all}. All coil sets have $N_c=5$ and the same coil limits. The table also lists design (c) of Fig.~\ref{fig:cold}, cold-started on the same problem, and its two-stage counterpart. In vacuum, the cold-started QSS design is close to the warm-started one in every plasma and coil property, and the two two-stage designs are close to each other.
All finite-pressure values of $\bn$ reported in this manuscript include the plasma field.

\begin{table*}[!t]
\caption{The six designs of the LP QA family, all with $N_{\mathrm{fp}}=2$ and $N_c=5$. The constraint column gives the vacuum and finite-pressure values. The cold-started designs are design (c) of Fig.~\ref{fig:cold} and its two-stage counterpart of Sec.~\ref{sec:cold_start_lpqa}.}
\label{tab:qa_all}
\begin{ruledtabular}
\footnotesize
\begin{tabular}{lccccccc}
 & constraint & \multicolumn{2}{c}{warm-start, vacuum} & \multicolumn{2}{c}{cold-start, vacuum} & \multicolumn{2}{c}{warm-start, $\beta=1.05\%$} \\
 & vacuum, $\beta$ & two-stage & QSS & two-stage & QSS & two-stage & QSS \\
\hline
start & & LP QA & LP QA & cold-start & cold-start & LP QA & LP QA \\
$\langle\beta\rangle$ [\%] & & -- & -- & -- & -- & 1.05 & 1.04 \\
$A$ / $E$ & 6 / $\le4.6$ & 6.01 / 4.59 & 6.00 / 3.78 & 6.01 / 4.74 & 6.00 / 3.99 & 6.00 / 4.42 & 6.00 / 3.91 \\
$\iota(0)$ / $\iota(1)$ & $\bar\iota=0.42$, 0.50 & 0.43 / 0.41 & 0.43 / 0.40 & 0.43 / 0.40 & 0.43 / 0.40 & 0.39 / 0.51 & 0.37 / 0.50 \\
$W$ [\%] & $\ge-3.0$, $\ge1.6$ & \best{$-2.8$} & $-3.0$ & $-2.9$ & \best{$-2.6$} & 2.3 & \best{2.6} \\
fixed-boundary $f_{\mathrm{QS}}$ [$10^{-5}$] & & \best{0.051} & 8.1 & \best{0.069} & 12 & 9.6 & \best{3.1} \\
$f_{\mathrm{QS}}$ [$10^{-3}$] & & 7.9 & \best{0.62} & 8.0 & \best{0.74} & 3.4 & \best{0.68} \\
$\delta_\mathrm{LCFS}$ mean / max [mm] & & 37 / 159 & \best{2.7} / \best{8.9} & 51 / 209 & \best{3.7} / \best{14} & 19 / 86 & \best{8.0} / \best{27} \\
$\alpha$-loss [\%] & & 20.6 & \best{6.1} & 20.6 & \best{8.0} & 16.3 & \best{3.3} \\
\hline
$\kappa_{\max}$ / $\kappa_\mathrm{MSC}$ [m$^{-1}$ / m$^{-2}$] & & 0.30 / 0.039 & \best{0.28} / 0.039 & 0.30 / 0.039 & \best{0.28} / 0.039 & 0.30 / 0.039 & 0.30 / 0.039 \\
$\bar\tau$ / $\gamma_c$ [m$^{-1}$ / $10^{-3}$~m$^{-2}$] & & 0.18 / 0.00 & \best{0.14} / 0.01 & 0.18 / 0.00 & \best{0.14} / 0.01 & 0.18 / 0.00 & 0.17 / 0.00 \\
$d_{cc}$ / $d_{cs}$ [m] & & \best{1.47} / 1.90 & 1.12 / \best{2.38} & \best{1.48} / 1.91 & 1.16 / \best{2.32} & \best{1.42} / 2.06 & 1.19 / \best{2.43} \\
$\bn$ [\%] & & 0.223 & \best{0.053} & 0.20 & \best{0.056} & 0.133 & \best{0.054} \\
\hline
\textit{shared coil limits} & \multicolumn{7}{c}{$M=5$, $L_i\le40.6$~m, $\kappa_{\max}\le0.30$~m$^{-1}$, $\kappa_\mathrm{MSC}\le0.039$~m$^{-2}$, $d_{cc}\ge1.01$~m, $d_{cs}\ge1.52$~m} \\
\end{tabular}
\end{ruledtabular}
\end{table*}

\begin{figure*}[!t]
\includegraphics[width=\textwidth]{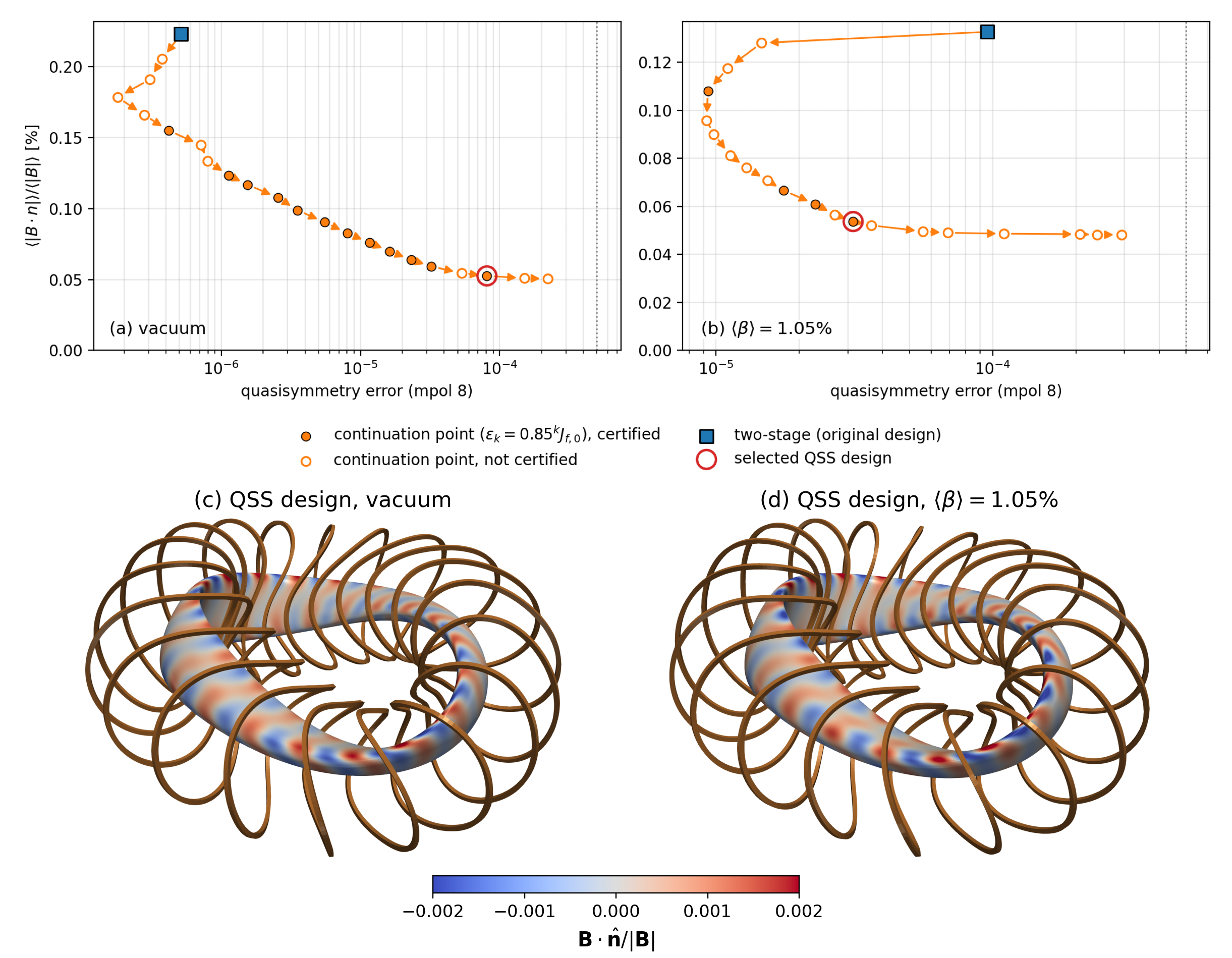}
\caption{Flux-bound continuations of LP QA with $N_c=5$ starting from the past coil designs, shown as squares, (a) in vacuum and (b) at $\beta=1.05\%$. Filled points are certified, open points are not. The circled points are the QSS designs, the lowest-flux certified points with a fixed-boundary quasisymmetry error of at most $5\times10^{-4}$ (dotted lines). (c), (d) The selected QSS designs in vacuum ($k=19$) and at $\beta=1.05\%$ ($k=12$).}
\label{fig:qa_homotopy}
\label{fig:fbqa_homotopy}
\end{figure*}

The two flux-bound continuations and the two selected designs are shown in Fig.~\ref{fig:qa_homotopy}. Their first steps lower the fixed-boundary quasisymmetry error and $\bn$ together, so even these highly optimized published designs are not optimal for coils. The normal field $\bn$ then falls steadily; comparing the QSS designs with the two-stage references in Table~\ref{tab:qa_all}, the QSS method reduces $\bn$ several-fold in vacuum and at finite pressure. All physics constraints are met and, since the length and mean-squared-curvature limits are active for every coil in all four designs, \textit{the improvement comes entirely from the plasma shape}. With the QSS coils, the free-boundary quasisymmetry error is lower by an order of magnitude in vacuum and several-fold at finite pressure, and the LCFS lies much closer to its target. In vacuum, the two-stage coils produce large islands just outside the target boundary, while the field lines of the QSS coils stay close to it (Appendix~\ref{app:desc}). A plasma shape that departs only slightly from LP QA thus produces up to an order of magnitude better free-boundary quasisymmetry if coils are jointly optimized.

A lower normal field is only useful if it translates into a better plasma. To check whether this is the case for fast particles, we trace fusion-born alpha particles in the free-boundary equilibria of the coil sets of Table~\ref{tab:qa_all}, as described in Appendix~\ref{app:alpha}. These equilibria must contain the ripple of the discrete coils, which lies around $n=2N_c$ per field period for $N_c$ coils per half period ($n=10$ for LP QA), in a broad band at about $n=9$--14. A VMEX free-boundary equilibrium at $n_\phi=8$ cannot represent it. We therefore trace the alphas in the DESC free-boundary equilibria, which resolve this ripple. 

The alpha losses by birth radius are given in Table~\ref{tab:alpha} of Appendix~\ref{app:alpha}. At equal coil limits, the QSS coils lose 3.4 times fewer alphas within 0.2~s than the two-stage coils in vacuum, and 4.9 times fewer at $\langle\beta\rangle=1.05\%$.
These results indicate that the QSS method works in an optimization with filamentary coils and that it can be used to design stellarators efficiently.

\section{Alpha-particle calculations}
\label{app:alpha}
This appendix describes how the alpha losses are computed, gives them by birth radius, and documents one detail of the field representation that affects the alphas born near the magnetic axis.

The alphas are born with a reactor D--T profile, with fuel densities $n_D=n_T\propto1-s^5$ and a temperature $T=11.5\,(1-s)$~keV, as in Ref.~\cite{Bader2021}, and the Bosch--Hale reactivity~\cite{BoschHale1992}. The orbits are followed in Boozer coordinates $(s,\theta_B,\phi_B)$~\cite{Boozer1981}. In these coordinates the field lines are straight and the field is $\mathbf B=G(s)\nabla\phi_B+I(s)\nabla\theta_B+B_s\nabla s$, where $2\pi G/\mu_0$ is the poloidal current outside the surface $s$ and $2\pi I/\mu_0$ the toroidal current inside it. The Jacobian of these coordinates is proportional to $|G+\iota I|/B^2$, with $\iota$ the rotational transform. For each equilibrium, 8{,}000 start points are drawn from the birth rate $n_Dn_T\langle\sigma v\rangle$ weighted by this Jacobian, with an isotropic pitch distribution. About 33, 35, 27 and 5\% of the alphas are born in $0\le s<0.1$, $0.1\le s<0.25$, $0.25\le s<0.5$ and $s\ge0.5$, respectively. Collisionless guiding-center orbits of 3.5~MeV alphas are then integrated for 0.2~s with CATAPULT, the GPU tracer of FIRM3D~\cite{FIRM3D}, in Boozer coordinates with $m,n\le32$ and 60 toroidal cells per field period (with 30 cells, the totals agree within the sampling error), and an alpha is counted as lost when it reaches $s=1$.

Table~\ref{tab:alpha} lists the fraction of the alphas lost within 0.2~s for the LP QA comparison of Appendix~\ref{sec:lpqa}, the reactor-like design and W7-X, in four bands of the birth radius. The totals also report the binomial sampling error of the 8{,}000 alphas traced in each equilibrium.

One detail of the field representation turned out to matter. FIRM3D extends the Boozer harmonics of $|B|$ to the magnetic axis by linear extrapolation from the two innermost surfaces, so that the $m\ge1$ harmonics do not vanish at $s=0$ and $|B|$ varies around the axis. Without correction, this loses a quarter or more of the alphas born at $s<0.011$, independently of the resolution of the field table. We therefore set the $s=0$ value of every $m\ge1$ harmonic to zero before building the interpolant, which changes the harmonics by less than $10^{-7}$~T at $s\ge0.2$. With this correction, CATAPULT agrees with SIMPLE~\cite{Albert_Kasilov_Kernbichler_2020} to within 5 particles in 1000 for 1000 alphas started on single surfaces ($s_0=0.1$--$0.75$) in two quasi-isodynamic equilibria derived from Wendelstein 7-X.

\begin{table*}[!t]
\caption{$\alpha$-loss by birth radius, in percent, for the warm-started LP QA coil sets of Table~\ref{tab:qa_all}, the reactor-like design of Sec.~\ref{sec:reactor} and W7-X of Table~\ref{tab:reactor}.}
\label{tab:alpha}
\begin{ruledtabular}
\begin{tabular}{lcccccc}
 & \multicolumn{2}{c}{LP QA, vacuum} & \multicolumn{2}{c}{LP QA, $\langle\beta\rangle=1.05\%$} & reactor-like & \\
birth $s$ & two-stage & QSS & two-stage & QSS & design & W7-X \\
\hline
0--0.1 & 16.5 & \best{0.5} & 11.0 & \best{0.0} & \best{0.7} & 22.9 \\
0.1--0.25 & 20.2 & \best{3.7} & 16.4 & \best{1.1} & \best{3.9} & 27.4 \\
0.25--0.5 & 24.8 & \best{12.7} & 20.4 & \best{7.6} & \best{8.3} & 30.8 \\
0.5--1 & 27.1 & \best{22.3} & 28.5 & \best{17.2} & \best{16.2} & 33.1 \\
\hline
total & $20.6\pm0.5$ & \best{$6.1\pm0.3$} & $16.3\pm0.4$ & \best{$3.3\pm0.2$} & \best{$4.7\pm0.2$} & $27.1\pm0.5$ \\
\end{tabular}
\end{ruledtabular}
\end{table*}

\begin{figure*}[!t]
\includegraphics[width=0.92\textwidth]{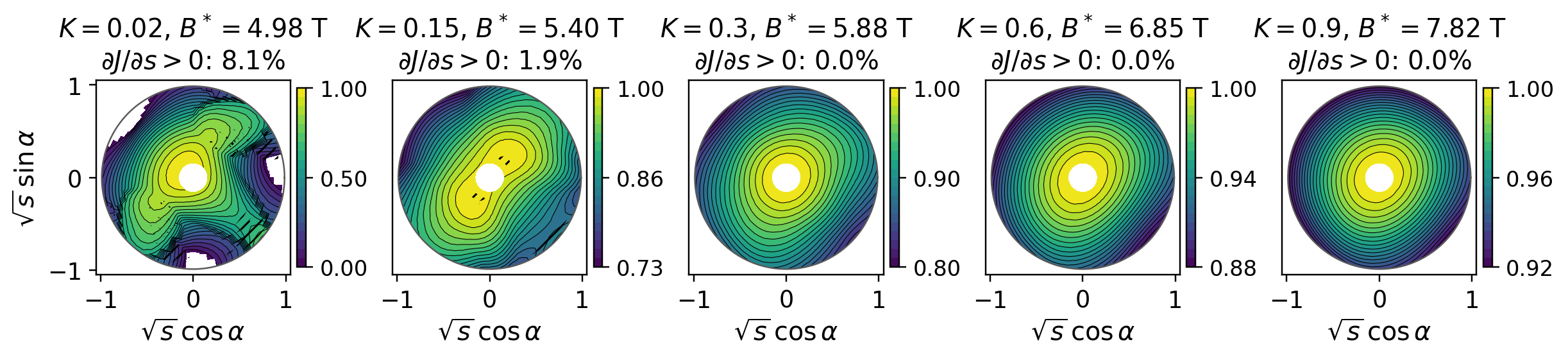}
\caption{Second adiabatic invariant $J(\sqrt s,\alpha)$ of the reactor-like design of Sec.~\ref{sec:reactor}, in the DESC free-boundary equilibrium of its certified coils, at fixed trapping depth $K$ of Appendix~\ref{app:errors}, for $K=0.02$, 0.15, 0.30, 0.6 and 0.9, normalized to its maximum in each panel. The radial coordinate is $\sqrt s$ and the angle is the field-line label $\alpha$. Hatched regions mark $\partial J/\partial s>0$, where the maximum-$J$ property is violated. The panel titles give their share of the cross-section for $0.05<s<0.9$.}
\label{fig:reactor_maxj}
\end{figure*}

\subsection{Trapped particles and stability of the reactor-like design}
\label{app:reactor}
This appendix shows the trapped-particle and stability properties of the reactor-like design that are summarized in Sec.~\ref{sec:reactor_physics}. Figure~\ref{fig:reactor_maxj} shows the second adiabatic invariant of the trapped particles in the free-boundary equilibrium of the certified coils. The optimization bounds the Mercier criterion from below by $s\hat D_\mathrm{Merc}\ge0.020$ on $0.15\le s\le0.95$, and the fixed-boundary design lies on this bound over most of that range. The effective ripple is computed with NEO\_JAX~\cite{NEOJAX} on the Boozer spectrum with $m,n\le32$.

\section{Toroidal-ordering constraint}
\label{app:slot}
The NCSX-like design (Sec.~\ref{sec:ncsx78}), the reactor-like design (Sec.~\ref{sec:reactor}) and the start-up coil set (Sec.~\ref{sec:startup}) use an additional coil constraint that keeps neighbouring coils in their toroidal order, so that they cannot cross. This is required as coils of QI configurations at finite $\beta$ tend to cross toroidally while keeping coil-to-coil distances satisfied. Each coil point is described by its toroidal angle $\phi$ and by a geometric poloidal angle $\vartheta$, which differs from the angle $\theta$ of the boundary parametrization, about the geometric axis of the boundary, the curve $R_0(\phi)=\sum_n r_{0n}\cos(nN_{\mathrm{fp}}\phi)$, $Z_0(\phi)=\sum_n z_{0n}\sin(nN_{\mathrm{fp}}\phi)$ traced by the poloidally averaged $R$ and $Z$ of the boundary at each $\phi$, that is, by its $m=0$ Fourier modes. For a pair of neighbouring coils, left (L) and right (R) in $\phi$, and every point $p$ of the left coil, the constraint asks
\begin{multline}
\Phi_p=\operatorname*{softmin}_{p'}\Big[\phi_{\mathrm R,p'}-\phi_{\mathrm L,p}\\
+\gamma\big(1-\cos(\vartheta_{\mathrm L,p}-\vartheta_{\mathrm R,p'})\big)\Big]\ge\Delta\phi_{\min},
\label{eq:slot}
\end{multline}
so that the right coil stays at least $\Delta\phi_{\min}=1.5^\circ$ further in $\phi$ than the left coil at the same poloidal angle. The term with $\gamma=20$ removes the points of the right coil at other poloidal angles from the minimum, and the soft minimum, a log-sum-exp with a sharpness of 500~rad$^{-1}$, lies below the hard minimum, so the constraint is conservative. The pairs are all neighbouring coils of a half field period and the first and last coils with their stellarator-symmetric images across $\phi=0$ and $\phi=\pi/N_{\mathrm{fp}}$. The rows are normalized by the toroidal width $\pi/(N_{\mathrm{fp}}N_c)$ available to each coil and reduced by a soft maximum to 16 segments per coil pair. Since the reference axis is built from the $m=0$ modes of the boundary, we fix the axis so that the constraint does not depend on the boundary through these modes, to avoid this dependence entering the envelope gradient of Appendix~\ref{app:envelope}.

\section{Derivation of the envelope gradient}
\label{app:envelope}
This appendix derives the gradient of the coil term $J_f^\star$ with respect to the plasma boundary, which the plasma problem needs at every iteration. The vacuum case is treated first. At finite pressure the plasma field adds a second term, which is derived afterwards.

\subsection{Vacuum fields}
\label{app:envelope_vac}
XCoil solves Eq.~\eqref{eq:inner} in the scaled form $\min_{\mathbf c}J_f(\mathbf c,\mathbf s)/J_{\mathrm{ref}}$ subject to $g_i(\mathbf c,\mathbf s)\le0$, $i\in\mathcal I$. The total current of the coils is fixed before the optimization and is not part of $\mathbf c$, so the problem has no equality constraints. Its Lagrangian is
\begin{equation}
\mathcal L(\mathbf c,\bm\lambda;\mathbf s)=\frac{J_f(\mathbf c,\mathbf s)}{J_{\mathrm{ref}}}+\sum_{i\in\mathcal I}\lambda_i\,g_i(\mathbf c,\mathbf s).
\label{eq:app_lagrangian}
\end{equation}
At a local solution $\mathbf c^\star(\mathbf s_0)$, the KKT conditions hold,
\begin{align}
\frac{1}{J_{\mathrm{ref}}}\nabla_{\mathbf c}J_f+\sum_{i}\lambda_i^\star\nabla_{\mathbf c}g_i&=0,\label{eq:app_stat}\\
g_i\le0,\quad\lambda_i^\star\ge0,\quad\lambda_i^\star g_i&=0.\label{eq:app_comp}
\end{align}
Let $\mathcal A=\{i:g_i(\mathbf c^\star,\mathbf s_0)=0\}$ be the active set. We assume (i) linear independence of the gradients $\nabla_{\mathbf c}g_i$, $i\in\mathcal A$, (ii) strict complementarity, $\lambda_i^\star>0$ for $i\in\mathcal A$, and (iii) the second-order sufficient condition on the tangent space of the active constraints. Under (i)--(iii), the implicit function theorem applied to the system \eqref{eq:app_stat} together with $g_i=0$ ($i\in\mathcal A$) gives a locally unique, continuously differentiable solution $(\mathbf c^\star(\mathbf s),\bm\lambda^\star(\mathbf s))$ with the same active set in a neighborhood of $\mathbf s_0$~\cite{MilgromSegal2002}.

Differentiate the value function $J_f^\star(\mathbf s)=J_f(\mathbf c^\star(\mathbf s),\mathbf s)$ along a direction $\mathbf d$, writing $\dot{\mathbf c}=D\mathbf c^\star[\mathbf d]$,
\begin{equation}
\frac{dJ_f^\star}{d\mathbf s}\cdot\mathbf d=\nabla_{\mathbf s}J_f\cdot\mathbf d+\nabla_{\mathbf c}J_f\cdot\dot{\mathbf c}.
\label{eq:app_chain}
\end{equation}
The second term involves the unknown coil response $\dot{\mathbf c}$. Stationarity \eqref{eq:app_stat} replaces $\nabla_{\mathbf c}J_f$ by the constraint gradients,
\begin{equation}
\nabla_{\mathbf c}J_f\cdot\dot{\mathbf c}=-J_{\mathrm{ref}}\sum_i\lambda_i^\star\nabla_{\mathbf c}g_i\cdot\dot{\mathbf c}.
\end{equation}
Each term vanishes or is known. For inactive constraints, $\lambda_i^\star=0$. The active constraints stay active along the solution path, $g_i(\mathbf c^\star(\mathbf s),\mathbf s)=0$, so their total derivative vanishes,
\begin{equation}
\nabla_{\mathbf c}g_i\cdot\dot{\mathbf c}+\nabla_{\mathbf s}g_i\cdot\mathbf d=0,\qquad i\in\mathcal A.
\end{equation}
Substituting into Eq.~\eqref{eq:app_chain},
\begin{equation}
\frac{dJ_f^\star}{d\mathbf s}=\nabla_{\mathbf s}J_f(\mathbf c^\star,\mathbf s)+J_{\mathrm{ref}}\sum_{i}\lambda_i^\star\,\nabla_{\mathbf s}g_i(\mathbf c^\star,\mathbf s),
\label{eq:envelope}
\end{equation}
which is the envelope gradient used in the plasma optimization. The coil response $\dot{\mathbf c}$ has dropped out, and only partial derivatives at fixed coils and multipliers remain. Equivalently, $dJ_f^\star/d\mathbf s=J_{\mathrm{ref}}\,\nabla_{\mathbf s}\mathcal L$ at the KKT point. Of the engineering limits, only the coil--surface distance depend on $\mathbf s$ (if one does not fix the axis, the toroidal-ordering constraint depends on $\mathbf s$ through the $m=0$ modes of the boundary). The coil-surface distance then enters the gradient through its multiplier, and its partial derivative with respect to $\mathbf s$ is obtained by automatic differentiation. When this constraint is inactive, the gradient reduces to $\nabla_{\mathbf s}J_f$ at fixed coils.

\paragraph{Remarks.}
(a) Assumption (ii) fails where a constraint enters or leaves the active set. The coils then respond with a kink, but the multiplier of that constraint vanishes there, so Eq.~\eqref{eq:envelope} is continuous across the change and $J_f^\star$ remains continuously differentiable under (i) and (iii)~\cite{BonnansShapiro2000}. The homotopy points of this paper have strictly positive multipliers on the active length and mean-squared-curvature rows. (b) If (i) fails, the multipliers are not unique, but $\nabla_{\mathbf s}\mathcal L$ is the same for every KKT multiplier vector as long as only $\mathbf s$-independent constraints are degenerate. (c) XCoil's objective also contains the arclength gauge term $w_{\mathrm{arc}}R_{\mathrm{arc}}(\mathbf c)$, which does not depend on $\mathbf s$. Repeating the argument with this term shows that Eq.~\eqref{eq:envelope} is exactly the gradient of $J_{\mathrm{ref}}$ times the solved problem's value, which differs from $J_f^\star$ by $w_{\mathrm{arc}}J_{\mathrm{ref}}R_{\mathrm{arc}}(\mathbf c^\star)$. The Taylor tests compare the gradient against $J_f^\star$ itself and show no discrepancy above their noise floors. (d) The coil problem is nonconvex, and Eq.~\eqref{eq:envelope} is the gradient of the local minimum that the coil solver follows. The warm-started solves follow this minimum continuously along the plasma optimization, and $J_f^\star$ is smooth along it. If the solver moves to a different local minimum, the value seen by the plasma optimizer has a jump or a kink, and Eq.~\eqref{eq:envelope} then gives the gradient of the new minimum. Such an event affects a single step of the trust-region method, which accepts a step only if the objective decreases. It does not affect the certified coils, which come from a cold re-solve on the final boundary and agree with the in-loop coils to within 1\%, as described in Sec.~\ref{sec:methods_coils}. This agreement is empirical and is no guarantee.

\subsection{Finite pressure}
\label{app:envelope_beta}
At finite $\beta$ the plasma field enters the coil problem through a vector $\mathbf b\in\mathbb R^{n_b}$ that is held fixed within it. Its entries are the cylindrical components of $\mathbf B_{\mathrm{pl}}$ on the virtual-casing grid of $32\times32$ points per field period, interpolated to XCoil's surface points at fixed angles $(\theta,\phi)$ by a Fourier series fitted on that grid. The grid is chosen once, so that $\mathbf B_{\mathrm{pl}}$ is accurate to six digits on the starting boundary, and is kept fixed throughout the optimization making $\mathbf B_{\mathrm{pl}}$ a smooth function of the boundary. The inner problem becomes
\begin{equation}
J_f^\star(\mathbf s,\mathbf b)=\min_{\mathbf c}J_f(\mathbf c,\mathbf s,\mathbf b)\quad\text{s.t.}\quad \mathbf g(\mathbf c,\mathbf s)\le0,
\label{eq:app_inner_beta}
\end{equation}
with $\mathbf b$ entering only through the residual $(\mathbf B_{\mathrm{coil}}+\sigma\mathbf B_{\mathrm{pl}})\cdot\hat{\mathbf n}/|\mathbf B_{\mathrm{coil}}+\sigma\mathbf B_{\mathrm{pl}}|$ of Sec.~\ref{sec:methods_coils}, not through the engineering limits. In the outer problem, $\mathbf b$ is not free. It is computed from the equilibrium,
\begin{equation}
\mathbf b=\mathbf B_{\mathrm{pl}}(\mathbf x,\mathbf p),\qquad \mathbf F(\mathbf x,\mathbf p)=0,\qquad \mathbf p=\mathbf p(\mathbf s),
\label{eq:app_chain_beta}
\end{equation}
where $\mathbf F$ is the VMEX force residual, $\mathbf x$ the converged state and $\mathbf p$ the run parameters, which contain the boundary. The quantity the outer problem constrains is the composite $\widehat J_f^\star(\mathbf s)=J_f^\star(\mathbf s,\mathbf b(\mathbf s))$.

\paragraph{Envelope in $(\mathbf s,\mathbf b)$.}
Treat $(\mathbf s,\mathbf b)$ jointly as the parameter of Eq.~\eqref{eq:app_inner_beta}. With the Lagrangian of Eq.~\eqref{eq:app_lagrangian}, $\mathcal L=J_f/J_{\mathrm{ref}}+\bm\lambda^\top\mathbf g$, and the assumptions (i)--(iii) of Appendix~\ref{app:envelope}, the argument there applies to every parameter component. The coil response drops out, and
\begin{equation}
\begin{aligned}
\frac{\partial J_f^\star}{\partial\mathbf s}&=J_{\mathrm{ref}}\,\nabla_{\mathbf s}\mathcal L\big|_{\mathbf c^\star,\bm\lambda^\star,\mathbf b},\\
\frac{\partial J_f^\star}{\partial\mathbf b}&=J_{\mathrm{ref}}\,\nabla_{\mathbf b}\mathcal L\big|_{\mathbf c^\star,\bm\lambda^\star}=\nabla_{\mathbf b}J_f(\mathbf c^\star,\mathbf s,\mathbf b),
\end{aligned}
\label{eq:app_env_beta}
\end{equation}
since $\mathbf g$ does not depend on $\mathbf b$. The first is Eq.~\eqref{eq:envelope} at fixed plasma field. The second is the cotangent of the plasma field $\mathbf b$, that is, the sensitivity of the optimal flux to the plasma field at fixed coils.

\paragraph{Chain rule through the equilibrium.}
Differentiating the composite along a direction $\mathbf d$,
\begin{equation}
\begin{aligned}
\frac{d\widehat J_f^\star}{d\mathbf s}\cdot\mathbf d&=\frac{\partial J_f^\star}{\partial\mathbf s}\cdot\mathbf d+\Big\langle\frac{\partial J_f^\star}{\partial\mathbf b},\;\dot{\mathbf b}\Big\rangle,\\
\dot{\mathbf b}&=\frac{\partial\mathbf B_{\mathrm{pl}}}{\partial\mathbf x}\dot{\mathbf x}+\frac{\partial\mathbf B_{\mathrm{pl}}}{\partial\mathbf p}\dot{\mathbf p},
\end{aligned}
\label{eq:app_total_beta}
\end{equation}
with $\dot{\mathbf p}=(\partial\mathbf p/\partial\mathbf s)\mathbf d$. The state tangent follows from differentiating the equilibrium condition $\mathbf F(\mathbf x(\mathbf s),\mathbf p(\mathbf s))=0$,
\begin{equation}
\frac{\partial\mathbf F}{\partial\mathbf x}\dot{\mathbf x}+\frac{\partial\mathbf F}{\partial\mathbf p}\dot{\mathbf p}=0
\;\Longrightarrow\;
\dot{\mathbf x}=-\Big(\frac{\partial\mathbf F}{\partial\mathbf x}\Big)^{-1}\frac{\partial\mathbf F}{\partial\mathbf p}\dot{\mathbf p},
\label{eq:app_ift}
\end{equation}
which is what the exact equilibrium Jacobian of Sec.~\ref{sec:equilibrium} computes, column by column, for every boundary coefficient. Inserting Eqs.~\eqref{eq:app_env_beta} and \eqref{eq:app_ift} into Eq.~\eqref{eq:app_total_beta} gives the finite-pressure gradient
\begin{equation}
\begin{aligned}
\nabla_{\mathbf s}\widehat J_f^\star=\;&J_{\mathrm{ref}}\,\nabla_{\mathbf s}\mathcal L\big|_{\mathbf b}\\
&+\Big\langle\nabla_{\mathbf b}J_f,\;\frac{\partial\mathbf B_{\mathrm{pl}}}{\partial\mathbf x}\frac{d\mathbf x}{d\mathbf s}+\frac{\partial\mathbf B_{\mathrm{pl}}}{\partial\mathbf p}\frac{d\mathbf p}{d\mathbf s}\Big\rangle ,
\end{aligned}
\label{eq:envelope_beta}
\end{equation}
whose first term is Eq.~\eqref{eq:envelope} at fixed plasma field. Again the coil response never appears. What the plasma field adds is the response of the equilibrium, which the outer problem already differentiates. The field term costs one vector--Jacobian product through the virtual casing, whatever the number of boundary coefficients. Since the precision plan of the virtual casing is fixed, $\mathbf B_{\mathrm{pl}}$ is a smooth function of the equilibrium state and this product is its exact derivative.

\paragraph{Remark.}
The interpolation from the grid to XCoil's surface points uses fixed angles $(\theta,\phi)$, which are the boundary's own angles in VMEX (the input parametrization is kept at $s=1$). The points themselves move with $\mathbf s$, but that motion acts on $J_f$ only through the surface geometry (points, normals and weights), which is already contained in $\partial J_f^\star/\partial\mathbf s$ at fixed $\mathbf b$. No cross term arises.

\begin{table*}[!t]
\caption{DESC free-boundary equilibria ($l_\rho=m_\theta=n_\phi=16$) of the coil sets of this paper. The $L^2$ force error is $\|\mathbf F\|_{L^2}/\|\nabla(B^2/2\mu_0)\|_{L^2}$, with $\mathbf F=\mathbf J\times\mathbf B-\nabla p$. The normalized residuals $\varepsilon_n$ and $\varepsilon_p$ are defined in Eq.~\eqref{eq:fb_residuals}. A dash marks the two-stage coils of design (f), whose free-boundary equilibrium did not converge.}
\label{tab:desc}
\begin{ruledtabular}
\begin{tabular}{lccccc}
coil set & Sec. & $L^2$ force error [$10^{-4}$] & $\varepsilon_n$ [$10^{-5}$] & $\varepsilon_p$ [$10^{-5}$] & $\langle|\mathbf B\cdot\hat{\mathbf n}|\rangle/\langle|\mathbf B|\rangle$ [$10^{-5}$] \\
\hline
(a) QA, $A=2.5$ & \ref{sec:lowA_tokamak} & 6.0 & 9.6 & 5.9 & 7.4 \\
(b) QA, bootstrap & \ref{sec:qa_nfp1} & 4.0 & 2.3 & 2.5 & 1.4 \\
(c) QA, two-stage & \ref{sec:cold_start_lpqa} & 2.6 & 4.3 & 0.88 & 2.9 \\
(c) QA, cold-start & \ref{sec:cold_start_lpqa} & 0.25 & 0.045 & 0.047 & 0.028 \\
(g) QH & \ref{sec:qh_cold} & 11 & 3.0 & 1.3 & 2.1 \\
(h) QH & \ref{sec:qh_cold} & 9.3 & 1.8 & 1.2 & 1.0 \\
(d) QI, $\langle\beta\rangle=2\%$ & \ref{sec:qi_cold} & 19 & 10 & 5.2 & 6.6 \\
(e) QI & \ref{sec:qi_cold} & 11 & 2.5 & 1.6 & 1.8 \\
(f) QI, two-stage & \ref{sec:qi_cold} & -- & -- & -- & -- \\
(f) QI, cold-start & \ref{sec:qi_cold} & 8.6 & 2.3 & 2.5 & 1.5 \\
NCSX-like & \ref{sec:ncsx78} & 7.5 & 3.1 & 5.0 & 2.2 \\
NCSX c09r00 & \ref{sec:ncsx78} & 12 & 8.9 & 9.9 & 5.8 \\
reactor-like QI & \ref{sec:reactor} & 15 & 7.4 & 6.2 & 4.6 \\
start-up, $\langle\beta\rangle=0$--4\% & \ref{sec:startup} & 7.1--15 & 1.4--6.8 & 1.3--5.3 & 0.99--4.1 \\
LP QA vac., two-stage & \ref{sec:lpqa} & 2.3 & 0.67 & 0.71 & 0.45 \\
LP QA vac., QSS & \ref{sec:lpqa} & 0.25 & 0.031 & 0.026 & 0.020 \\
LP QA $\beta$, two-stage & \ref{sec:lpqa} & 2.3 & 0.92 & 0.72 & 0.71 \\
LP QA $\beta$, QSS & \ref{sec:lpqa} & 1.9 & 0.35 & 0.50 & 0.27 \\
LP QA, 3 coils, AL & \ref{sec:al3} & 0.042 & 0.0074 & 0.0068 & 0.0050 \\
LP QA, 3 coils, QSS & \ref{sec:al3} & 0.040 & 0.0065 & 0.0061 & 0.0043 \\
\end{tabular}
\end{ruledtabular}
\end{table*}

\section{Taylor tests}
\label{app:taylor_tests}

\begin{figure*}[!t]
\includegraphics[width=\textwidth]{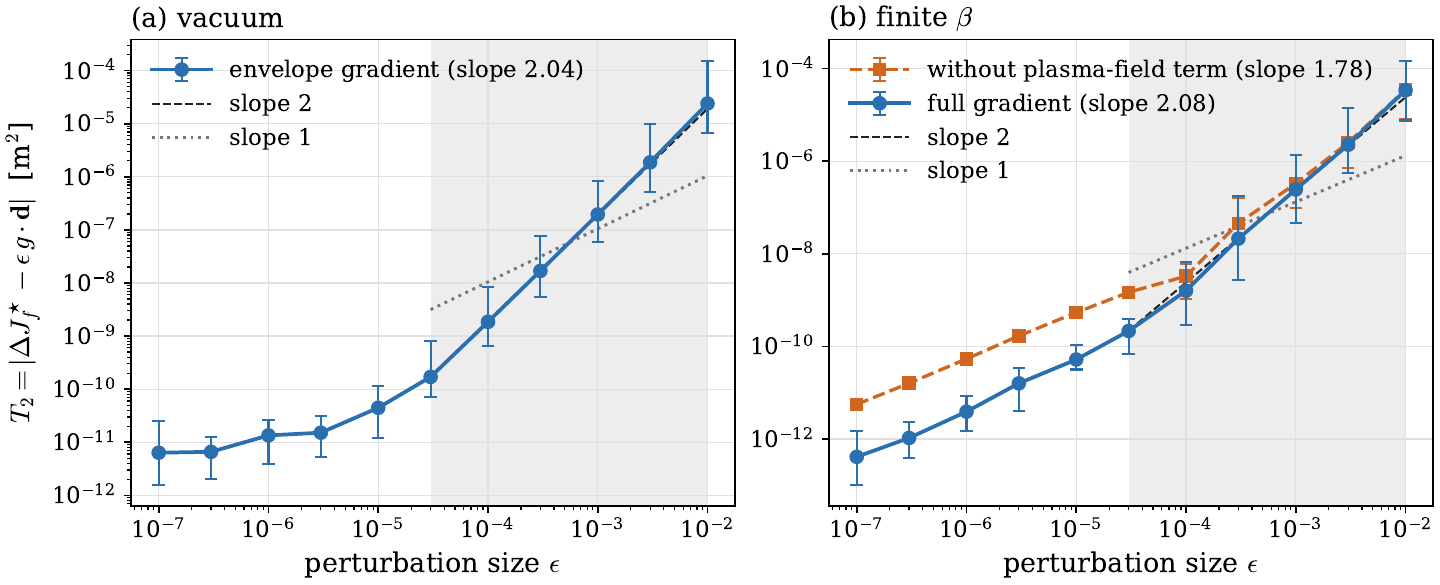}
\caption{Taylor tests of the envelope-theorem gradient along three random draws of a smooth boundary direction $\mathbf d$ without $m=1$ modes. Symbols show the geometric mean of the remainder $T_2$ over the three draws, and the error bars span the three values. Every point re-solves the equilibrium and the coils. The slopes are fitted in the shaded window, and the lines show slope-2 and slope-1 references. (a) Vacuum, at the two-stage optimum of the published LP QA boundary, the vacuum two-stage design of Table~\ref{tab:qa_all}. (b) Finite $\beta$, at the two-stage optimum of the LP QA variant at $\beta=1.05\%$ with a Redl bootstrap current, which is re-optimized for quasisymmetry and is not one of the designs of Table~\ref{tab:qa_all}, with the full gradient of Eq.~\eqref{eq:envelope_beta} and without its plasma-field term.}
\label{fig:taylor}
\label{fig:taylor_beta}
\end{figure*}

We verify Eq.~\eqref{eq:envelope} with a Taylor test at the two-stage optimum of LP QA (Appendix~\ref{sec:lpqa}), where the length and mean-squared-curvature limits are active. We make three random draws of the direction $\mathbf d$ in the space of the boundary coefficients. Each draw is a smooth perturbation of the boundary without the $m=1$ modes because of a technical detail regarding the gauge choice in VMEX. For every draw and every step size $\epsilon$, the equilibrium and the coils are solved again, and we form the remainder $T_2=|J_f^\star(\mathbf s_0+\epsilon\mathbf d)-J_f^\star(\mathbf s_0)-\epsilon\,g\cdot\mathbf d|$, with $g=\nabla_{\mathbf s}J_f^\star$. Figure~\ref{fig:taylor}(a) shows the mean of $T_2$ over the three draws, with error bars that span the three values. For $3\times10^{-5}\le\epsilon\le10^{-2}$, the remainder decreases with a slope of two, for the mean and for each draw, as expected for an exact gradient, whereas an inexact gradient would give a slope of one. As is expected, at smaller $\epsilon$, numerical roundoff begins to contribute significantly. The envelope gradient of Eq.~\eqref{eq:envelope} therefore passes the test. It is exact to the accuracy of the coil solves, and it can be used in the plasma optimization without differentiating through the coil solver.

The vacuum Taylor test of Fig.~\ref{fig:taylor}(a) exercises only the first term of Eq.~\eqref{eq:envelope_beta}. When finite pressure is included, the second term is not small. Along the test directions, it amounts to 7--40\% of $g\cdot\mathbf d$, with the opposite sign to the geometric term. We therefore repeat the Taylor test on a finite-pressure variant of LP QA~\cite{LandremanBullerDrevlak2022}, with $\beta=1.05\%$, $p\propto1-s$ and a self-consistent Redl bootstrap current~\cite{Redl2021} of $-1.15$~MA, at the two-stage optimum on its boundary. Figure~\ref{fig:taylor}(b) shows the result for three random draws of the same kind. For $3\times10^{-5}\le\epsilon\le10^{-2}$, the mean of $T_2$ decreases with a slope of 2.08, against 1.78 without the field term. The three draws separately give slopes of 2.07, 1.96 and 2.21, against 1.46, 1.70 and 2.19 without the field term. The finite-pressure gradient is therefore exact to the accuracy that the coil solves allow. 

\section{Free-boundary equilibria with DESC}
\label{app:desc}
To verify that the free-boundary equilibria of this paper are converged and that their last closed flux surfaces are indeed flux surfaces of the field produced by the coils, we computed every one of them with the free-boundary solver of DESC~\cite{desc_code}. Each equilibrium uses the certified coil set and its currents, the pressure profile and the toroidal flux of the design, with a Fourier--Zernike resolution $l_\rho=m_\theta=n_\phi=16$. Each free-boundary solve starts from the fixed-boundary DESC equilibrium of the corresponding target at the same resolution.

For each equilibrium, Table~\ref{tab:desc} lists the normalized force error reported by DESC, the two normalized free-boundary residuals of Eq.~\eqref{eq:fb_residuals} and the mean normal field on the free-boundary LCFS. 
The mean normal field is evaluated on a $64\times64$ grid per field period. In every case, the LCFS is a flux surface of the total field to a mean relative normal field of $7.4\times10^{-5}$ or better, and the force errors lie between $4.0\times10^{-6}$ and $1.9\times10^{-3}$ in the $L^2$ normalization of Table~\ref{tab:desc}. 

The free-boundary residual of DESC has two parts, both evaluated pointwise on the LCFS. The first is the normal component of the external field, $\mathbf B_{\mathrm{out}}\cdot\hat{\mathbf n}$. The second is the jump of the total pressure across the LCFS, $B_{\mathrm{in}}^2+2\mu_0p-B_{\mathrm{out}}^2$. Here $\mathbf B_{\mathrm{in}}$ is the field of the equilibrium and $\mathbf B_{\mathrm{out}}$ the field of the coils and the plasma currents on the boundary. We  report the area-weighted root-mean-square value of each part, divided by the corresponding mean field quantity on the LCFS,
\begin{equation}
\begin{aligned}
\varepsilon_n&=\frac{\big\langle(\mathbf B_{\mathrm{out}}\cdot\hat{\mathbf n})^2\big\rangle^{1/2}}{\langle|\mathbf B_{\mathrm{in}}|\rangle},\\
\varepsilon_p&=\frac{\big\langle(B_{\mathrm{in}}^2+2\mu_0p-B_{\mathrm{out}}^2)^2\big\rangle^{1/2}}{\langle B_{\mathrm{in}}^2\rangle},
\end{aligned}
\label{eq:fb_residuals}
\end{equation}
where $\langle\cdot\rangle$ is the mean over the LCFS. Both residuals are dimensionless and do not depend on the size of the device or on the evaluation grid. The first measures how closely the LCFS is a flux surface of the external field. The second measures how well the magnetic and plasma pressures balance across it. In every equilibrium of Table~\ref{tab:desc}, both normalized residuals are at most $1.0\times10^{-4}$, and the pressure-balance residual is of the same order as the normal-field residual.

These free-boundary equilibria are the most direct test of how well the coils reproduce a flux surface. The lowest normal fields of Table~\ref{tab:desc} are those of the vacuum LP QA design and of the three-coil design of Appendix~\ref{sec:al3}, and at finite pressure that of the LP QA design at $\langle\beta\rangle=1.05\%$. The lowest comparable values that we found in the literature are about $10^{-5}$. Hudson \emph{et al.}~\cite{Hudson2025shafranov} report a mean normalized normal field of $1.0\times10^{-5}$ on the free-boundary LCFS of a vacuum configuration computed from filamentary coils with VMEC, DESC and SPEC. Their finite-pressure equilibria are compared only with each other. Conlin \emph{et al.}~\cite{conlin2024high} compare the free-boundary residuals of DESC and VMEC at finite pressure only in relative terms. The free-boundary examples distributed with DESC~\cite{desc_code} reach mean normalized normal-field errors of $1.4\times10^{-5}$ in vacuum and $3.6$--$4.8\times10^{-5}$ at finite pressure. We found no finite-pressure free-boundary equilibrium in the literature with a normal-field error below about $3\times10^{-5}$. To the best of our knowledge, the free-boundary residuals reported here are therefore the lowest reported for a stellarator equilibrium computed from a filamentary coil set. They are lower by factors of 50 to 200 in vacuum and, for LP QA at $\langle\beta\rangle=1.05\%$, by more than a factor of ten at finite pressure. The definitions of the normal-field error differ slightly between these works. Most published free-boundary runs also interpolate the coil field from a grid, which can limit the accuracy to about $10^{-5}$, whereas we evaluate the field of the coils directly by Biot--Savart integration.

\begin{figure*}[!t]
\centering
\includegraphics[width=0.8\textwidth]{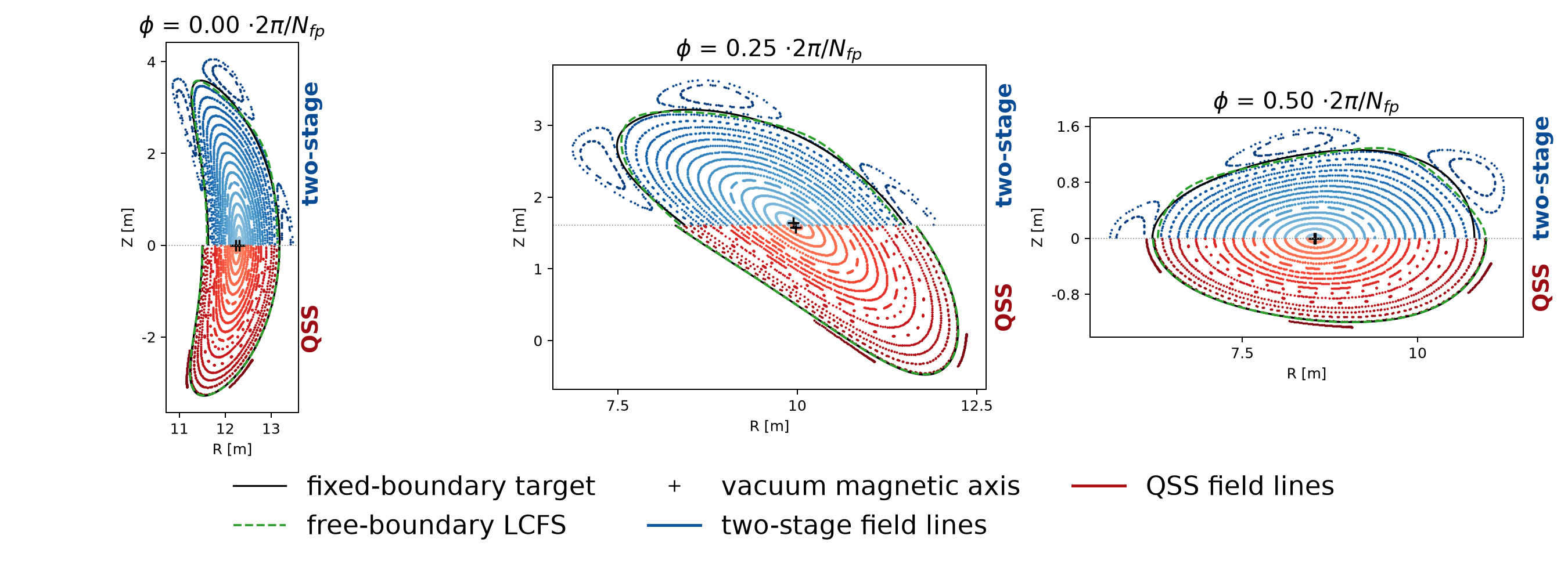}
\caption{Poincar\'e sections of the vacuum coil fields of LP QA at the toroidal angles $\phi=0$, $1/4$ and $1/2$ of a field period. Upper halves, blue: the two-stage coils of the published LP QA boundary. Lower halves, red: the QSS coils of the vacuum design $k=19$. Both are the vacuum coil sets of Table~\ref{tab:qa_all}.}
\label{fig:qa}
\end{figure*}

The Poincar\'e sections of the vacuum coil fields of the warm-started LP QA designs are shown in Fig.~\ref{fig:qa}.

\section{Benchmarks}
\label{sec:benchmarks}
This appendix compares the method with single-stage optimization and with augmented Lagrangian coil design, and compares its computational cost with that of other coupled methods.

\subsection{Comparison with single-stage optimization}
\label{sec:jorge}
We repeat the two vacuum problems of the single-stage optimization of Jorge \emph{et al.}~\cite{Jorge_2023}, a quasi-axisymmetric stellarator with $N_{\mathrm{fp}}=3$, $A=6$ and two coils of 56.2~m per half field period, and a quasi-helically symmetric one with $N_{\mathrm{fp}}=4$, $A=7$ and three coils of 41.7~m, with their coil limits, imposed here as hard constraints. In the workflow of Jorge \emph{et al.}, a stage-one optimization of the boundary and a stage-two optimization of the coils on that boundary are run first, and the single-stage optimization follows. Their single-stage runs therefore start from an equilibrium that is already quasisymmetric and at its targets, and from coils that are already fitted to it. The cost of these two earlier stages is not part of the walltimes we report here. Our optimizations cold-start, with the whole boundary-mode continuation inside the reported time. In free boundary, $f_{\mathrm{QS}}$ is $4.5\times10^{-3}$ against $9.1\times10^{-3}$ for the quasi-axisymmetric case and $15\times10^{-3}$ against $20\times10^{-3}$ for the quasi-helically symmetric one. Appendix~\ref{sec:speedup} compares the time of one fixed-weight solve, the single-stage solve of Jorge \emph{et al.}, against one outer solve of our method at the same boundary resolution. Even with a cold-start, each of our solves is several times shorter. Their weighted formulation meets the limits on the coil--coil distance and the mean-squared curvature approximately, while the hard constraints used here meet all limits.

\subsection{Comparison with augmented Lagrangian coils}
\label{sec:al3}
We also compare the method with the augmented Lagrangian coil design of Refs.~\cite{auglag_prl, auglag_pre}, which fits three coils per half field period to the original LP QA boundary. 
Starting from a cold-start, in a single run, we impose the same coil count, total length, and curvature, the mean-squared curvature and coil distances reached by those coils as limits, and a limit on the coil force at the value they reach. The plasma is constrained only by quasisymmetry, $A=6$ and the mean rotational transform of LP QA, without any shape constraint. The first boundary modes are optimized with a loose bound on the coil fit, which lets the plasma settle into a less elongated quasi-axisymmetric shape, and the bound is then tightened at the last boundary mode. Table~\ref{tab:lpqa3} compares the two designs. 
At the same length, curvature and force, and with fewer degrees of freedom (a coil Fourier order of 5 against 16), the QSS coils are equal or better in every coil metric of Table~\ref{tab:lpqa3}, while confining the alphas equally well within the sampling error and exhibiting similar free-boundary quasisymmetry errors. The fixed-boundary error of the augmented Lagrangian design is orders of magnitude lower but it degrades dramatically during the free-boundary calculation. With three coils, i.e. in a coil ripple dominated regime, the free-boundary quasisymmetry is set by the field error of the coils rather than by the target plasma.

\begin{table}[!t]
\caption{Comparison with the augmented Lagrangian (AL) coils of Refs.~\cite{auglag_prl, auglag_pre} for LP QA with three coils per half field period, at the same total length, curvature limit, coil force and total current.}
\label{tab:lpqa3}
\begin{ruledtabular}
\begin{tabular}{lcc}
 & AL & QSS \\
\hline
start & LP QA & cold-start \\
$N_c$ / total coils / Fourier order $M$ & 3 / 12 / 16 & 3 / 12 / \best{5} \\
$\bn$ [\%] & 0.082 & \best{0.049} \\
$\kappa_{\max}$ / $\kappa_\mathrm{MSC}$ [m$^{-1}$ / m$^{-2}$] & 0.53 / 0.051 & 0.53 / \best{0.040} \\
$\bar\tau$ / $\gamma_c$ [m$^{-1}$ / $10^{-3}$~m$^{-2}$] & 0.99 / 11 & \best{0.37} / \best{5.7} \\
$d_{cc}$ / $d_{cs}$ [m] & 1.48 / 2.74 & 1.48 / 2.76 \\
fixed-boundary $f_{\mathrm{QS}}$ [$10^{-5}$] & \best{0.06} & 53 \\
$f_{\mathrm{QS}}$ [$10^{-3}$] & \best{1.0} & 1.2 \\
$\delta_\mathrm{LCFS}$ mean / max [mm] & 8.0 / 31.7 & \best{3.2} / \best{9.6} \\
$\alpha$-loss [\%] & $11.2\pm0.4$ & $12.0\pm0.4$ \\
\end{tabular}
\end{ruledtabular}
\end{table}

\subsection{Speed-ups}
\label{sec:speedup}
As mentioned previously, one of the contributions of this work is a solver that is able to rapidly and simultaneously optimize the coils and the equilibrium. 
The complete vacuum designs of Appendix~\ref{sec:jorge}, which start from a cold-start and include the whole boundary-mode continuation and the final coil stage, take 0.95--1.1 GPU-hours. In our reruns of their code, the method of Jorge \emph{et al.} takes 2.3--3.5 CPU core-hours, starting from an equilibrium that an earlier stage-one optimization had already made quasisymmetric and from coils that a stage-two optimization had already fitted to it. The cost of these two stages is not included. The global-optimization workflow of Ref.~\cite{Giuliani_2024} takes 24--48 CPU core-hours per design. These figures compare GPU and CPU hours for different problems and resolutions, and indicate orders of magnitude only. We make no claims that the other methods cannot be made more computationally efficient, or that we have compared these codes at optimal settings.

\bibliography{qss}

\end{document}